\documentclass[%
aip,
 amsmath,amssymb,
reprint,%
]{revtex4-1}

\usepackage{graphicx}
\usepackage[utf8]{inputenc}
\usepackage[T1]{fontenc}
\usepackage{hyperref}
\usepackage[a4paper]{geometry}
\usepackage{units}
\usepackage{textcomp}
\usepackage{mathrsfs}
\usepackage{stackrel}
\usepackage{wasysym}
\usepackage{comment}
\usepackage{babel}
\usepackage[mathlines]{lineno}% Enable numbering of text and display math
\makeatletter
\def\@email#1#2{%
 \endgroup
 \patchcmd{\titleblock@produce}
  {\frontmatter@RRAPformat}
  {\frontmatter@RRAPformat{\produce@RRAP{*#1\href{mailto:#2}{#2}}}\frontmatter@RRAPformat}
  {}{}
}%
\makeatother

\begin{document}
%\preprint{AIP/123-QED}

\title[Second sound resonators and tweezers as vorticity or velocity probes]{Second sound resonators and tweezers as vorticity or velocity probes : fabrication, model and method}
% Force line breaks with \\

\author{Eric Woillez}
 \altaffiliation[present affiliation: ]{CEA-Liten, Grenoble.}%Lines break automatically or can be forced with \\
\author{Jérôme Valentin}%
 \altaffiliation[present affiliation: ]{Observatoire de Paris - PSL, CNRS, LERMA, F-75014, Paris, France.}%Lines break automatically or can be forced with \\
 \author{Philippe-E. Roche}%
 \altaffiliation[Corresponding author]{}%Lines break 
\affiliation{ 
Univ. Grenoble Alpes, CNRS, Institut NEEL, F-38042 Grenoble, France%\\This line break forced with \textbackslash\textbackslash
}%
\date{\today}% It is always \today, today,

\begin{abstract}

An analytical model of open-cavity second sound resonators is presented and validated against simulations and experiments in superfluid helium using a new resonator design that achieves unprecedented resolution. The model incorporates diffraction, geometrical misalignments, and flow through the cavity, and is validated using cavities with aspect ratios close to unity, operated up to their 20th resonance in superfluid helium.

An important finding of this study is that resonators can be optimized to selectively sense either the quantum vortex density carried by the throughflow -as typically done in the literature- or the mean velocity of the throughflow. We propose two velocity probing methods: one that takes advantage of geometrical misalignments between the tweezers plates, and another that drives the resonator non-linearly, beyond a threshold that results in the self-sustainment of a vortex tangle within the cavity.

A new mathematical treatment of the resonant signal is proposed to adequately filter out parasitic signals, such as temperature and pressure drift, and accurately separate the quantum vorticity signal. This \emph{elliptic method} consists in a geometrical projection of the resonance in the inverse complex plane. Its effectiveness is demonstrated over a wide range of operating conditions.

The resonator model and elliptic method are being utilized to characterize a new design of second-sound resonator with high resolution thanks to miniaturization and design optimization. These \emph{ second-sound tweezers} are capable of providing time-space resolved information similar to classical local probes in turbulence, down to sub-millimeter and sub-millisecond scales. The principle, design, and micro-fabrication of second sound tweezers are being presented in detail, along with their potential for exploring quantum turbulence.

\end{abstract}

\maketitle

\tableofcontents{}

\section{Introduction to second sound resonators}

\subsection{Quantum fluids and second sound}

Below the so-called lambda transition, liquid $^{4}$He undergoes a quantum state change to He\:II. This transition occurs at around $T_{\lambda}\simeq2.18\:K$ under saturated vapor conditions. According to the Tisza and Landau's \textit{ two-fluid model}, the hydrodynamics of He\:II can be described as the hydrodynamics of two interpenetrating fluids, namely the superfluid component and the normal fluid component \cite{balibar2007discovery,Griffin2009}.

The superfluid density $\rho_{s}$ vanishes immediately below the transition and increases as temperature decreases, while the normal fluid density $\rho_{n}$ exhibits the opposite behavior and vanishes in the zero-temperature limit. The properties of the two fluids differ markedly. The superfluid has zero viscosity and entropy and the circulation of its velocity field is quantized in units of $\kappa = h/m \simeq0.997.10^{-7}m^{2}/s$, where $h$ is the Planck constant and $m$ is the atomic mass of $^4$He. This quantization constraint results in the formation of filamentary vortices with Ångström-scale diameters, later referred to as superfluid or quantum vortices \cite{DonnellyLivreVortices}. In contrast, the normal fluid follows classical viscous dynamics and carries all the entropy of He\:II \cite{putterman1974,Khalatnikov:Book}.

The presence of two distinct velocity fields $\mathbf{v}_{s}$
and $\mathbf{v}_{n}$ in the superfluid and normal fluid respectively, leads to the existence of two independent sound modes in He\:II. This can be demonstrated by linearizing the equations of motion \cite{putterman1974,Khalatnikov:Book,donnellyPhysToday2009}. The ``first sound'' mode corresponds to a standard acoustic wave, where both fluids oscillate in phase ($\mathbf{v}_{s}=\mathbf{v}_{n}$), resulting in oscillations of the local pressure and density $\rho=\rho_{s}+\rho_{n}$. On the other hand, the ``second sound'' mode corresponds to both fluids oscillating in antiphase with no net mass flow, i.e., $(\rho_{s}\mathbf{v}_{s}=-\rho_{n}\mathbf{v}_{n})$. As a result, the \emph{relative} densities of the superfluid and normal fluid locally oscillate along with the entropy and temperature.

\subsection{Generation and detection of second sound waves}

Experimentally, two techniques are commonly used for generating and detecting second sound in He\:II: mechanical and thermal. In principle, they could be combined for generating and detecting, although we are not aware of any composite configuration reported in the literature. Other techniques, such as optical scattering and acoustic detection above the liquid-vapor interface or within the flow itself, have been occasionally used but will not be discussed here (e.g., \onlinecite{peters1970brillouin,Lane1947,heiserman1976acoustic}).

The \emph{mechanical technique} involves the excitation and sensing of one component of He\:II exclusively at the transducer surface, which can be either its superfluid component or the normal fluid component. This boundary condition involves first and second sound simultaneously, and their combination produces an exact compensation of the motion of the component that remains static at the transducer surface.
Given that the velocity of first sound is typically an order of magnitude larger than that of second sound\cite{donnelly1998observed}, their effects can be distinguished, and the entanglement of both sounds is generally not considered an issue in second sound probes.
In practice, the selective displacement of the superfluid component at the transducer surface is achieved using \textit{Peshkov transducers}, which consist of a standard acoustic transducer and a fixed porous membrane-filter. The tiny pores of the membrane act as viscous dampers for the normal fluid but are transparent to the superfluid (\textit{``superleaks''})\cite{Peshkov:1948,Henjes:1988aa}. Alternatively, selective displacement of the normal fluid component is achieved using \textit{oscillating superleak transducers}. These are based on a vibrating porous membrane that is coupled to the motion of the normal fluid through viscous forces but decoupled from the inviscid superfluid. They can be made by replacing the membrane of a loudspeaker or microphone with a millipore or nucleopore sheet\cite{WILLIAMS:1969,Sherlock:1970p374,DHumieres:1980p448}.

The \emph{thermal technique} for generating and detecting second sound involves inducing second sound by the Joule effect and detecting it with a thermometer. Depending on the temperature range and practical considerations, various types of thermometers may be suitable for detecting second sound waves. Given the vast literature, we only list a few thermistor materials and bibliographic entry points. Materials with negative temperature coefficients\footnote{Phosphor bronze wire, a positive temperature coefficient thermometer, has also been used in the early days \cite{Peshkov1946}.} include carbon in various forms (\textit{aquadag} paint, fiber, pencil graphite, etc.) \cite{Hilton2001}, doped Ge \cite{Snyder:RSI1962}, RuOx \cite{yang2018decay}, ZrN${x}$/Cernox \cite{yotsuya1997,Fuzier:2004p373}, and Ge-on-GaAs \cite{Mitin:2007p528,midlik2021transition}. Transition edge superconductor thermometers are  preferred when large sensitivity or low resistivity is important, for instance Au${_2}$Bi \cite{NotarysPhD}, PbSn \cite{crooks1983,Rinberg:2001}, granular Al \cite{Cohen1968,Mathieu1976}, and AuSn \cite{NotarysPhD,laguna1976photolithographic,borner1983AuSn}. The second sound tweezers presented in this study utilize this thermal technique, with a AuSn thermometer. More information will be provided on this material in section \ref{subsec:thermometrie}

\subsection{From macroscopic second sound sensors to microscopic tweezers}

In the presence of superfluid vortices, there is mutual coupling between the superfluid and normal fluid, leading to the damping of second sound waves \cite{DonnellyLivreVortices}. This attenuation has been extensively utilized as a tool for exploring the properties of He\:II flows over the past 60 years \cite{Vinen:1957p399}, especially in the field of quantum turbulence (see, for example, \onlinecite{varga2019}).  Mechanical second sound transducers have been successfully employed to sense the turbulence of He\:II in the wake of a grid by groups in Eugene, Prague, and Tallahassee \cite{stalp2002,Babuin:EPL2014,mastracciRSI2018}. Meanwhile, thermal second sound transducers successfully used to sense turbulence of He\:II flows have been described, for instance, by groups from Paris, Tallahassee, Grenoble, and Gainesville\cite{Wolf:Couette1981,Holmes1992,roche2007vortex,yang2018decay}.

A specific type of probe allows for very sensitive probing of the density of quantum vortices in He\:II flow: standing-wave second sound resonators. Such resonators consist of two parallel plates facing each other, with one plate functioning as a second sound emitter and the other as a receiver. The emitter excites the cavity at resonance to benefit from the amplification of the cavity. The characteristics of the standing wave between the plates provide information on the properties of the fluid and the flow between the plates, particularly the density of vortex lines, which affects the amplitude of the standing wave. In addition to vortex density measurements, second sound can also provide information on the fluid temperature, as the second sound velocity depends on it, and on the velocity of the background He\:II flow when it induces a phase shift or Doppler effect on the second sound (see, for example, \onlinecite{Dimotakis:1977p408,Wolf:Couette1981,woillez2021vortex}).

\begin{figure}
%\begin{centering}
\includegraphics[width=0.5\textwidth]{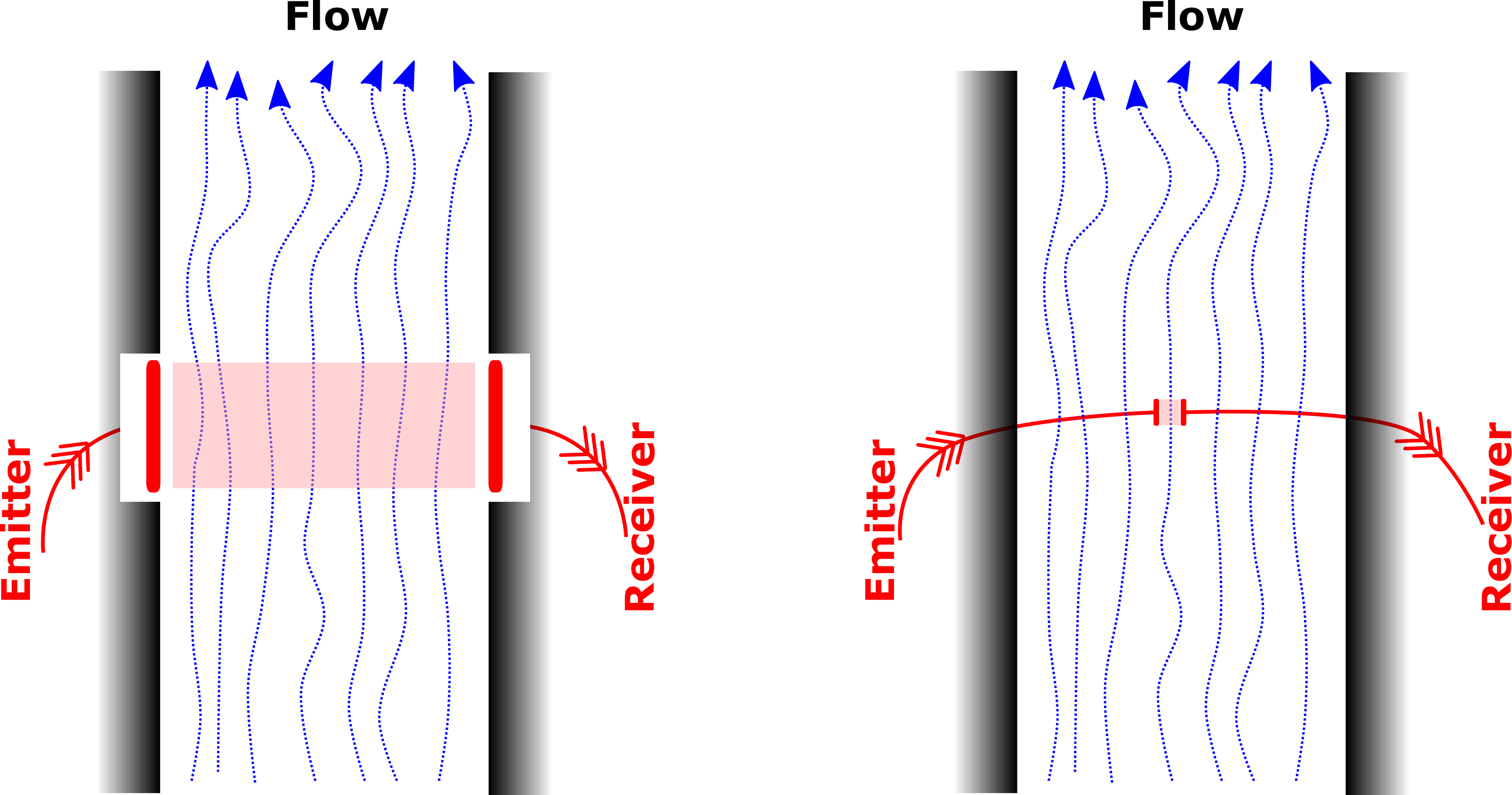}%[height=6cm]
%\par\end{centering}
\caption{\textbf{Left:} A macroscopic second sound resonator, embedded in the sidewall, is used to sense the averaged flow properties in the shaded region. \textbf{Right:} Second sound tweezers. In contrast to the macroscopic design of the left schematic, this miniaturized resonator, positioned within the flow, allows for space and time resolution of the flow variations. \label{fig:intro}}
\end{figure}

The characteristics of standard (macroscopic) second sound resonators listed above also apply to their miniaturized counterparts, named second sound tweezers. In addition to their smaller size, a key feature of tweezers is their minimal impact on the flow when positioned in its core. This is illustrated in Figure \ref{fig:intro}, which highlights the differences between standard resonators and tweezers. Standard resonators provide information on the averaged properties of the flow, while tweezers enable space and time-resolved measurements. As such, tweezers function as local probes, similar to hot-wire anemometers or cold-wire thermometers used in turbulence studies.

The present design for tweezers utilizes thermal actuation, which is preferred over mechanical actuation due to its compatibility with the constraints of miniaturization and reduced flow blockage.

\subsection{Overview of the manuscript}

The following sections cover distinct topics.

In Section \ref{sec:Models-of-second}, we present a comprehensive model of second-sound resonators that accounts for plate misalignment, advection, finite size, and near-field diffraction. Diffraction, which was previously neglected in quantitative models, is shown to be a significant source of degradation of the quality factor in our case studies. We also consider applications of this model for the measurement of vortex concentration or velocity.

In Section \ref{sec:Measurements-with-second}, we discuss existing methods for processing the signal from second-sound resonators and their limitations. To overcome these limitations, we introduce a new general approach, called the \emph{elliptic method}, based on mathematical properties of resonance. This method enables us to dynamically separate the amplitude variations of the standing wave due to variations of vortex density or velocity from the phase variations, such as those resulting from variations of the second sound velocity due to a temperature drift.

In Section \ref{sec:Design}, we report on the design, clean-room fabrication, and operation of miniaturized second-sound resonators called. These \emph{tweezers} allow us to probe the throughflow of helium with unprecedented spatial and time resolution.

To ensure clarity, we present the second-sound tweezers first to illustrate the topics on modeling and methods with a practical case. However, we emphasize that the modeling and methods introduced in this article are general and relevant to second-sound resonators, regardless of their size, including the macroscopic sensors embedded in parallel walls that are encountered in the literature.

\section{\label{sec:Design}Design, fabrication and mode of operation of second
sound tweezers}

\begin{figure*}
\includegraphics[width=0.9\textwidth]{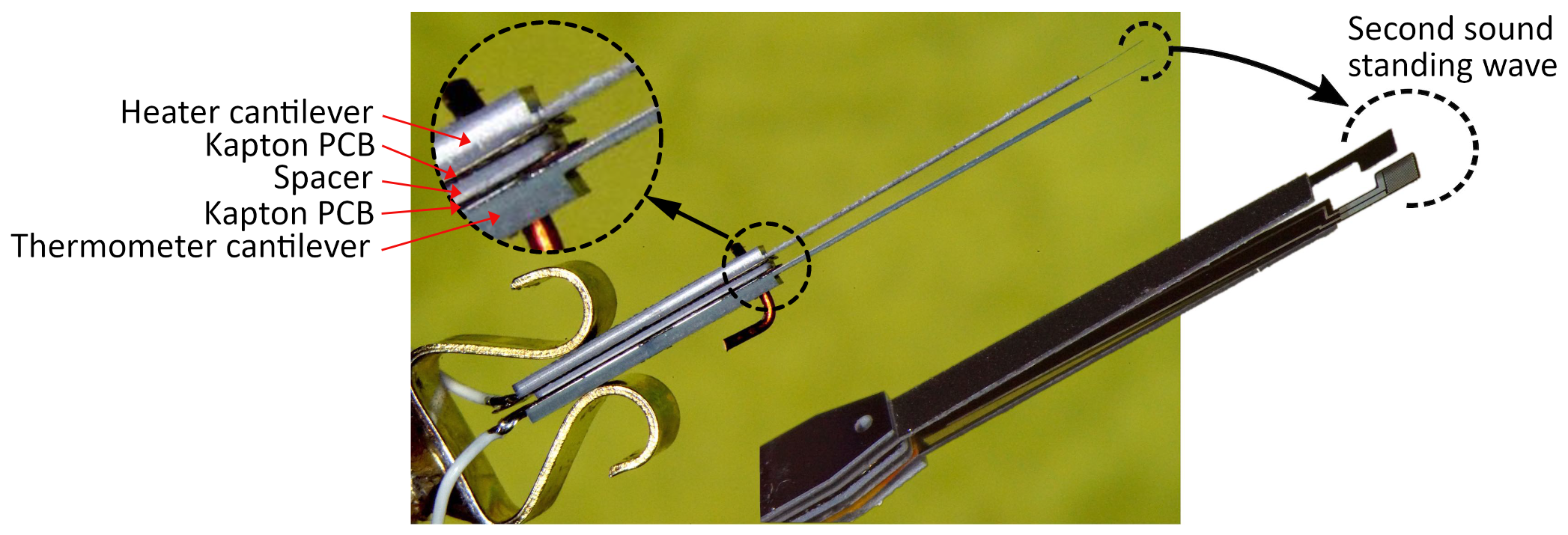}
\caption{Second sound tweezers, pictured from two angular perspectives.
\textbf{Center}:
The probe is seen in the direction of the mean flow (or nearly so). The second sound cavity is localized at the end of the probe tip, in the rightmost encircled areas. The bent copper wire passing through the probe is a temporary \textit{joystick} used for precise alignment of the cavity plates under the microscope. The two coaxial cables for heating and thermometry are visible in the lower left corner of the picture.
\textbf{Top left}:
Magnified detail of the stack.
\textbf{Right}:
The picture insert shows the same probe after rotation, and with the \textit{joystick} removed, revealing the through-hole across the silicon stack. The two staggered notches used for thermal confinement of the standing wave in the cavity are clearly visible.
\label{fig:MechanicalDesign_Global}}
%\begin{centering}
%\includegraphics[width=10cm]{figure_Schema_tweezer.pdf}
%\par\end{centering}
%\caption{Schematic side view of the constitutive stack of second sound tweezers  (the pieces are shown separated for illustration, their thicknesses are exaggerated). The active areas, the emitter and receiver plates, are constituents of the tip. \label{fig:jerome2}}
\end{figure*}

\begin{figure}
%\begin{centering}
\includegraphics[width=0.5\textwidth]{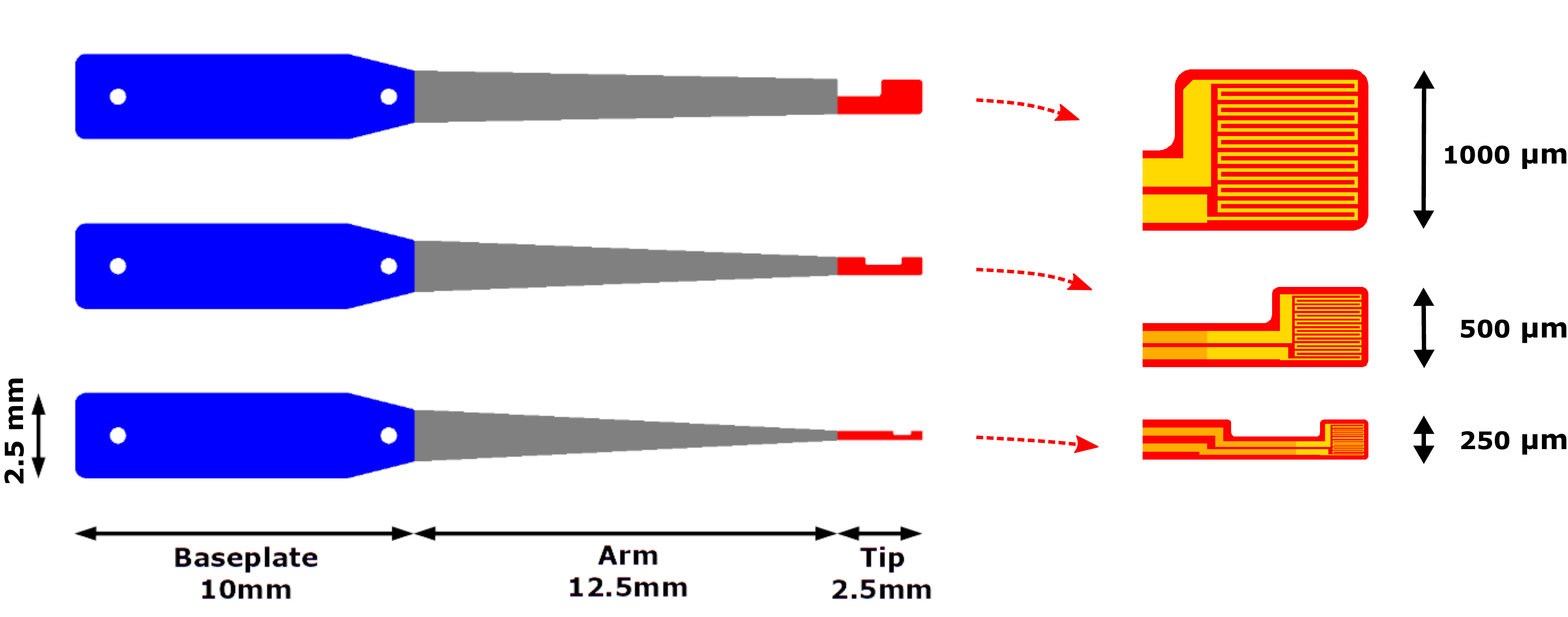}
%\hspace{1.0cm}
%\par\end{centering}
\caption{Top view of the 3 cantilever types. The tips widths are respectively 1000\,µm, 500\,µm and 250\,µm.
\textbf{ Left}: Mechanical structures, all parts are silicon made, different thicknesses are represented by different colors. The baseplate width is 2.5\,mm for all types.
\textbf{ Right}:
Electrical path on each tip type. Yellow areas are a deposition of TiPt for heaters and AuSn for thermometers. Orange areas are a thick AuPt deposition for current leads.
 \label{fig:jerome3}
}
\end{figure}

The basic component of second sound tweezers consists of a stack comprising a heating cantilever and a thermometer cantilever, separated by a spacer. Additionally, two Kapton films with golden copper tracks are inserted in contact with the tracks of the heater and thermometer (see Fig. \ref{fig:MechanicalDesign_Global}). The heaters and thermometers cantilevers are composed of a baseplate, an elongated arm, and a tip (see Fig. \ref{fig:jerome3}). The baseplate is the thickest part, while the tip is the thinnest. The active areas, which are the emitter and receiver plates, are located on the tips. The distance between the plates is set by the spacer, composed of one or several micro-machined silicon elements. For a given device, the heater and thermometer have identical mechanical structures, with the only difference being the chemical elements used in the serpentine electrical path deposited on the tip. Three cantilever types were fabricated to allow the assembly of resonators with three different tip sizes (see Fig. \ref{fig:jerome3}). The tip widths are 1000\,µm, 500\,µm, and 250\,µm. The resulting assembly is clamped with a standard picture clip, which was downsized in width by electro-wire erosion, and soldered to the head of a mounting screw. An improvement compared to the clamping technique introduced in \onlinecite{roche2007vortex} is the possibility to insert a temporary "joystick" through the entire assembly to allow for precise alignment (or offsetting) of the cavity plates under a microscope.

Next subsection \ref{subsec:Mechanical} presents the considerations that have
prevailed in the mechanical design of the second sound tweezers.  The following section, \ref{subsec:Second-sound-detection}, discusses the detection (thermometry) and generation (heating) of second sound by the tweezers. The last two subsections present the microfabrication techniques (section \ref{subsec:Microfabrication-and-assembling}) and the electrical circuitry used to operate the probes (section \ref{subsec:Electric-circuit}).

\subsection{\label{subsec:Mechanical}Mechanical design}

%\textbf{Resolution}. The tweezers space-resolution $L_{\text{res}}$ is set by the largest dimension of its cavity, which can be either the inter-plates distance $D$, also called the ``gap'', or the side length $L$ of the plates here assumed to be squared shaped. The present study mostly focuses on cavities with an aspect ratio of order 1, to benefit from optimal space averaging of the signal at given space-resolution. The tweezers time-resolution $\tau_{\text{res}}$ is set by the decay-time of a wave bouncing between the plates. In section \ref{par:Empirically-modified-Fabry=002013Perot}, we introduce and validate a simple model accounting for dissipation in the cavity due to diffraction loss and residual inclination of the plates. An upper bound for $\tau_{\text{res}}$ is obtained  from the diffraction loss term: $\tau_{\text{res}}\simeq L^{2}f/bc_{2}^{2}$, where $b\approx0.38$, $c_{2}$ is the second sound velocity and $f$ is the wave frequency which can be approximated as $nc_{2}/2D$ for the $n^{th}$ mode of resonance (see eq.\ref{eq:classical Fabry-Perot}). Thus, the tweezers time-resolution due to diffraction loss can be estimated as
\textbf{Resolution}. The space resolution of the tweezers, denoted as $L_{\text{res}}$, is determined by the largest dimension of its cavity, which can be either the inter-plates distance $D$, also known as the "gap", or the side length $L$ of the plates which are assumed to be square-shaped. This study mostly focuses on cavities with an aspect ratio of order 1, which allows for optimal space averaging of the signal at a given space resolution. The time resolution of the tweezers, denoted as $\tau_{\text{res}}$, is set by the decay time of a wave bouncing between the plates. In Section \ref{subsec:Analytical-approximations}, we introduce and validate a simple model that accounts for dissipation in the cavity due to diffraction loss and residual inclination of the plates. An upper bound for $\tau_{\text{res}}$ is obtained from the diffraction loss term: $\tau_{\text{res}}\simeq L^{2}f/bc_{2}^{2}$, where $b\approx0.38$, where $c_{2}$ is the second sound velocity, and $f$ is the wave frequency, which can be approximated as $n c_{2}/2D$ for the $n^{th}$ mode of resonance (see Eq. \ref{eq:classical Fabry-Perot}). Thus, the tweezers time resolution due to diffraction loss can be estimated as
\[
\tau_{\text{res.}}\simeq\frac{L^{2}f}{bc_{2}^{2}}\simeq n\frac{L^{2}}{Dc_{2}}
\]

The ratio $L_{\text{res}}/\tau_{\text{res}}$ of the space resolution and time resolution defines a characteristic velocity for which the probe optimally averages the space-time fluctuations. For instance, in cavities of aspect ratio one ($D=L$) operated on its $n^{th}$ resonance, the nominal velocity is estimated as $c_{2}/n$. These estimates show that cavities with an aspect ratio of order unity are appropriately sized for second-sound-subsonic flow, with a mean velocity of a few meters per second.

\textbf{Blocking effect}. The aforementioned considerations regarding space resolution are pertinent only if the flow being measured is not disturbed by the probe support. The current design adheres to the empirical $10\times$ rule, which mandates that components of the support that impede the flow on a length scale $\mathcal{X}$ must be situated at least $10\mathcal{X}$ away from the measurement region. Accordingly, the cavity is located at the end of elongated arms, and the cantilevers exhibit decreasing widths and thicknesses along their length of 25 mm, as depicted in Figure \ref{fig:MechanicalDesign_Global}. The thickness successive values are around 520$\mu m$, 170$\mu m$ and 20$\mu m$ while the width decreases from 2.5mm to $145\:\mu m$ in the narrowest zone (resp. $275\:\mu m$ and $500\:\mu m$) for cavities with $L=250\,\mu m$ (resp. $L=500\,\mu m$ and $L=1000\,\mu m$).

\textbf{Wave confinement}: The spatial resolution of the tweezers would be degraded if the second sound standing wave spread out of the $L\times L\times D$ cavity due to reflections between the supporting arms. To confine the standing wave in the cavity region, a design trick was implemented by breaking the mirror symmetry between the two cantilevers. As shown in Fig. \ref{fig:MechanicalDesign_Global}, anti-symmetric notches in the tips prevent the second sound from escaping by bouncing away from the cavity, at least in the geometric-optic approximation where diffraction is neglected.

\textbf{Mechanical resonances}. In addition to the $\times10$ rule, these dimensions are chosen to push the mechanical vibrations of the arm to around $1$\:kHz or higher. The fundamental resonance frequency of the trapezoidal-shaped arm in vacuum was estimated using the analytical formula in \onlinecite{lobontiu2007dynamics} (section 1.3.1.1).

\[
f_{0}=\frac{8.367}{2\pi}\frac{e}{t^{2}}\sqrt{\frac{E_{Si}(3w_{2}+w_{1})}{\rho_{Si}(49w_{2}+215w_{1})}}
\]

We obtain $f_{0}=2195\:$Hz (resp. 1889\:Hz and 1569\:Hz) using the material properties $E_{Si}=140$\:GPa, $\rho_{Si}=2330\:$kg$/$m$^{3}$, and the dimensions of the intermediate section of the arm having thickness $e=172\:\mu$m, length $t=12.5\:$mm, and width decreasing from $w_{2}=1.5\:$mm to $w_{1}=250\:\mu$m (resp. to $500\:\mu$m and $1000\:\mu$m). An experimental validation was conducted at room temperature in air with an arm having $w_{1}=1000\:\mu$m. Its mechanical vibration frequency spectrum was measured using a photoreceptor that detected a laser beam reflecting off the arm. The mechanical excitation was provided either by tapping the table supporting the set-up with a small hammer or by directing a jet of compressed air toward the arm. In both cases, the fundamental mechanical resonance frequency was found to be 1215\:Hz, in reasonable agreement with the predicted value of 1569 Hz given the uncertainty in the Young's modulus $E_{Si}$ and deviations from the trapezoidal shape. As discussed in Section \ref{subsec:Filtering-the-vibration}, indirect measurements of the resonance frequency were conducted in a superfluid flow with a velocity of $1.2$\:m$/$s, and gave $f\approx825$\:Hz, $f\approx1050$\:Hz, and an amplitude of vibration smaller than $1\:\mu$m. The decrease in frequency compared to the room-temperature measurement is interpreted as being mostly due to a fluidic added mass effect \cite{Sader_JApplPhys98}.

\textbf{Deflection of the tips' ends.}

The thicknesses of the tweezers parts are chosen such that the mechanical deflection at the tip endpoint remains significantly lower than the inter-plate distance under typical operating conditions.

The deflection at the tip endpoint can be estimated by considering separately the arm deflection (with thickness 172$\:\mu$m) and the tip deflection (with thickness 20$\:\mu$m).
As a first approximation, both arm and tip are considered as cantilever beams of uniform width submitted to a uniformly distributed load, and having one embedded end and one free end. This geometrical approximation overestimates the deflection of the arm, as its endpoint is narrower than its base, and it underestimates the deflection of the tip, as the notch is ignored. Nevertheless, this provides order-of-magnitude estimates. The load is estimated as the dynamic pressure of a liquid helium flow impinging the tweezers in the transverse direction at a velocity $U=0.1\:$m/s, which is 10\% of the typical longitudinal flow velocity of 1 m/s. The dynamic pressure $P$ is given by:
\[
P=\frac{1}{2}\rho U^2
\]

where the liquid helium density is $\rho \simeq 140\:$kg/m$^3$. According to Euler-Bernouilli beam theory, the free end deflection $\delta_{\text{max}}$ of the cantilever is given by:
\[
\delta_{\text{max}} = \frac{3}{2} \frac{P t^4}{E_{Si}.e^3}
\]

The total deflection of the tweezers (arm and tip) can be upper-bounded by considering the sum of the deflection of a 2.5\:mm long tip and a 15\:mm (not 12.5\:mm) long arm. This takes into account the small angle generated on the tip by the arm deflection. Using the values $t=15\: \text{mm}$ (length), $e=172\:\mu\text{m}$ (thickness), and $E_{Si}=140\:\text{GPa}$ (Young's modulus), the deflection of the arm endpoint is found to be 75\:nm, while the deflection of the tip endpoint with $t=2.5\:\text{mm}$ and $e=20\:\mu\text{m}$ gives a deflection of 37\:nm. Thus, the total mechanical deflection of the tweezers tip due to a steady lateral flow of 0.1\:m/s is a fraction of a micron, which is decades smaller than the inter-plate distance.

The mechanical resonance of the tweezers arm and tip discussed above could lead to deflections larger than those due to steady forcing. The amplitude of these mechanical oscillations was measured in a turbulent He\:II flow, up to velocities exceeding 1\:m/s, taking advantage of the dependence of the second sound resonance with respect to the cavity gap. The measured signal will be presented to illustrate the efficiency of the elliptic projection method in separating the fluctuations of the acoustical path of the cavity and the fluctuations of the bulk attenuation of second sound between plates. The mechanical oscillations of the cavity gap are found to be typically $0.5\:\mu\text{m}$ (around 1 kHz). As expected, such a deflection is decades lower than the interplate distance (1.3\:mm in this case) and the second sound wavelength.

\textbf{Boundary layer}. In the presence of a mean flow through the cavity, a velocity boundary layer will develop along the tweezers' plates. In principle, this boundary layer could contribute to the measured signal and alter the measurement of the incoming flow. For instance, it could increase the density of superfluid vortices in the cavity and therefore affect second sound attenuation. However, as illustrated later, the second sound standing wave that settles between the plates has nodes of velocity near the plates while the sensitivity of second sound to vortices arises in antinodal regions of velocity. As long as the boundary layer thickness is thin enough, say within a fraction of $\lambda/4$ ($\lambda=c_{2}/f$ is the second sound wavelength), it is not expected to significantly alter the measured signal.

A first requirement for this condition is that the mean flow direction is parallel to the plates, so that the flow penetrates through the cavity with minimal deflection. A consequence of this is that the plates should be widely separated when operated in flows with undefined or zero mean velocity, such as the core of a mixing layer.

A second requirement is that the plate thickness is much thinner than $\lambda/4$. The current plates are 20\:microns thin, which is much smaller than $\lambda/4\simeq D/2n$ for the $n^{th}$ mode of resonance. For instance, with $D=500\:\mu m$ and $n=3$, the condition $20\ll\lambda/4\simeq83\:\mu$m is indeed  satisfied.

A third condition pertains to the downstream development of the boundary layer thickness, which should also remain within $\lambda/4$. The physics of boundary layers in He\:II is not yet well-understood \cite{stagg2017}, but existing experiments (e.g. \onlinecite{Smith1999}) suggest that classical hydrodynamic phenomenology could remain valid in the high-temperature limit. In classical hydrodynamics, the so-called \textit{displacement thickness} of a laminar Blasius boundary layer at a distance $L$ from its origin is given by
\[
\delta_{\text{{bl}}}=1.73\sqrt{\frac{L\nu}{U}}
\]

where $U$ is the mean velocity far from the boundary layer and $\nu$ is the kinematic viscosity of the fluid. In He\:II, several diffusive coefficients could arguably play the role of $\nu$, including the quantum of circulation around a quantum vortex and the kinematic viscosity associated with the dynamics viscosity of the normal fluid normalized either by the normal fluid density or by the total density. In the temperature range of interest, all these diffusive coefficients are within one order of magnitude, typically $10^{-8}-10^{-7}$m$^{2}$/s. Taking $\nu=3.10^{-8}\:$m$^{2}$/s, $L=1000\:\mu$m, and $U=0.5\:$m/s, one finds $\delta_{\text{bl}}=13\:\mu$m, and a boundary layer Reynolds number $\delta_{\text{bl}}U/\nu=217$ consistent with the laminar picture. This thickness estimate, similar in magnitude to the plate thickness, satisfies the third requirement $\delta_{\text{bl}}\ll\lambda/4$.

\subsection{\label{subsec:Second-sound-detection}Second sound detection and
generation}

\subsubsection{Thermometry\label{subsec:thermometrie}}

The temperature-sensitive material used in the present study is AuSn, which fulfills two requirements: (1) it is compatible with the microfabrication process and (2) it can be tuned to become temperature-sensitive over a range of interest to quantum turbulence studies \cite{woillez2021vortex}, from 1.5\:K up to the superfluid transition temperature $T_{\lambda}\simeq2.18$\:K in saturated vapor conditions. Other materials may be more suitable for other conditions; for example, Al was previously used for tweezers operated around 1.5\:K in \onlinecite{roche2007vortex}.

\begin{figure}
\includegraphics[width=0.5\textwidth]{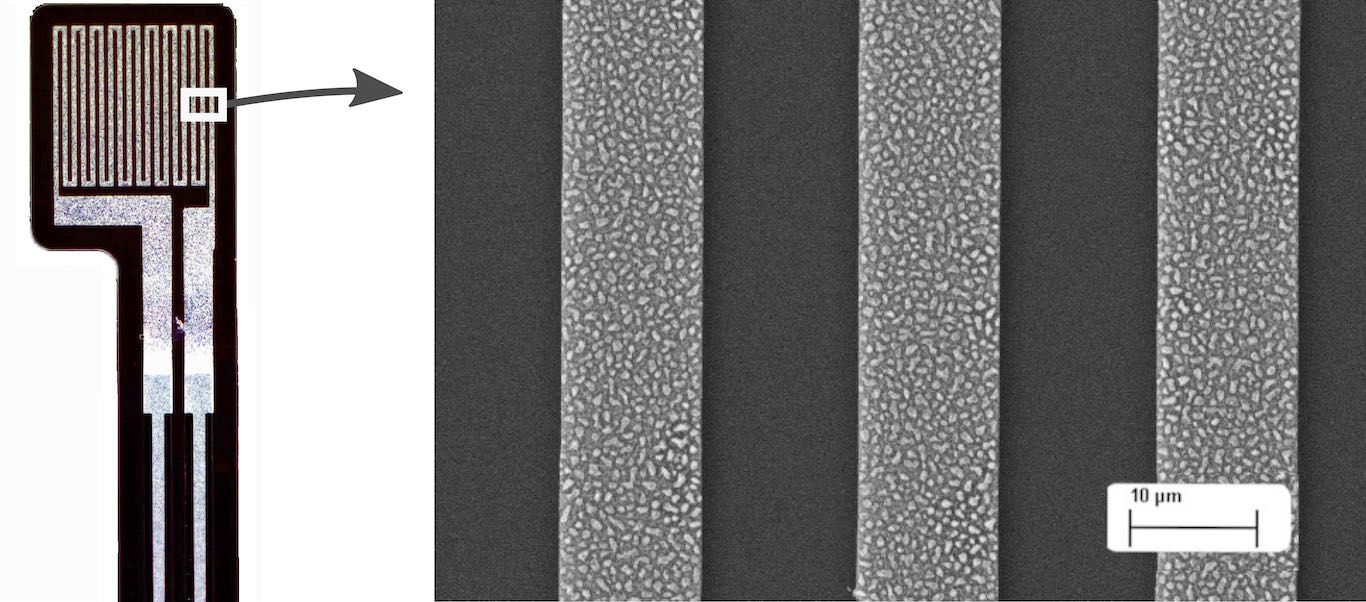}
\caption{Scanning electron microscope view of a tip. \textbf{ Left}: Tip frontside view. The overlaps between the serpentine path material and the current leads are visible on the lower part of the picture.
\textbf{ Right}: Detail of AuSn layer showing its granular aspect. Depending on tweezers models (see Fig. \ref{fig:jerome3}), the width of the AuSn track is 4, 11 or 24\:$\mu$m. \label{fig:AuSn}}
\end{figure}

The gold-tin AuSn thermometer is a metal-superconductor composite material, with superconducting Sn islands electrically connected by a gold layer. This granular structure is shown by electron microscopy in Fig. \ref{fig:AuSn} (right). The temperature dependence phenomenology can be interpreted simply. Indeed, by proximity effect, the gold in contact with tin behaves as a superconductor over a spatial extent that depends on temperature. By adjusting the characteristic length scales and thicknesses of the granular pattern, the temperature response of the material can be tuned.

\begin{figure}
%\begin{centering}
\includegraphics[width=0.45\textwidth]{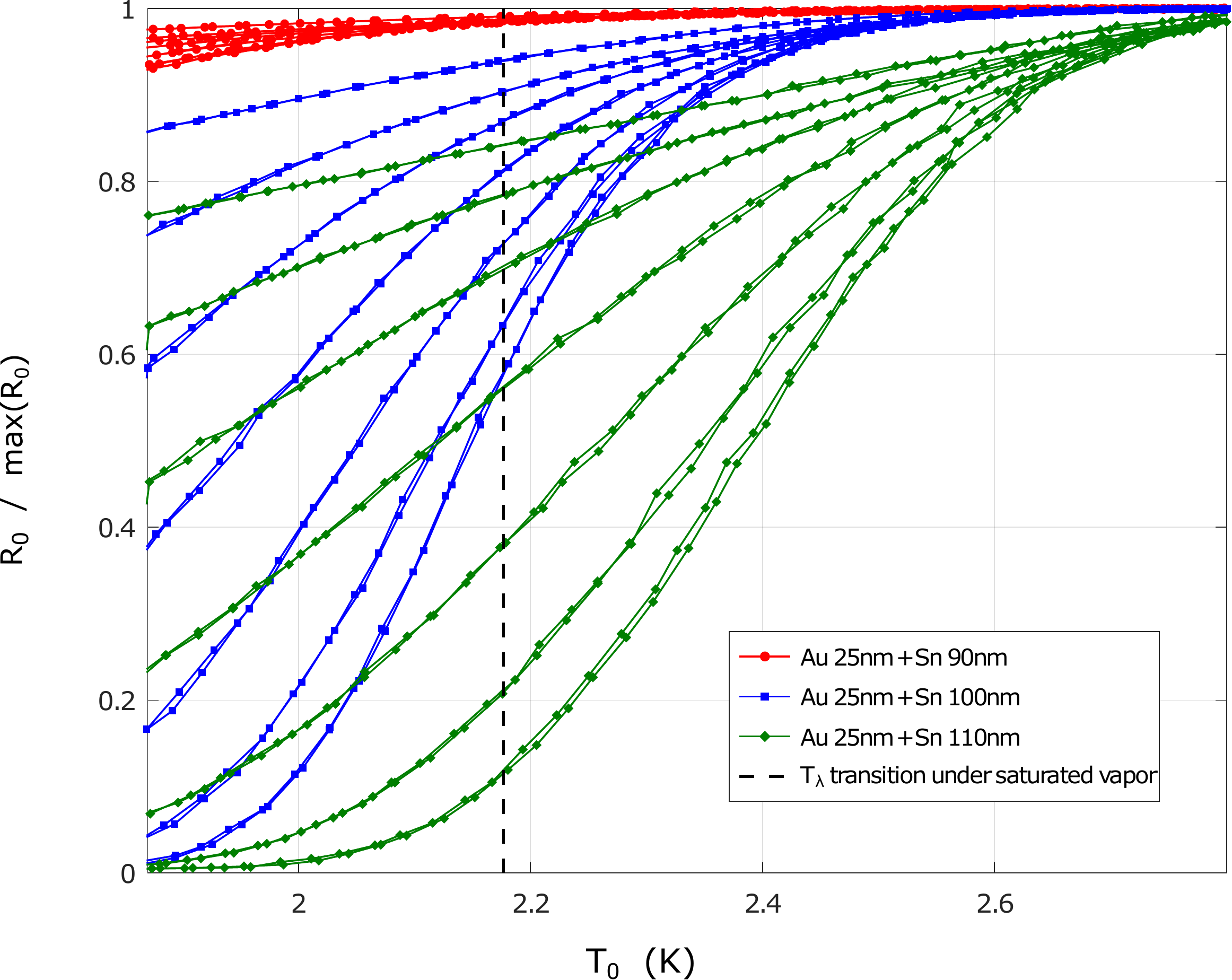}
\,\,\,\,\,\,
\includegraphics[width=0.45\textwidth]{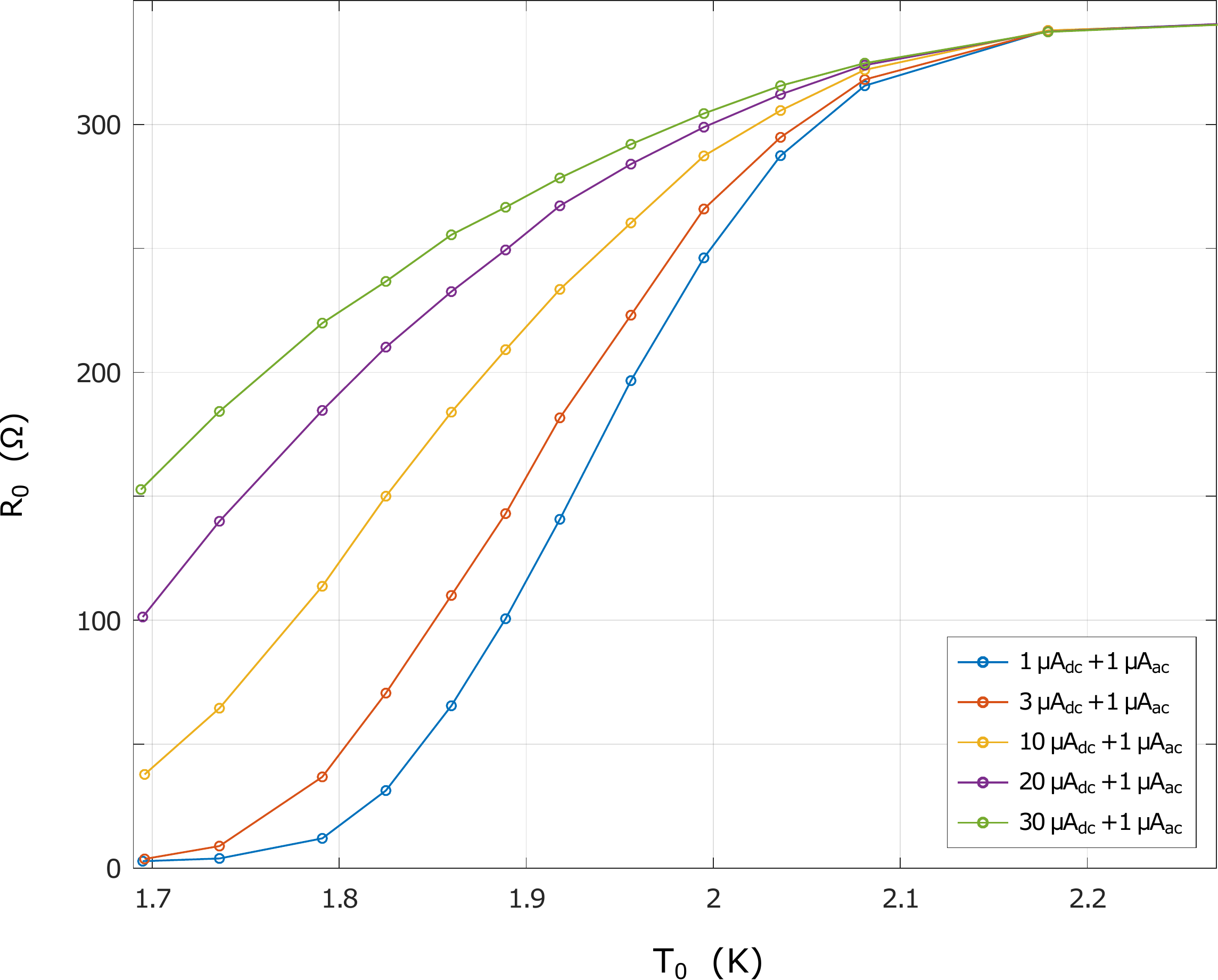}
\caption{\textbf{ Top}: Example of temperature and current dependence of AuSn layers with different Sn thicknesses deposited on a track of width $50\:\mu$m. The current setpoint spans from $1\:\mu$A up to $1$\:mA. 
\textbf{ Bottom}: Temperature response of the AuSn track of small tweezers for different electrical currents. The thickness of the tin layer is 100\:nm of Tin, and the width of the track is $4\:\mu$m. The small difference between both plots -when compared at similar current densities- is compatible with the uncertainty on the layer thicknesses. This good agreement indicates that AuSn properties are robust to the full  fabrication process of the tweezers.  \label{fig:resistance vs Temperature vs Courant}}
\end{figure}

As a preliminary study, the temperature dependence of the resistance of a 100-square-long AuSn track was measured for three different tin thicknesses, as shown in the top plot of Fig. \ref{fig:resistance vs Temperature vs Courant}. The track resistance is directly proportional to its number of squares (length-to-width ratio of the track) with the sheet resistance as a multiplying factor. A description of the conduction mechanism in AuSn is presented in \onlinecite{borner1983AuSn}.

In the present study, the total resistance at superfluid temperatures does not exceed a few hundred ohms. This value was chosen to be much larger than the resistance of the leads, but small enough to prevent parasitic effects from the leads' capacitance (typically a few hundred picoFarads) up to the highest frequencies of operation.

To achieve resistance values in this range, the meander length was fixed at around 700 squares for all tip sizes. Depending on the tip size, the track width was adjusted so that the serpentine shape occupies the entire available area on the tip. At room temperature, the AuSn layer resistance was found to drift from low values to a final value over the course of a few days (less than one week) after deposition. After this period, the resistance was found to be stable for at least a few years.

Figure \ref{fig:resistance vs Temperature vs Courant} (bottom plot) shows a typical resistance-temperature curve $R(T_{0})$ for an AuSn thermometer at different direct currents $I$. Regarding the temperature dependence of resistance, the current density is a more significant parameter than the total current. Thus, the comparison between the top and bottom plots of Figure \ref{fig:resistance vs Temperature vs Courant} should be made at constant values of the ratio of current to track width.
At low current densities ($I\leq10\mu A$), the sensitivity exceeds $1\:\varOmega.$mK$^{-1}$. At larger current densities, the current-induced magnetic field significantly shifts the superconducting-metal transition to lower temperature and broadens it, allowing the measurement range to be extended down to 1.6\:K and below. In the range of currents explored in Figure \ref{fig:resistance vs Temperature vs Courant}, the reduction in sensitivity in $\varOmega.$K$^{-1}$ at larger currents $I$ is more than compensated by the larger sensitivity in V$.$K$^{-1}$ units across the thermistor. Most measurements presented hereafter are performed with a measuring current $I\simeq27\:\mu$A.

\subsubsection{Heating}

The heater has the same meander length as the thermometer, which is close to 700 squares for all tip sizes.

As shown in Figures \ref{fig:jerome3} and \ref{fig:AuSn}, a buffer zone was designed between the gold tracks and the meander. In this zone, the electrical path is wide, but the material is the same as in the meander (platinum for the heater). The buffer zone's length is approximately 20\:squares, aiming to provide some thermal insulation between the meander and the gold track.

Numerous resistive materials are suitable for this purpose. For instance, chrome was used for the tweezers in \onlinecite{roche2007vortex}, and platinum was used in \onlinecite{woillez2021vortex}. The present data were obtained with platinum to benefit from the temperature-independence of its resistivity at superfluid temperatures \cite{poker1982} and also allow reusing these miniature heaters as miniature thermometers or hot-film anemometers in experiments conducted at higher temperatures where Pt regains temperature dependence \cite{PlatiniumITS90}. A 5\:nm titanium layer was deposited before platinum as an adhesion layer.

The thickness of the Pt layer, around 80\:nm, was chosen to ensure that the electrical resistance of the heater at superfluid temperatures is on the order of a few hundred ohms, similar to the maximum resistance of the thermometer and for the same reasons.

The heater is driven with a sinusoidal current at a frequency of $f/2$. The resulting Joule effect can be separated into a constant mean heating and a sinusoidal heat flux at frequency $f$, which drives the second sound resonance. One advantage of this $f/2$ excitation is that the signal detected by the thermometer, centered around $f$, is not affected by spurious  electromagnetic coupling at $f/2$ from the excitation circuitry. Thus, no special care is needed to minimize the electromagnetic cross-talk between the electrical tracks of the heater and the electrical tracks of the thermometer, despite their proximity.

The non-zero mean heating results in a steady thermal flux in He\:II, with the corresponding entropy carried away from the heater in the form of steady normal fluid flow. This outgoing normal flow is balanced by an opposite steady mass flow of superfluid toward the heater. Such cross-flows are referred to as \textit{counterflows} in the quantum fluid literature \cite{Tough1982, Nemirovskii1995}. This steady counterflow adds up to a pure second sound generated by the heater, but contrary to it, its effects are not amplified by resonance in the cavity.

\begin{figure*}
\includegraphics[width=0.45\textwidth]{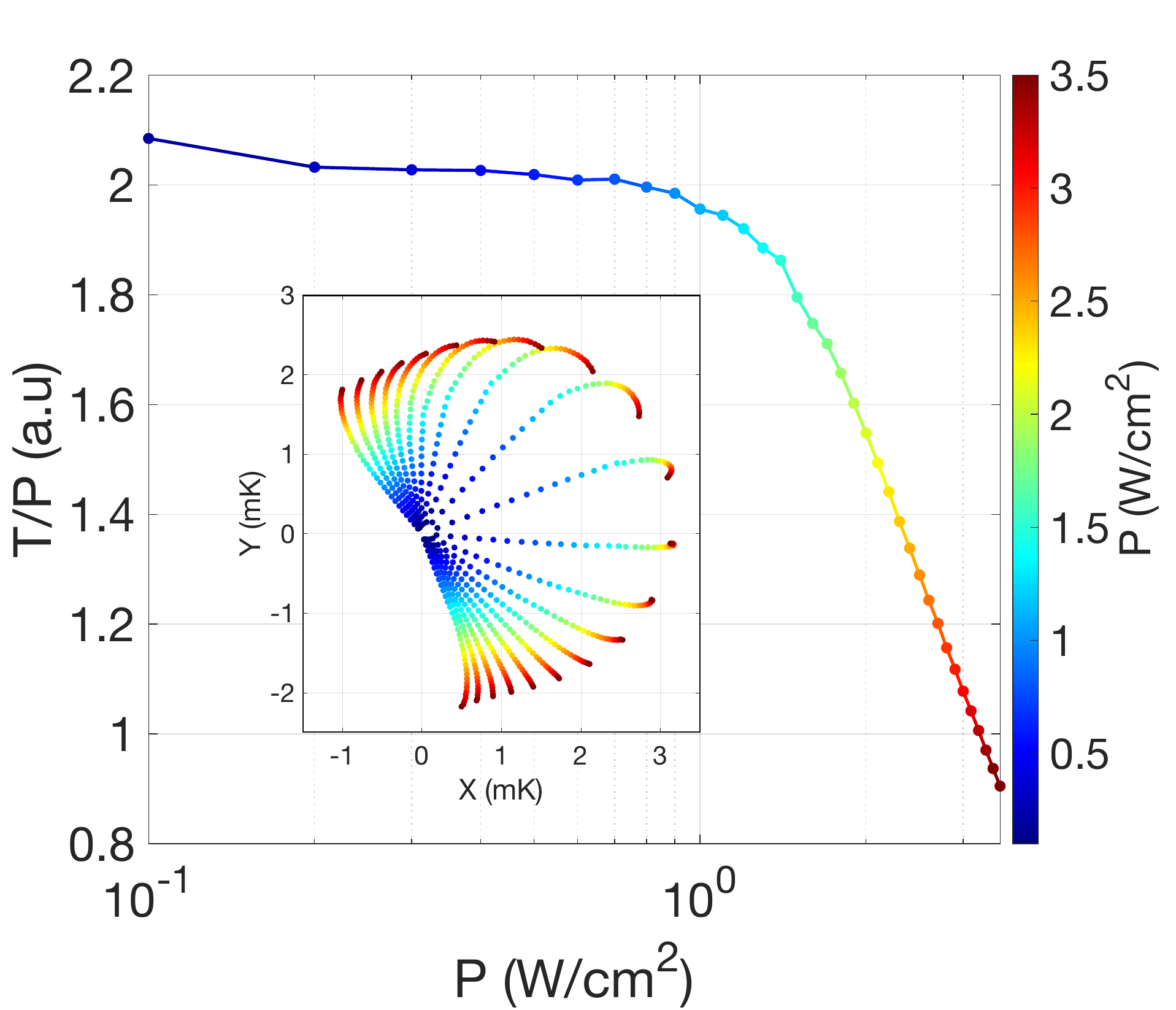}
\hspace{1cm}
\includegraphics[width=0.45\textwidth]{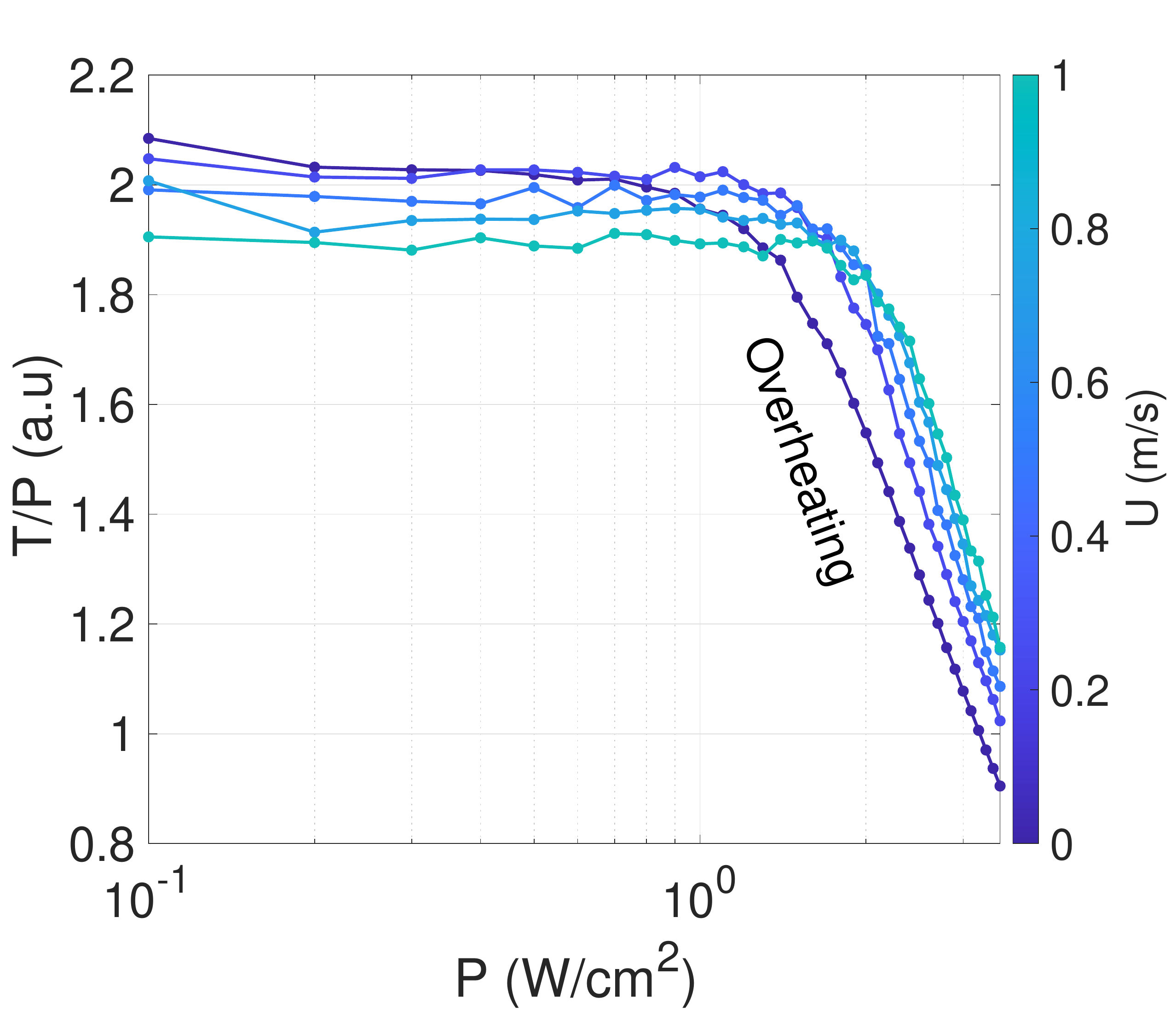}
\caption{
\textbf{Left:} The normalized amplitude $T/P$ of the temperature standing wave versus heating power $P$ is shown for second sound tweezers in quiescent He\:II around 1.6\:K. The transition around $P=1$\:W/cm$^{2}$ is interpreted as the development of a self-sustained vortex tangle within the probe. The inset shows the amplitude of the temperature standing waves in the complex plane for a subset of frequencies belonging to the same second sound resonance. In this representation, the extra-dissipation associated with the self-sustained vortex tangle results in a curvature of the iso-frequency radial "lines", revealing the broadening of the resonance.
\textbf{Right:} The same quantity is shown for second sound tweezers swept by turbulent flows of different mean velocities but similar turbulence intensity. In the linear regime $P\lesssim1$\:W/cm$^{2}$, the velocity dependence is opposite to that in the non-linear regime, $P\gtrsim2$\:W/cm$^{2}$, demonstrating respectively vortex and velocity sensing by the probe.
 \label{fig:NonLinearite}
}
\end{figure*}

\textbf{Quasi-linear vs non-linear regimes:} The second sound resonators are operated with standing waves of low amplitude, typically around $\sim100\:\mu$K. In this regime, the amplitude $T$ of the temperature standing wave responds nearly linearly to the heating power $P$. However, for larger heating power, the ratio $T/P$ decreases with $P$, indicating a turbulent transition within the tweezers that leads to a dense tangle of quantum vortices dissipating the second sound wave\footnote{In one dataset, a small discontinuity in the $T/P$ versus $P$ dependence around $P^{\star}\simeq10^{-2}\:$W/cm$^{2}$ was observed in the quasi-linear region of quiescent superfluid around 1.5\:K (not shown), suggesting another flow transition, but this effect was not detectable in other datasets. The Reynolds number of this possible transition can be defined using the transverse characteristic length scale $L\simeq1\:$mm, the quantum of circulation $\kappa\simeq0.997.10^{-7}\:$m$^{2}$/s, and the counterflow superfluid velocity towards the heater $v_{s}\simeq0.3\:$mm/s (amplification of velocity by the quality factor of the cavity has not been taken into account). One finds 
\begin{equation*}
Re_{s}=v_{s}L/\kappa\simeq3.
\end{equation*}
Critical Reynolds numbers $Re_{s}$ of a few units have already been reported to characterize the threshold of the appearance of a few superfluid vortices across the section of pipes that are closed at one of their ends with a heating plug (see Fig. 3 in \cite{Bertolaccini:PRF2017}), a transition referred to as the T1-transition in the counterflow literature\cite{Tough1982,Nemirovskii1995}. By analogy, this could suggest that the discontinuity at $P^{\star}$ might be associated with the appearance of a sparse tangle of quantum vortices near the heater, which density is expected to increase at larger $P$. Such vortices would damp the standing wave, but no such effect has been detected. Indeed, the observed damping of the standing wave in quiescent He\:II can be accounted for by the sole effect of diffraction (as shown later), indicating that all the other sources of loss are comparatively small. Loss due to such ``counterflow'' vortices would make the $T(P)$ dependence sub-linear rather than linear, which is not clearly observed. Since no quantitative evidence of these vortices could be clearly identified, this effect was not further explored.
}. The crossover from the quasi-linear to non-linear response of $T(P)$ is shown in Figure \ref{fig:NonLinearite} (left plot) for tweezers at 1.6 K in the absence of external flow. In these conditions and for these tweezers, the transition occurs around $P\simeq1$\:W/cm$^{2}$, where $P$ is the total Joule power normalized by the heating surface. In other conditions, the transition was observed at smaller power densities, but no systematic study has been carried out to determine the threshold value.

\subsubsection{Digression on the operation in the non-linear heating regime}

The current study primarily focuses on the linear regime of heating, but a brief investigation of higher powers reveals an interesting property of non-linear operation and supports the above interpretations of the nature of the non-linear regime. Figure \ref{fig:NonLinearite}-right displays the amplitude of the normalized temperature standing wave $T/P$ versus $P$ in flows with different mean velocities $U$ and turbulence intensity of a few percent. In the linear regime ($P\apprle1\:$W/cm$^{2}$ in the conditions of Fig. \ref{fig:NonLinearite}), the plateaus of $T/P$ decrease as $U$ increases, in accordance with the classical interpretation (discussed later) that the standing wave $T$ is damped by the vortices present in the external flow, whose concentration increases with $U$. Interestingly, in the non-linear regime (around $P\apprge2\:$W/cm$^{2}$ in Fig. \ref{fig:NonLinearite}), the dependence of $T/P$ on $U$ is opposite. The interpretation is that the extra damping of the standing wave $T$ is mainly due to the vortices generated within the tweezers by the heating itself. This vortex density decreases at higher $U$ because vortices are more effectively swept out of the tweezers. Thus, in the non-linear regime, the second-sound tweezers behave as local anemometers. In the linear regime, we will demonstrate that second-sound tweezers can not only act as vortex probes (as shown in Fig. \ref{fig:NonLinearite}-right), but also as anemometers through a mechanism discussed later.

\subsubsection{Thermal load of the tip on the fluid}

The specific heat of the cantilever tip, mostly made of silicon \cite{desai1986thermodynamic}, is much smaller than the specific heat of a similar volume of liquid helium at the temperature of interest. For instance, at the intermediate temperature of 1.8\:K, they differ by more than 5 orders of magnitude\cite{olson1994kapitza,donnelly1998observed} with $C_p^{Si}\simeq 3.5$\:J.K$^{-1}$.m$^{-3}$  while $C_p^{LHe} \simeq 4.3\times10^5$\:J.K$^{-1}$.m$^{-3}$.

The thermal resistance of the interface between the tip and the fluid can be estimated from the literature on the Kapitza resistance (\onlinecite{Pollack1969_Kapitza,ramiere2016thermal} and references within).
At a given temperature, this resistance depends on the tip surface material (Pt, AuSn, SiO$_2$ or Si), on the surface roughness and cleanliness, and on the normal/superconducting state of the material. Based on values reported in \cite{Pollack1969_Kapitza,johnson1963experiments,olson1994kapitza}, the tip-helium thermal resistance at 1.8\:K is estimated to be within $1-6$\:cm$^2$.K.W$^{-1}$ on each side.

The characteristic response frequency $\left( 2\pi RC \right)^{-1}$, where $R= 3$\:cm$^2$.K.W$^{-1}$ is the typical Kapitza resistance and $C=e \cdot C_p^{Si}$ with $e=20$\:$\mu$m is the thermal inertia of the cantilever tip per surface unit, exceeds 7\:MHz at 1.8\:K. This frequency is significantly larger than the largest frequencies of the second sound considered here. As the Kapitza resistance roughly scales as $T^{-3}$ and the tip inertia as $T^3$, this characteristic frequency does not strongly depend on temperature.

The estimates above illustrate the negligible thermal load of the probe compared to the surrounding fluid and its ability to respond to rapid environmental fluctuations. Previous studies have reported measurements extending over a bandwidth of 1\:MHz or higher, not only in superfluids (e.g. see \onlinecite{Cummings:1978p376}) where helium's high thermal conductivity benefits the fluid-probe system's dynamics, 
 but also in gazeous helium\cite{Castaing1992_HotWire,Chanal:EPJB2000,Pietropinto:2003p331},  
 where the system's thermal inertia is determined by the fluidic boundary layer\cite{GauthierEPL2009}. 

Deriving the full transfer functions for the coupling between the second sound standing wave and the probe would require a detailed thermal analysis beyond the scope of this paper. Instead, our approach was to assume an ideal response of the probe and to show that the resulting analytical predictions closely match experimental measurements.

\subsection{\label{subsec:Microfabrication-and-assembling}Microfabrication and
assembling}

\begin{figure}
\includegraphics[width=0.35\textwidth]{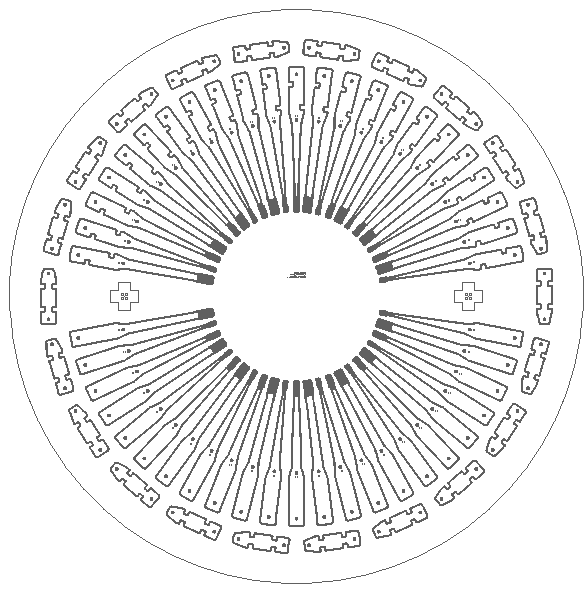}
\caption{Overview of mask design. The disk diameter is 100\:mm. \label{fig:jerome4}}
\end{figure}

The mechanical structures of the cantilevers and spacers are made of silicon. The cantilevers were fabricated by processing SOI (Silicon On Insulator) wafers by microelectronic techniques. SOI wafers consist of a thin silicon layer (known as the device layer) separated from a thick silicon substrate by an insulator layer, which in this case is a buried oxide with a thickness of 1\:µm. The device layer has a thickness of 20\:µm, while the silicon substrate layer is 500\:µm thick. Standard photolithography was used to create the metal and silicon shapes. The diameter of the wafers was 100\:mm. As shown in Fig. \ref{fig:jerome4}, the patterns for the 46 cantilevers on each wafer were arranged radially, with the cantilever tips positioned close to the wafer center to ensure higher reproducibility of the tip properties. The wafers were double-side polished and oxidized to create a 100\:nm thick SiO$_2$ layer on both sides.

A simplified version of the cantilever fabrication process is presented in Table \ref{tab:jerome5}, with full details provided in the appendix. The serpentine electrical path (colored red) was deposited first on the frontside of the SOI wafer. For heaters, the evaporation sequence Ti 5\:nm + Pt 80\:nm was used, while for thermometers (assuming a hypothetical thickness for a planar - not granular - tin layer), Au 25\:nm + Sn 100\:nm was used. During a second sequence, current leads (colored orange) were deposited, with an evaporation sequence of Ti 5\:nm + Au 200\:nm + Ti 5\:nm + Pt 50\:nm. The use of a platinum layer was found to facilitate lift-off and may also be useful for brazing purposes.

The cantilevers' 3D shape was achieved through the use of an STS HRM deep reactive ion etching (DRIE) tool, employing recipes based on the technology known as "Bosch process". This tool is capable of etching silicon to a depth of several hundred microns with perpendicular sidewalls. Non-etched regions were protected by resist or aluminum layers. By using various etching recipes and protecting layers (resist, aluminum, and silicon oxide), the silicon was initially etched from the backside with two different mask shapes, followed by the frontside to achieve silicon plate piercing. The resulting pieces had tip areas made up of the original 20 µm thick SOI device layer, with the 1\:µm thick oxide removed from the tip backside to avoid bending due to oxide mechanical stress. While some spacers were fabricated together with the cantilevers, most of them were made separately from two silicon wafers with thicknesses of 300\:µm and 525\:µm, coated by thin dielectric layers on both sides.

\begin{table*}
\addtolength{\tabcolsep}{-1pt}
\begin{tabular}{|l|l|}
\hline
Step & Cut view\tabularnewline
\hline

\hline
\begin{tabular}{@{}lc@{}}
1. SOI wafer (surface oxide 0.1\:µm / silicon device 20\:µm / \\ buried oxide 1\:µm / silicon substrate 500\:µm / surface oxide 0.1\:µm). \\ { }\end{tabular}
& \includegraphics[width=5cm]{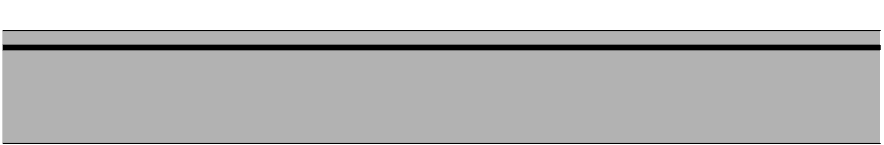} \tabularnewline

\hline
\begin{tabular}{@{}lc@{}}
2. Electrical circuitry fabrication. \\ { }\end{tabular}
& \includegraphics[width=5cm]{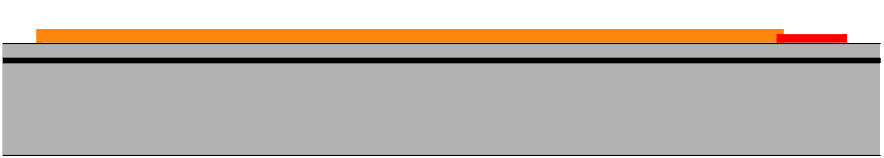} \tabularnewline

\hline
\begin{tabular}{@{}lc@{}}
3. Backside etching 1.\\ { }\end{tabular} & \includegraphics[width=5cm]{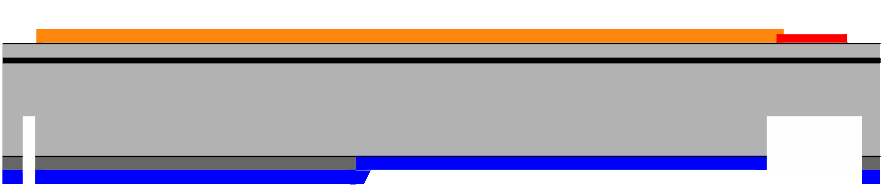}\tabularnewline

\hline
\begin{tabular}{@{}lc@{}}
4. Backside etching 2.\\ { }\end{tabular}
  & \includegraphics[width=5cm]{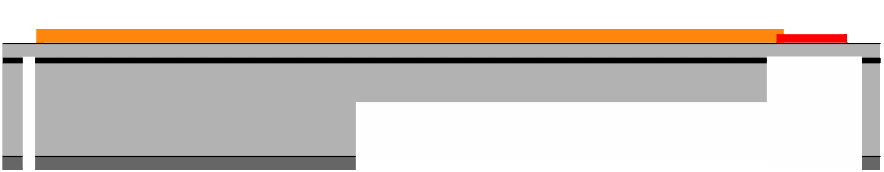} \tabularnewline

\hline
\begin{tabular}{@{}lc@{}}
5. Frontside etching mask fabrication.\\ { }\end{tabular}
& \includegraphics[width=5cm]{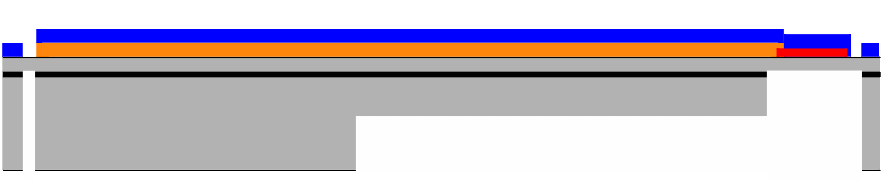} \tabularnewline

\hline
\begin{tabular}{@{}lc@{}}
6. Frontside etching and mask removal.\\ { }\end{tabular}
& \includegraphics[width=5cm]{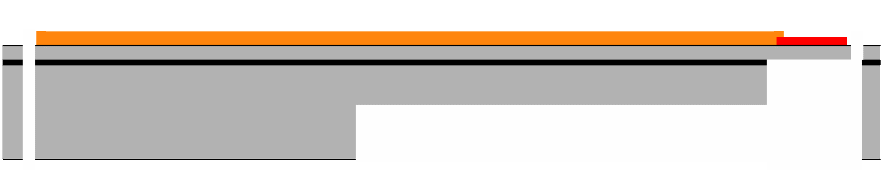} \tabularnewline

\hline 
\end{tabular}
\caption{\label{tab:jerome5}Cantilevers fabrication process.}

\end{table*}

% _-----------------------------------------------_
% _-----------------------------------------------_
% _-----------------------------------------------_
% _-----------------------------------------------_

\subsection{\label{subsec:Electric-circuit}Electric circuit}

\begin{figure}
\includegraphics[width=0.5\textwidth]{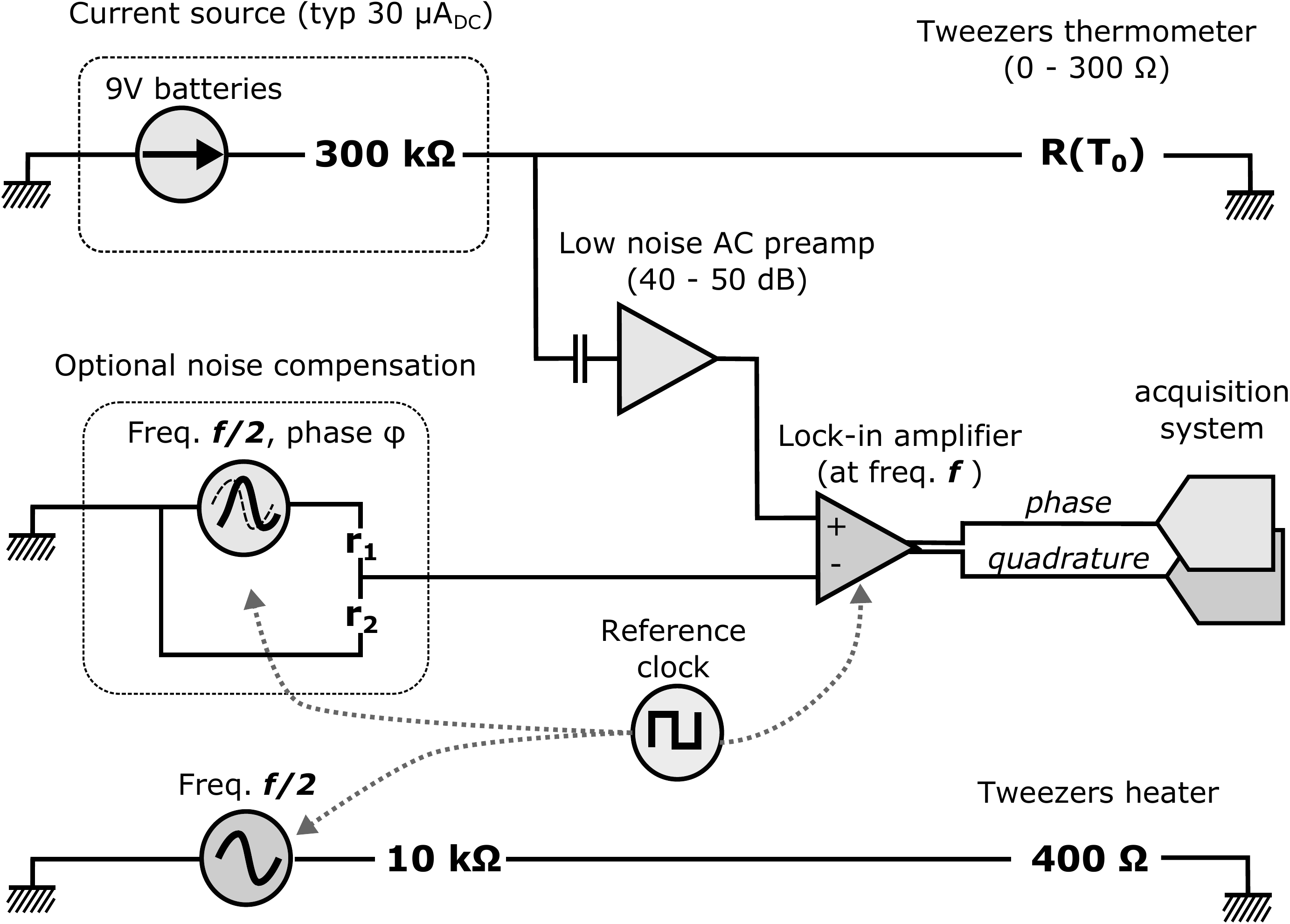}
\caption{Example of 
circuitry for measurements with a high dynamical reserve. \label{fig:circuit}}
\end{figure}

Figure \ref{fig:circuit} depicts the circuit utilized for time-resolved measurements using second-sound tweezers, including example values for resistances and gain. The time-resolved data presented in this paper were obtained using this circuit and the following equipment. The front-end  preamplifier is either the Celian EPC1-B model or the NF SA-400F3 model when exploring frequencies above 100\:kHz. The lock-in amplifier is the NF LI-5640 model or the SignalRecovery Model 7280 above 100\:kHz. In most cases, the lock-in amplifier's built-in internal generator provides both the drive for the tweezers' heater (at frequency $f/2$) and the reference frequency to detect the temperature signal (at frequency $f$). The acquisition system is based on the National Instrument PXI-4462 analog input cards, and it records both the in-phase ($X$) and quadrature ($Y$) signals from the lock-in amplifier's analog outputs.

In certain situations, the temperature signal at the lock-in input is overwhelmed by a much larger electromagnetic parasitic signal at $f/2$, and it cannot be accurately resolved by the limited voltage dynamic range of the lock-in amplifier. This can happen when the tweezers are operating far from resonance, where the second-sound signal is small, or when the tweezers are operating at very high frequencies (e.g., > 100 kHz), as electromagnetic parasitic coupling increases with frequency. The magnitude of this parasitic coupling is determined by the tweezers and cable's geometrical and electrical characteristics. Within the range of parameters investigated in this study, the order of magnitude of the parasitic voltage induced across the thermometer resistor, normalized by the voltage applied across the heating resistor, is given by
\[
0.5\% \times  \frac{{f}/{2}}{100\:\text{kHz}}
\]

Such situations are handled thanks to the differential input of the lock-in amplifier, which removes a signal mimicking the parasitic one. In such cases, a two-channel waveform generator is used: one channel drives the heater (at frequency $f/2$), another channel mimics the parasitic signal (at frequency $f/2$, with manually tuned amplitude and phase shift), and the "sync" output of the generator synchronizes the lock-in demodulation (at frequency $f$). The Agilent 33612A generator is used for this purpose. Alternatively, the compensation signal can be generated directly from the lock-in internal generator, completed with a simple RC phase shifter and, eventually, a ratio transformer.

In principle, any thermistor with a positive temperature coefficient, such as an transition-edge thermometer, that is not well thermalized with the fluid can become unstable when driven by a current source. An infinitesimal thermistor fluctuation from $T_{0}$ to $T_{0}+\delta T_0$ leads to a resistance variation of $\delta R=\frac{\partial R}{\partial T_{0}}.\delta T_0>0$, resulting in an excess of Joule dissipation $\delta R.I^{2}$ for a constant current drive $I$. Let $\mathcal{R}_{th}$ be the thermal resistance of the thermistor-fluid interface, then this extra Joule dissipation results in an overheating of $\mathcal{R}_{th}\times\delta R.I^{2}$, which could lead to a thermal instability. The stability condition is difficult to predict for a spatially distributed thermistor deposited on a Si crystal and immersed in superfluid. Thus, initial tests were done with a voltage drive before empirically validating the stability of our current drive.

The frequency bandwidth of the measurements is arbitrarily set by the integration time constant of the lock-in amplifier. In practice, the circuit's performance is limited by the input voltage noise of the EPC1-B pre-amplifier, which is $0.65\:$nV/$\sqrt{\text{Hz}}$. For a current drive of $27\:\mu\text{A}$, a thermometer sensitivity of $0.5\:\Omega\cdot\text{mK}^{-1}$, and a demodulation bandwidth of 10\:Hz or 1000\:Hz, the temperature resolution $T_{rms}$ is given by:
\begin{equation*}
T_{rms}=\frac{\sqrt{10}.0.65}{0.5\times27}\mu \:\text{K}\simeq150\:\text{nK}
\end{equation*}
\noindent for a 10\:Hz measurement bandwidth, or 
\begin{equation*}
T_{rms}=\frac{\sqrt{1000}.0.65}{0.5\times27}\mu \text{K}\simeq1.5\:\mu \text{K}
\end{equation*}
\noindent for a 1\:kHz measurement bandwidth

These resolutions are sufficient under standard conditions. They are three and two orders of magnitude smaller, respectively, than the typical amplitude of second sound at resonance. Reaching the same temperature resolution at a significantly larger bandwidth would be futile, given the spatiotemporal resolution of the probe itself. If necessary, better resolution could be achieved with a larger current across the thermometer or by using a cryogenic amplifier (e.g., see \onlinecite{DongHEMT2014} and http://cryohemt.com) before being limited by the thermal noise floor of the thermistor (typically $0.15\:$nV/$\sqrt{\text{Hz}}$ for $200\:\varOmega$ at 2\:K).

\section{\label{sec:Models-of-second}Models of second sound resonators}

The second sound equations within the linear approximation can be
written in terms of the temperature fluctuations $T$ and the velocity
of the normal component $\mathbf{v}_{n}$ as

\begin{align}
\partial_{t}\mathbf{v}_{n} & +\frac{\sigma\rho_{s}}{\rho_{n}}\nabla T=0,\label{eq:normal velocity vs T}\\
\partial_{t}T & +\frac{\sigma T_{0}}{c_{p}}\nabla.\mathbf{v}_{n}=0.\nonumber 
\end{align}
with $\sigma$ the entropy per unit of mass, $c_{p}$ the heat capacity,
and $\rho_{s}$ , $\rho_{n}$ are the densities of the superfluid
and normal components respectively. All along the present section,
$T_{0}$ is the notation for bath temperature far away from the tweezers. $T$
denotes the local temperature fluctuations, that depend both on space and time. From this definition, we obviously have $\left\langle T\right\rangle =0$ where $\left\langle \right\rangle$ is the time average.

We introduce the second sound velocity $c_{2}$ defined by the relation
\begin{equation}
c_{2}^{2}=\frac{\rho_{s}}{\rho_{n}}\frac{\sigma^{2}T_{0}}{c_{p}}.
\label{eq:second sound def}
\end{equation}
It can be deduced from Eqs. (\ref{eq:normal velocity vs T}) that
both the temperature $T$ and the normal velocity $\mathbf{v}_{n}$
follow the wave equation 
\begin{equation}
\partial_{t}^{2}T-c_{2}^{2}\Delta T=0.
\label{eq:wave eq for T}
\end{equation}
We explain in the present section how Eqs. (\ref{eq:normal velocity vs T}-\ref{eq:second sound def}-\ref{eq:wave eq for T})
can be used to build quantitative models of second sound resonators.
We first focus on phenomenological aspects in sec. \ref{subsec:phenomenology}.
Then, we give analytical approximations in sec. \ref{subsec:Analytical-approximations}
and an accurate numerical model in sec. \ref{subsec:Numeric-algorithm}.
Finally, we discuss the model quantitative predictions in secs. \ref{subsec:Quantitative-predictions}
and \ref{subsec:Summary-of-the}.

\subsection{Resonant spectrum of second sound resonator: phenomenological aspects\label{subsec:phenomenology}}

The basic idea of second sound resonators is to create a second sound
resonance between two parallel plates facing each other. A second
sound wave is excited with a first plate, while the magnitude and
phase of the temperature oscillation is recorded with the second plate,
used as a thermometer. For
simplicity, we assume from now on that the second sound wave is excited
by a heating, but the whole discussion can be straightforward extended
to nucleopore mechanized resonators. The temperature oscillations
within the cavity are coupled to normal fluid velocity oscillations
according to the second sound equations (\ref{eq:normal velocity vs T}).

We note $j_{Q}=j_{0}e^{2i\pi ft}$ the periodic component of the heat
flux emitted from the heater. We assume throughout the present article
perfectly insulating plates, which means that the boundary conditions
for the second sound wave are
\begin{equation}
\mathbf{v}_{n}.\mathbf{n}=\begin{cases}
0 & \text{for }z=D\\
\frac{j_{Q}}{\rho\sigma T_{0}} & \text{for }z=0
\end{cases},\label{eq:boundary conditions}
\end{equation}
where $\mathbf{n}$ is the unit vector directed inward the cavity
and normal to the plates. The second equation in (\ref{eq:boundary conditions})
reflects the fact that the normal component carries all the entropy
in the fluid. According to
the first relation in Eq. (\ref{eq:normal velocity vs T}), the boundary
conditions (\ref{eq:boundary conditions}) for the normal velocity
translate into the following boundary conditions for the temperature
field 
\begin{equation}
\nabla T.\mathbf{n}=\begin{cases}
0 & \text{for }z=D\\
-\frac{\rho_{n}\partial_{t}j_{Q}}{\rho\rho_{s}\sigma^{2}T_{0}} & \text{for }z=0
\end{cases}.\label{eq:boundary conditions_T}
\end{equation}
\\

\begin{figure}
\includegraphics[width=0.5\textwidth]{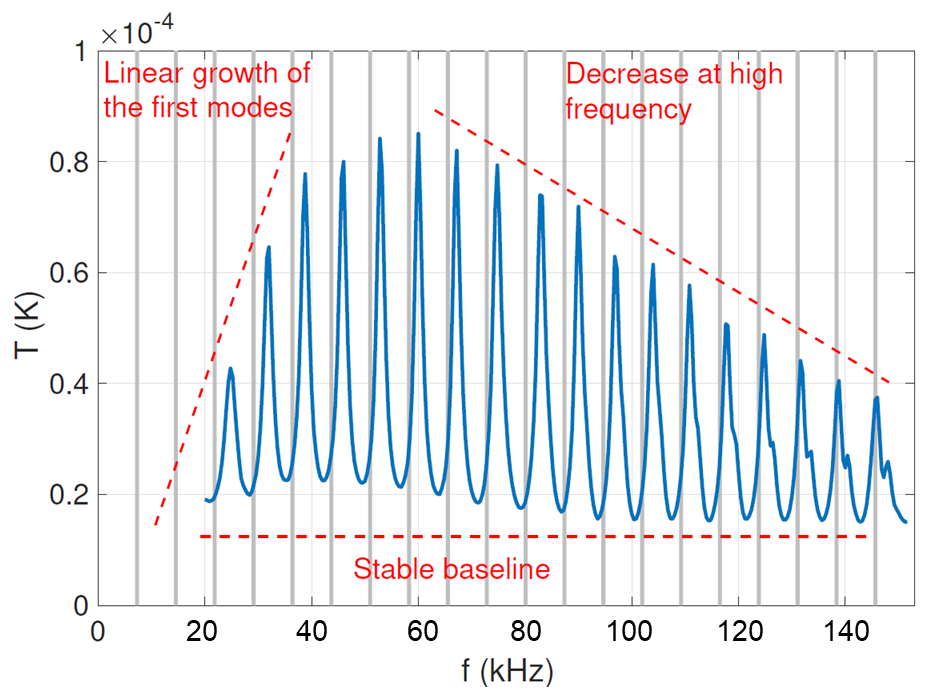}
\caption{Experimental spectrum of second sound tweezers of lateral size $L=1$
mm and gap $D\approx1.435$ mm. $f$ is the heating frequency, and
$T$ is the thermal wave magnitude. The figure displays the main characteristics
of tweezers typical spectrum: first, the resonant frequencies are
not located at the values $f_{n}$ given by $\frac{2\pi f_{n}D}{c_{2}}=n\pi,(n\in\mathbb{N})$,
displayed by the gray vertical lines. The amplitudes of the resonant
modes first increases linearly with $f$ until they saturate and eventually
decrease at high frequency. The baseline level only weakly depends
on $f$.\label{fig:Experimental-spectrum_sec3.1}}
\end{figure}

We display in Fig. \ref{fig:Experimental-spectrum_sec3.1} a typical
experimental spectrum of second sound tweezers, that is, the temperature
magnitude averaged over the thermometer plate, as a function of the
heating frequency $f$. The spectrum is reminiscent of that of a Fabry--Perot
resonator:
it displays very clear resonant peaks that are almost equally spaced, and
a stable non-zero minimum at non-resonant frequencies. However,
the spectrum of Fig. \ref{fig:Experimental-spectrum_sec3.1} displays
three major characteristics that can be observed for every tweezers'
spectrum. First, the locations of the resonant frequencies are slightly
shifted compared to the standard values $f_{n}$ given by $\frac{2\pi f_{n}D}{c_{2}}=n\pi,(n\in\mathbb{N})$
expected for an ideal Fabry--Perot resonator. Only for large mode
numbers do the resonant peaks again coincide with the expected values.
Second, the temperature magnitude vanishes in the zero frequency limit,
and the first modes of the spectrum roughly grow linearly with $f$. In between,
the resonant amplitudes saturate and then slowly decrease at high
frequency. 

These latter peculiarities of the frequency response were not described
in previous references about second sound resonators. This prompted
us to study different models for second sound resonators, including
the finite size effects and near-field diffraction phenomena. We first
describe analytical approximations in sec. \ref{subsec:Analytical-approximations},
then we develop in sec. \ref{subsec:Numeric-algorithm} a numerical
algorithm based on the exact solution of the wave equation (\ref{eq:wave eq for T}).
The numerical scheme can be adapted for various types of planar second
sound resonators. We then give quantitative predictions specifically
for the response of second sound tweezers without and in the presence
of a flow in sec. \ref{subsec:Quantitative-predictions} and a summary
of the main results in sec. \ref{subsec:Summary-of-the}. 

\subsection{Analytical approximations\label{subsec:Analytical-approximations}}

The starting point to build our model of second sound tweezers is
to assume that all zeroth order physical effects observed with the
tweezers are geometrical effects of diffraction. This means in particular
that we assume perfectly reflecting resonator plates, and we also
neglect bulk attenuation of second sound waves when the fluid is at
rest \cite{crooks1983,Mehrotra:1984p375}. These assumptions turn
to be self-consistent, because the predictions of the model developed
in sec. \ref{subsec:Numeric-algorithm} reproduce the main features
observed in experiments.

\paragraph{Second sound resonators embedded in infinite walls}

\begin{figure}
\includegraphics[width=0.5\textwidth]{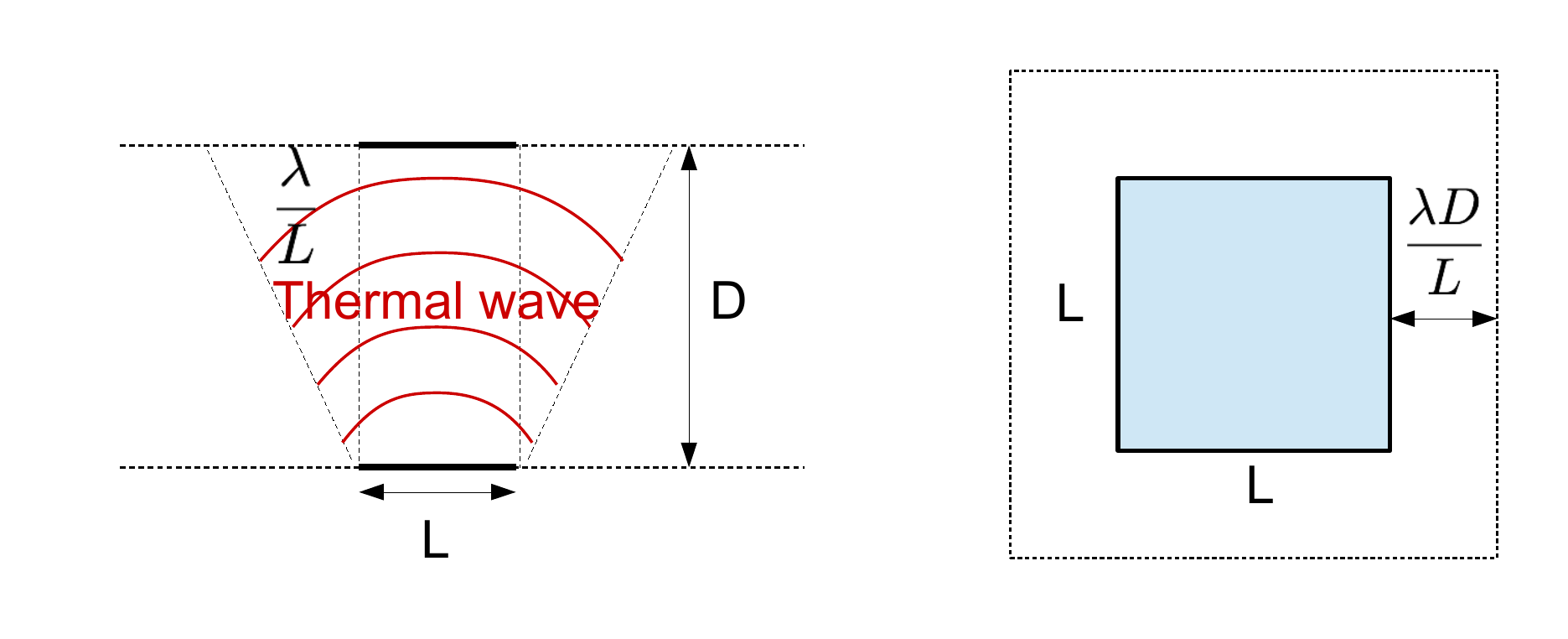}
\caption{Representation of the wave dispersion. The left picture is a top view parallel to the plates, and the right figure represents a front view of a tweezer's plate The energy loss is controlled
by the non-dimensional number $\frac{\lambda}{L}$, according to standard
diffraction theory. In this section, we discuss both the case of tweezers
embedded in walls, and the case of free tweezers in open space.\label{fig:ND number1}}
\end{figure}

We first consider a 
finite-size heater and a thermometer of size $L$ embedded in two parallel
and infinite walls facing each other. This geometry is most commonly
encountered in the literature. With such a configuration, the thermal
wave is not a plane wave any more because it is emitted by a finite
size heater. The model thus contains diffraction effects. An illustration of
the model setup is displayed in Fig. \ref{fig:ND number1}. An exact
solution of the wave equation Eq. (\ref{eq:wave eq for T}) can be
found using the technique of image source points. Let $\Sigma_{1}$ be the
heating plate and $\Sigma_{2}$ be the thermometer plate, and we assume
that the thermometer is sensitive to the average temperature over
$\Sigma_{2}$. Then the response of the tweezers is given by $T(t)=\mathcal{R}e\left(\overline{T}e^{2i\pi ft}\right)$
with
\[
\overline{T}=\frac{ikj_{0}}{2\pi\rho c_{p}c_{2}}\frac{1}{L^{2}}\iint_{\Sigma_{2}}{\rm d}^{2}\mathbf{r}_{2}\iint_{\Sigma_{1}}{\rm d}^{2}\mathbf{r}_{1}G\left(\mathbf{r}_{2}-\mathbf{r}_{1}\right),
\]
with the Green function $G(\mathbf{r})$ defined for every vector
$\mathbf{r}$ in the $(x,y)$ plane \cite{goodman2005introduction}
\[
G(\mathbf{r})=2\stackrel[n=0]{+\infty}{\sum}\frac{1}{\left|(2n+1)D\mathbf{e}_{z}+\mathbf{r}\right|}e^{-ik\left|(2n+1)D\mathbf{e}_{z}+\mathbf{r}\right|}.
\]
Such a model correctly predicts that the tweezers' spectrum vanishes
when the heating frequency $f$ goes to zero. Yet, it does not reproduce
a linear increase of the resonant magnitude of the first modes,
neither the decrease of the resonant peaks at large frequency observed
in experiments with second sound tweezers. This means that other effects
have to be taken into account to model a fully-immersed open resonant cavity, such
as non-perfect plates alignment and energy loss by diffraction outside
the cavity when the latter is not embedded in infinite walls. % XXXX\\

\paragraph{\label{par:Empirically-modified-Fabry=002013Perot}Empirically modified
Fabry--Perot model}

The Fabry--Perot model corresponds to a one-dimensional resonator
composed of two infinite parallel plates separated by a gap $D$. In that case, the wave Eq. (\ref{eq:wave eq for T}) together with
the boundary conditions Eqs. (\ref{eq:boundary conditions_T}) can
be solved exactly, for a periodic heating $j_{Q}=j_{0}e^{2i\pi ft}$ to find the (complex) temperature $\overline{T}$ at the thermometer plate \cite{bennett2022principles}
\begin{equation}
\overline{T}=\frac{A}{\sinh\left(i\frac{2\pi fD}{c_{2}}+\xi D\right)},\label{eq:classical Fabry-Perot}
\end{equation}
where $\xi$ (in m$^{-1}$) is an empirical dissipation coefficient and $A=-\frac{j_{0}}{\rho c_{p}c_{2}}$. An illustration of a Fabry--Perot
spectrum is displayed in grey in Fig. \ref{fig:Analytical-models sec3.2},
with $\xi D=0.15$ and $A=1$. We introduce the wave number $k=\frac{2\pi f}{c_{2}}$. For the
simple Fabry--Perot model of Eq. (\ref{eq:classical Fabry-Perot}),
all the resonant peaks have equal height and are uniformly separated.
Therefore, some main features of experimental spectra are missing,
an indication that important other physical effects have to be included
in the model.
\\

Contrary to a Fabry--Perot resonator composed of infinite plates,
second-sound resonators are built with plates of finite size $L$,
approximately of the same order as the gap $D$ between them. Those
finite size effects are important as they introduce a frequency-dependent
energy diffracted outside the cavity. This mechanism is sketched in
Fig. \ref{fig:ND number1}. According to standard diffraction theory,
a finite wave initially of size $L$ with a wavelength $\lambda=\frac{c_{2}}{f}$
spreads with a typical opening angle given by $\frac{\lambda}{L}$.
By this geometrical effect, a part of the wave energy is lost as the
wave reaches the other side of the cavity. The energy loss is roughly
proportional to the surface of the wave cross-section that ``misses''
the reflector (see the right panel of Fig. \ref{fig:ND number1}).
Therefore, the fraction of energy lost at the wave reflection is controlled
by the ratio
\begin{align}
\frac{\left(L+\frac{2\lambda D}{L}\right)^{2}-L^{2}}{\left(L+\frac{2\lambda D}{L}\right)^{2}} & \approx4\frac{\lambda D}{L^{2}},\label{eq:ND number1}\\
 & \approx\frac{4}{N_{F}},\nonumber 
\end{align}
where we have introduced the Fresnel number 
\begin{equation}
N_{F}=\frac{L^{2}}{\lambda D}.
\end{equation}

\begin{figure}
\includegraphics[width=0.3\textwidth]{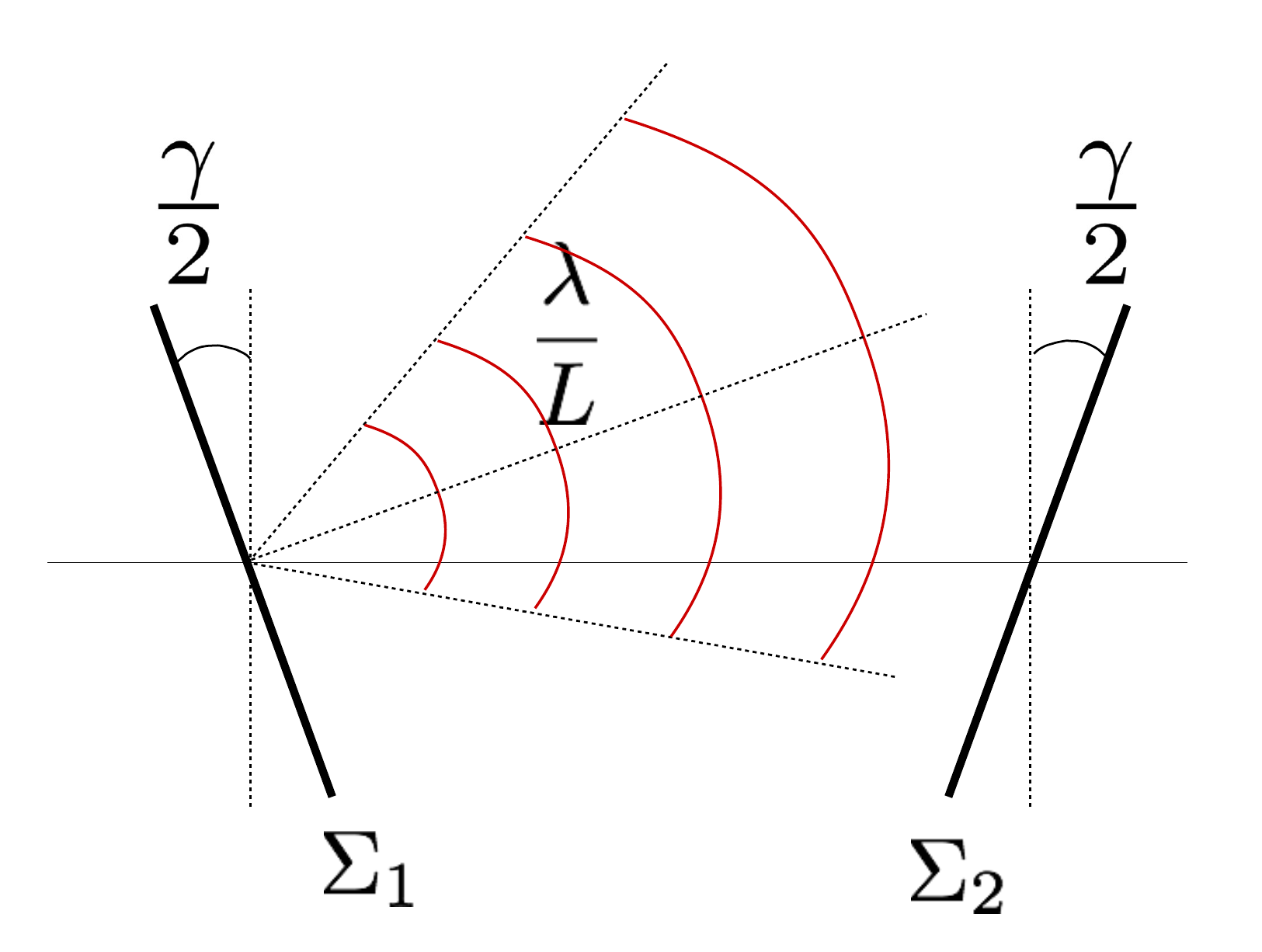}
\caption{Effect of the plates' inclination $\gamma$. Inclination creates an
additional energy loss mechanism controlled by the second non-dimensional
number $\frac{\lambda}{\gamma L}$.\label{fig:ND number2}}
\end{figure}

\begin{figure}
\includegraphics[width=0.5\textwidth]{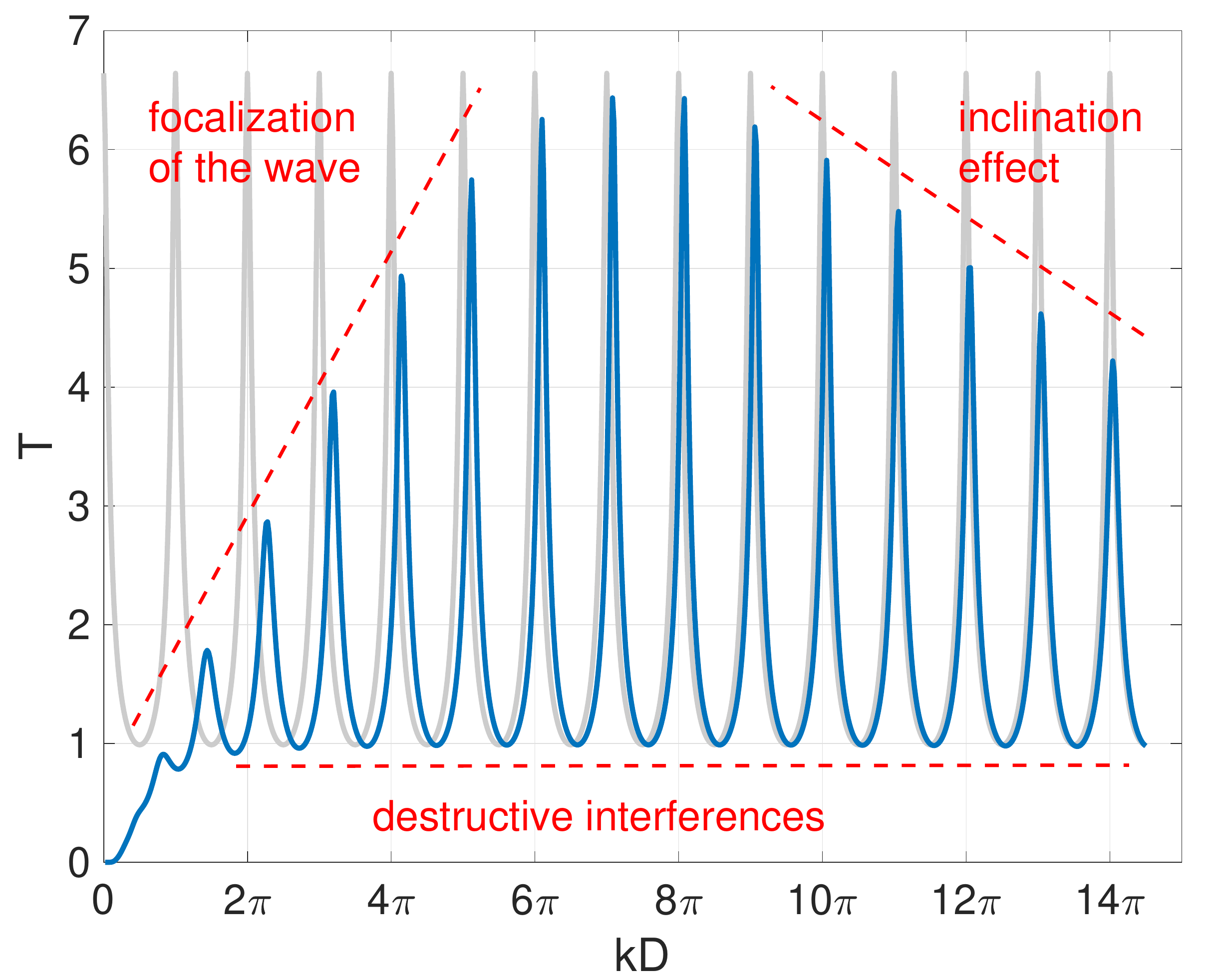}
\caption{Analytical models of second-sound tweezers. The grey curve represents
the standard Fabry-Perot spectrum. The blue curve represents the empirical
correction of the Fabry-Perot formula using the two non-dimensional
numbers $\frac{\lambda}{L}$ and $\frac{\lambda}{\gamma L}$. The
modified spectrum displays the characteristic features of a tweezers
experimental spectrum, as displayed in Fig. \ref{fig:Experimental-spectrum_sec3.1}.\label{fig:Analytical-models sec3.2}}

\end{figure}

The tweezers plates are mounted at the top of arms of a few millimeters.
The perfect parallelism of the plates is usually not reached for our
tweezers, but a small inclination $\gamma$ of the order of a few
degrees can be observed instead. A relative inclination $\gamma$
-even small- of both plates creates an additional energy loss mechanism
(see Fig. \ref{fig:ND number2}). Intuitively, this second mechanism
is controlled by the non-dimensional number
\begin{equation}
N_{i}=\frac{\lambda}{\gamma L}.\label{eq:ND number2}
\end{equation}

We assume that the Fabry--Perot model (\ref{eq:classical Fabry-Perot})
can be corrected using the two non-dimensional numbers $N_{F}=\frac{L^{2}f}{c_{2}D}$
in Eq. (\ref{eq:ND number1}) and $N_{i}=\frac{c_{2}}{\gamma fL}$
in Eq. (\ref{eq:ND number2}). More precisely, based on empirical
observations, we find that second-sound tweezers spectra can be accurately
represented by the formula
\begin{equation}
\overline{T}=\frac{A}{\sinh\left(i\left(\frac{2\pi fD}{c_{2}}-a\frac{c_{2}D}{L^{2}f}\right)+b\frac{c_{2}D}{L^{2}f}+c\left(\frac{\gamma fL}{c_{2}}\right)^{2}\right)},\label{eq:modified Fabry-Perot}
\end{equation}
where $a,b$ and $c$ are empirical coefficients. Based on comparison with the full numerical model, we find that the values $a\approx0.95$,
$b\approx0.38$ and $c\approx1.3$ give accurate spectra predictions.
An illustration of a modified Fabry--Perot spectrum with Eq. (\ref{eq:modified Fabry-Perot})
is given in Fig. \ref{fig:Analytical-models sec3.2}. The linear amplitude
growth of the first resonant peaks can be interpreted as a progressive
focalization of the wave, and is thus controlled by the Fresnel diffraction
number $N_{F}$ in Eq. (\ref{fig:ND number1}). The shift proportional
to $\frac{1}{f}$ in peaks frequency positions, observed in the experimental
spectra, is also controlled by $N_{F}$. The decrease in resonant
magnitude for large mode numbers can be interpreted as a wave deflection
outside the cavity, after back and forth propagation between the plates.
This latter effect is controlled by the second non-dimensional number
$N_{i}$ in Eq. (\ref{eq:ND number2}).

The major interest of the Fabry--Perot model is to offer an analytical
expression to fit locally a resonant peak of a second sound resonator
spectrum. The local fit of a peak is of particular interest to interpret
the experimental data, as will be explained in sec. \ref{sec:Measurements-with-second}.
Based on Eq. (\ref{eq:modified Fabry-Perot}), given a measured resonant
frequency $f_{0}$, we will look for a fitting expression
\begin{equation}
\overline{T}=\frac{A}{\sinh\left(i\frac{2\pi(f-f_{0})D}{c_{2}}+\xi_{0}D\right)},\label{eq:local fit}
\end{equation}
valid for second sound frequencies $f$ close to $f_{0}$. In that
expression, $\xi_{0}$ encapsulates the different geometrical mechanisms
responsible for energy loss when the fluid is at rest. $A$ and $\xi_{0}$
are thus fitting parameters that can be found easily with the experimental
data obtained by varying $f$ in the vicinity of $f_{0}$.\\

\subsection{Numeric algorithm\label{subsec:Numeric-algorithm}}

We develop in the present section a numerical algorithm, based
on the exact resolution of the wave equations with the particular
tweezers' geometry, with and without flow. The algorithm could be extended
to any second sound resonator with a planar geometry. As will become
clear in the following, this numerical model allows going far beyond
the approximate models of sec. \ref{subsec:Analytical-approximations}.

\subsubsection{For a backgroud medium at rest\label{subsec:backgroud at rest}}

\begin{figure*}
\includegraphics[width=0.45\textwidth]{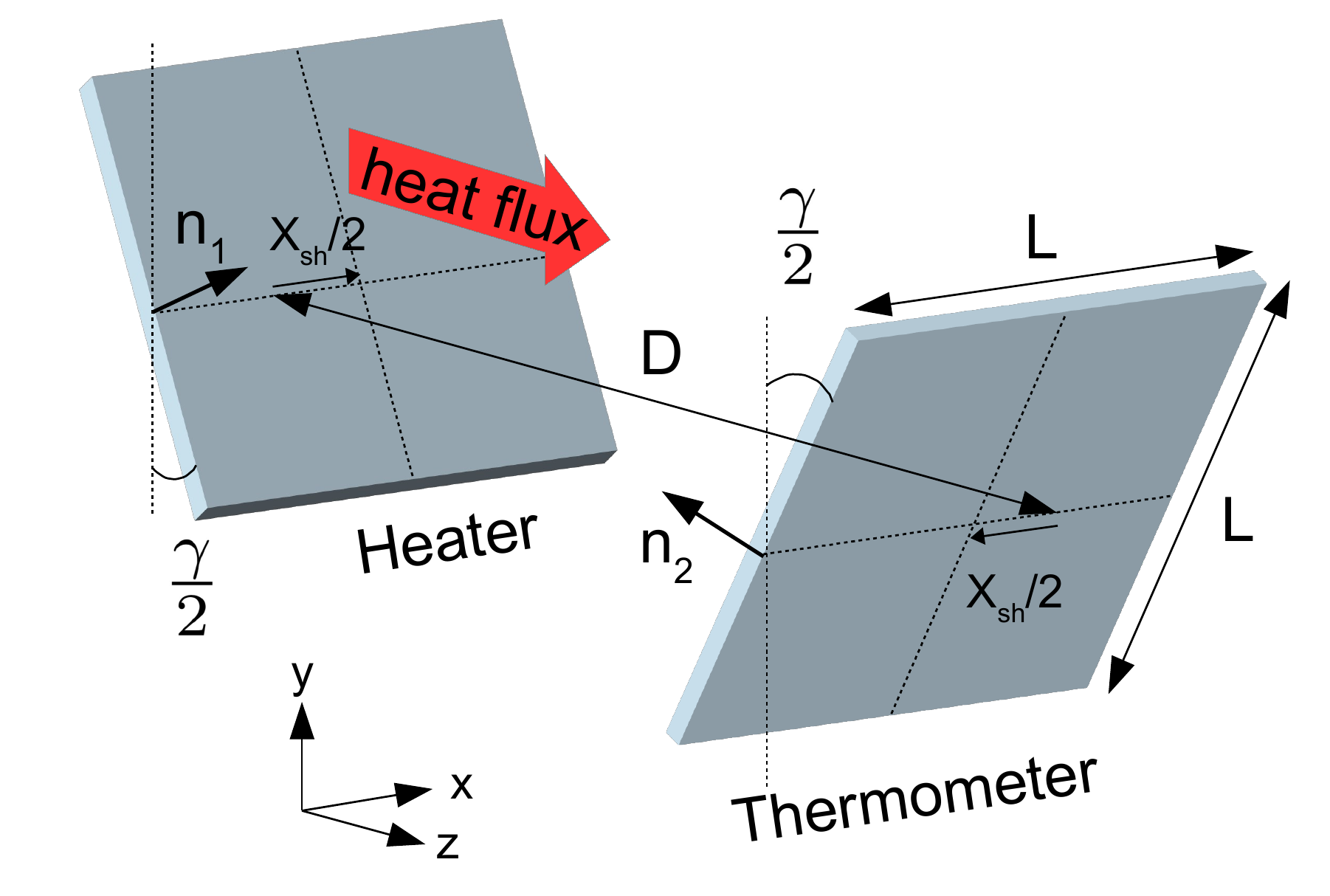}
\hspace{1cm}
\includegraphics[width=0.3\textwidth]{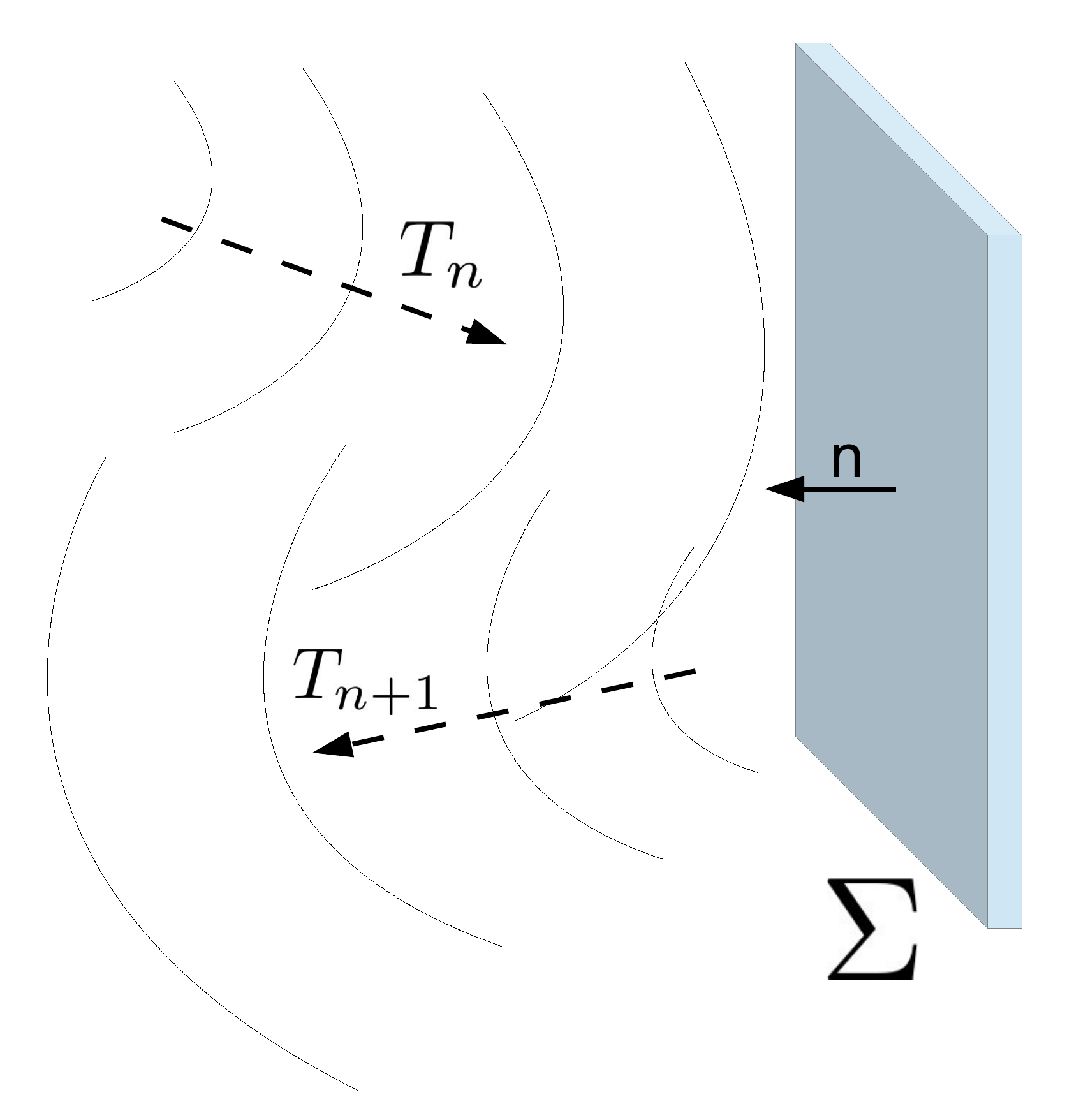}
\caption{\textbf{Left: }Geometrical setup of the numerical algorithm and notations.
\textbf{Right: }Representation of an incoming and outcoming wave at
the n$^{th}$ reflection. \label{fig:notations}}

\end{figure*}

The aim of the present section is to build a numerical algorithm to
solve the wave equation (\ref{eq:wave eq for T}) for a periodic heating
$j_{Q}=j_{0}e^{2i\pi ft}$. We look for a solution with the ansatz
$T(\mathbf{r},t)=\mathcal{R}e\left(\overline{T}(\mathbf{r})e^{2i\pi ft}\right)$.
Then, the wave equation for $\overline{T}$ is
\[
\Delta\overline{T}+k^{2}\overline{T}=0,
\]
 where we have introduced the wave number $k=\frac{2\pi f}{c_{2}}$.
The boundary conditions are 
\begin{equation}
\begin{cases}
\nabla\overline{T}(\mathbf{r}).\mathbf{n}_{1}=-\frac{ikj_{0}}{\rho c_{p}c_{2}} & \text{for }\mathbf{r}\in\Sigma_{1}\\
\nabla\overline{T}(\mathbf{r}).\mathbf{n}_{2}=0 & \text{for }\mathbf{r}\in\Sigma_{2}
\end{cases},\label{eq:boundary conditions_algorithm}
\end{equation}
where $\Sigma_{1}$ is the heater plate and $\Sigma_{2}$ the thermometer
plate. The temperature fluctuations have to vanish far away from the tweezers, which implies $T\rightarrow 0$ when $\mathbf{r}\rightarrow \infty$. The notations are given in Fig. \ref{fig:notations}. We propose
the method described below, based on the Huyggens--Fresnel principle.
The principle states that every point of the wave emitter can be considered
as a point source. The linearity of the wave equation can then be
used to reconstruct the entire wave by summation of all point source
contributions. The Huyggens--Fresnel principle has been widely used
in the context of electromagnetism, for example to compute diffraction
patterns produced by small apertures, or interference patterns...
The major difficulty in the context of second sound tweezers is that
none of the standard approximations of electromagnetism can be done,
neither the far-field approximation nor the small wavelength approximation.
This explains why numerical resolution is very useful in this context.

We neglect the tweezers arms, which means that both plates are considered
as freestanding, infinitely thin and perfectly insulating plates.
We allow a relative inclination $\gamma$ around the $x$-axis and
a possible relative lateral shift $X_{sh}$ of one plate with respect
to the other along the $x$-axis. We assume that the thermometer is
sensitive to the temperature averaged over $\Sigma_{2}$.\\

Let us introduce the Green function 
\begin{equation}
G(\mathbf{r})=\frac{1}{\left|\mathbf{r}\right|}e^{-ik\left|\mathbf{r}\right|},\label{eq:Green function}
\end{equation}
which is the fundamental solution of the wave equation 
\begin{equation}
\Delta G+k^{2}G=4\pi\delta(\mathbf{r}).\label{eq:wave eq with sources}
\end{equation}
Let $\Sigma$ be one of our two square plates, and $U(\mathbf{r}')$
be a smooth function defined over $\Sigma$. We introduce the wave
defined by
\begin{equation}
\overline{T}(\mathbf{r})=\frac{-1}{2\pi}\iint_{\Sigma}G(\mathbf{r}-\mathbf{r'})U(\mathbf{r'})\;{\rm d^{2}\mathbf{r}'}.\label{eq: diffracted wave}
\end{equation}
By linearity, $\overline{T}$ is a solution of Eq. (\ref{eq:wave eq for T}),
for all $\mathbf{r}\notin\Sigma$, because $G$ is a solution. An asymptotic
calculation in the vicinity of $\Sigma$ then shows that $\overline{T}$
satisfies the boundary condition
\begin{equation}
\nabla\overline{T}(\mathbf{r}).\mathbf{n}\underset{\mathbf{r}\rightarrow\mathbf{r}_{0}\in\Sigma}{\longrightarrow}U(\mathbf{r}_{0}),\label{eq:boundary convergence}
\end{equation}
where $\mathbf{n}$ is the unit vector normal to $\Sigma$ and directed
inward the cavity (see Fig. \ref{fig:notations}). We are going to
use Eqs. (\ref{eq: diffracted wave}) and (\ref{eq:boundary convergence})
as the two fundamental relations to build our algorithm. We will compute
the solution of the wave equation as an infinite summation of all
the emitted and reflected waves in the cavity.\\

The first wave $\overline{T}_{1}$ is emitted by the heating plate
$\Sigma_{1}$ and satisfies the first relation in Eq. (\ref{eq:boundary conditions_algorithm})
\[
\nabla\overline{T}_{1}(\mathbf{r}).\mathbf{n}_{1}=-\frac{ikj_{0}}{\rho c_{p}c_{2}}\quad\text{for }\mathbf{r}\in\Sigma_{1}.
\]
Given Eq. (\ref{eq: diffracted wave}) and (\ref{eq:boundary convergence}),
it is clear that the first wave is given by 
\begin{equation}
\overline{T}_{1}(\mathbf{r})=\frac{ikj_{0}}{2\pi\rho c_{p}c_{2}}\iint_{\Sigma_{1}}G(\mathbf{r}-\mathbf{r'})\;{\rm d^{2}\mathbf{r}_{1}}.\label{eq:first wave}
\end{equation}
Then each time a wave denoted $\overline{T}_{n}$ hits a plate $\Sigma$
($\Sigma_{1}$ or $\Sigma_{2}$), it produces a reflected wave $\overline{T}_{n+1}$
to satisfy the boundary condition
\begin{equation}
\nabla\left(\overline{T}_{n}(\mathbf{r})+\overline{T}_{n+1}(\mathbf{r})\right).\mathbf{n}=0.\label{eq:reflection boundary}
\end{equation}
The situation is sketched in the left panel of Fig. (\ref{fig:notations}).
If we choose for $\overline{T}_{n+1}$ the expression 
\begin{equation}
\overline{T}_{n+1}(\mathbf{r})=\frac{1}{2\pi}\iint_{\Sigma}G(\mathbf{r}-\mathbf{r'})\left[\nabla\overline{T}_{n}(\mathbf{r}').\mathbf{n}\right]\;{\rm d^{2}\mathbf{r}'},\label{eq:reflected wave}
\end{equation}
then Eq. (\ref{eq:boundary convergence}) shows that $T_{n+1}$ satisfies
the boundary condition 
\[
\nabla\overline{T}_{n+1}(\mathbf{r}).\mathbf{n}\underset{\mathbf{r}\rightarrow\mathbf{r}_{0}\in\Sigma}{\longrightarrow}-\nabla\overline{T}_{n}(\mathbf{r}_{0}).\mathbf{n},
\]
which is exactly Eq. (\ref{eq:reflection boundary}). Eqs. (\ref{eq:first wave})
and (\ref{eq:reflected wave}) define our recursive algorithm. Eq.
(\ref{eq:reflected wave}) shows that the reflected wave is generated
by the gradient of the incident wave. Practically, the recursive computation
of all forth and back reflected waves thus requires at each step $n$
the computation of $\nabla\overline{T}_{n}$ only on the plates, rather
than $\overline{T}_{n}$. For a reflection at (say) $\Sigma_{1}$,
we have:
\begin{widetext}
\begin{equation}
\nabla\overline{T}_{n+1}(\mathbf{r}).\mathbf{n}_{2}=\frac{-1}{2\pi}\iint_{\Sigma_{1}}G(\mathbf{r}-\mathbf{r}_{1})\left[\frac{1}{\left|\mathbf{r}-\mathbf{r}_{1}\right|}+ik\right]\mathbf{n}_{2}.\frac{\mathbf{r}-\mathbf{r}_{1}}{\left|\mathbf{r}-\mathbf{r}_{1}\right|}\left[\nabla\overline{T}_{n}(\mathbf{r}_{1}).\mathbf{n}_{1}\right]\;{\rm d}^{2}\mathbf{r}_{1},\label{eq:calcul gradient}
\end{equation}
\end{widetext}
The solution of the wave equation is finally given by the superposition
of all waves $\overline{T}_{n}$, that is
\begin{align*}
\overline{T}(\mathbf{r})= & \underset{n=1}{\stackrel{+\infty}{\sum}}\overline{T}_{n}(\mathbf{r}),\\
= & \frac{1}{2\pi}\iint_{\Sigma_{1}}G(\mathbf{r}-\mathbf{r}_{1})\underset{n=0}{\stackrel{+\infty}{\sum}}\left[\nabla\overline{T}_{2n+1}(\mathbf{r}_{1}).\mathbf{n}_{1}\right]\;{\rm d^{2}\mathbf{r}_{1}}\\
 & +\frac{1}{2\pi}\iint_{\Sigma_{2}}G(\mathbf{r}-\mathbf{r}_{2})\underset{n=1}{\stackrel{+\infty}{\sum}}\left[\nabla\overline{T}_{2n}(\mathbf{r}_{2}).\mathbf{n}_{2}\right]\;{\rm d^{2}\mathbf{r}_{2}}.
\end{align*}
and the thermometer response is given by 
\[
\left\langle \overline{T}\right\rangle _{\Sigma_{2}}=\frac{1}{L^{2}}\iint_{\Sigma_{2}}\overline{T}(\mathbf{r})\;{\rm d^{2}\mathbf{r}}.
\]

\begin{figure}
\includegraphics[width=0.5\textwidth]{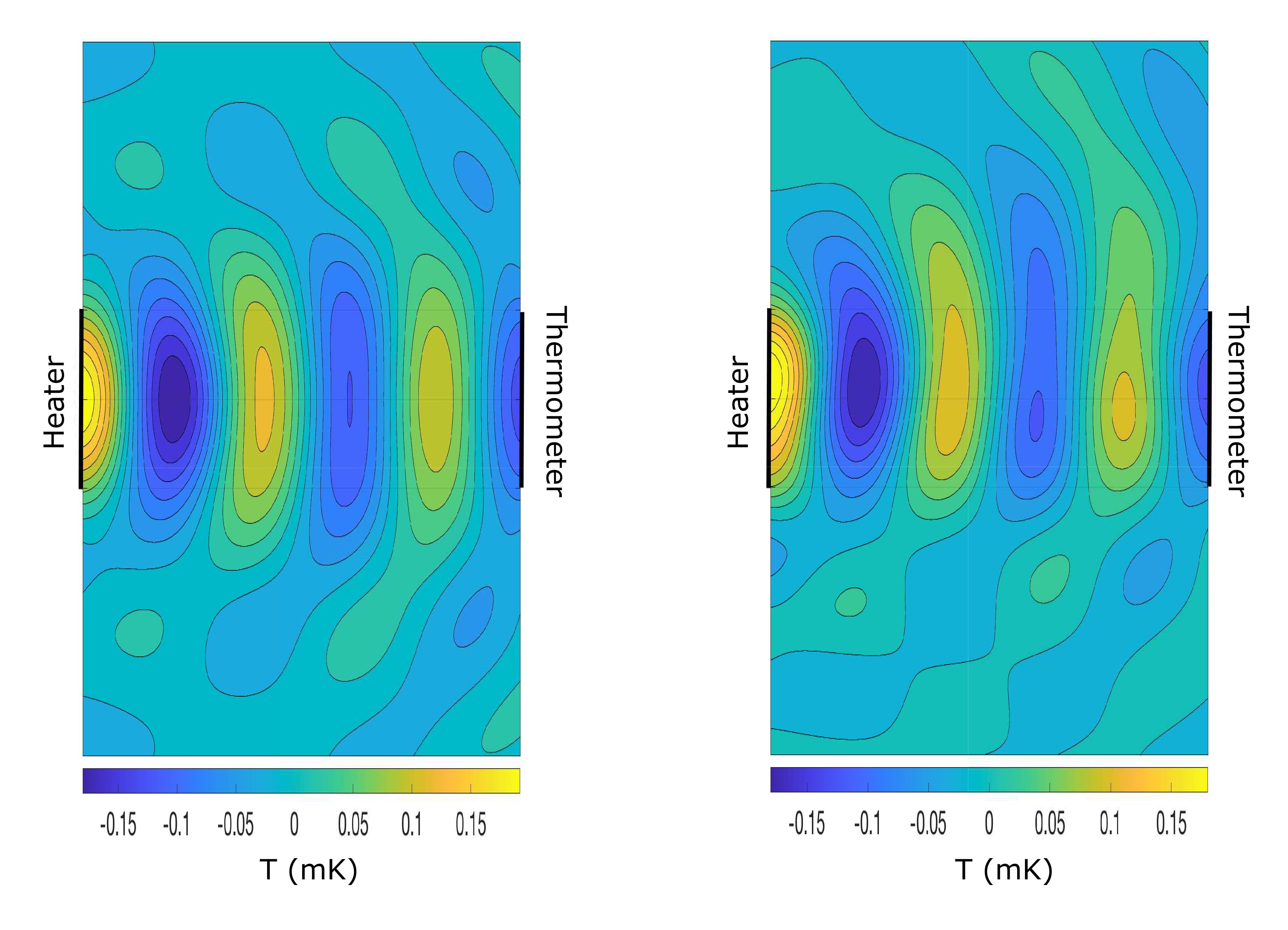}
\caption{\textbf{ Left}:
Temperature field fluctuations of the $5^{th}$ resonant mode of
second sound tweezers with aspect ratio $\frac{L}{D}=0.4$, and heating
power $j_{Q}=0.185$ W/cm$^{2}$, at the bath temperature $T_{0}=2K$.\textbf{
Right}: Temperature field fluctuations for the same conditions with
an additional flow of velocity $\frac{U}{c_{2}}=0.18$ directed upward
\label{fig:Temperature-field_sec3.3}. The nodes of the temperature
standing wave correspond to antinodes of second sound velocity, and
vice-versa.}
\end{figure}

A simulation of the temperature field at $t=0$ of the $5^{th}$ resonant
mode of second sound tweezers with aspect ratio $\frac{L}{D}=0.4$,
without lateral shift nor inclination of the plates, is displayed
in Fig. \ref{fig:Temperature-field_sec3.3}. It can be clearly seen
in particular that the amplitude of the temperature field decreases
along the $z$-axis, contrary to a Fabry--Perot resonator. This symmetry
breaking is due to the diffraction effects associated with the finite
size of the plates. 

A bulk dissipation can be included in the algorithm, for example, to
account for quantum vortex lines inside the cavity. In that case,
let $\xi$ be the second-sound attenuation coefficient (in m$^{-1}$),
the wave number $k=\frac{2\pi f}{c_{2}}$ of the Green function (\ref{eq:Green function})
should be replaced by
\begin{equation}
k=\frac{2\pi f}{c_{2}}-i\xi.\label{eq:wave number}
\end{equation}

\subsubsection{In the presence of a turbulent flow\label{subsec:In-the-presence of flow}}

One of the aims of second-sound resonator modelling is to understand
their response in the present of a flow $\mathbf{U}$ sweeping the
cavity. One effect of the flow is to advect the second sound wave.
In the present section, we explain how the algorithm of sec. \ref{subsec:backgroud at rest}
should be modified to account for this effect. We assume in the following
that the inequality $\left|\mathbf{U}\right|<c_{2}$ is strictly satisfied,
which means that the flow is not supersonic for second sound waves.

In the presence of a non-zero flow $\mathbf{U}$, the Green function
(\ref{eq:Green function}) becomes
\begin{equation}
G(\mathbf{r},t)=\frac{e^{-2i\pi ft^{*}}}{\left|\mathbf{r}-\mathbf{U}t^{*}\right|\left(1+\frac{\mathbf{U}}{c_{2}}.\frac{\mathbf{r}-\mathbf{U}t^{*}}{\left|\mathbf{r}-\mathbf{U}t^{*}\right|}\right)},\label{eq:exact Green}
\end{equation}
 where $t^{*}$ is the time shift corresponding to the signal propagation
from the source
\begin{equation}
\left|\mathbf{r}-\mathbf{U}t^{*}\right|=c_{2}t^{*}.\label{eq:instant retard}
\end{equation}
In practice, the flow velocity range reached in quantum turbulence
experiments is most often much lower than the second sound velocity,
with $\left|\mathbf{U}\right|$ hardly reaching a few m/s. Most experiments
are done in the temperature range where $10<c_{2}<20$ m/s. We thus
introduce the small parameter $\mathbf{\beta}=\frac{\left|\mathbf{U}\right|}{c_{2}}\ll1$.
Similarly to the standard approximations of electromagnetism, we assume
that the effect of $\beta$ is mostly concentrated in the phase shift
$e^{-2i\pi ft^{*}}$ of Eq. (\ref{eq:exact Green}). We use the approximation
$\left|\mathbf{r}-\mathbf{U}t^{*}\right|\left(1+\frac{\mathbf{U}}{c_{2}}.\frac{\mathbf{r}-\mathbf{U}t^{*}}{\left|\mathbf{r}-\mathbf{U}t^{*}\right|}\right)\approx\left|\mathbf{r}\right|$,
and we solve Eq. (\ref{eq:instant retard}) to obtain $t^{*}$ to
leading order in $\beta$. The Green function then becomes 
\begin{equation}
G(\mathbf{r},t)=\frac{e^{-ik\left|\mathbf{r}\right|\Gamma\left(\mathbf{r},\mathbf{U}\right)}}{\left|\mathbf{r}\right|},\label{eq:modified Green}
\end{equation}
where as previously $k=\frac{2\pi f}{c_{2}}$ and
\begin{align*}
\Gamma\left(\mathbf{r},\mathbf{U}\right) & =1-\frac{\mathbf{U}}{c_{2}}.\frac{\mathbf{r}}{\left|\mathbf{r}\right|}.
\end{align*}
The algorithm detailed in sec. \ref{subsec:backgroud at rest} can
be applied straightforward with the Green function Eq. (\ref{eq:modified Green}).
In particular, Eq. (\ref{eq:calcul gradient}) becomes
\begin{widetext}
\begin{equation}
\nabla\overline{T}_{n+1}(\mathbf{r}).\mathbf{n}_{2}=\frac{-1}{2\pi}\iint_{\Sigma_{1}}G\left(\mathbf{r}-\mathbf{r}_{1},\mathbf{U}\right)\left[\left(\frac{1}{\left|\mathbf{r}-\mathbf{r}_{1}\right|}+ik\right)\frac{\mathbf{r}-\mathbf{r}_{1}}{\left|\mathbf{r}-\mathbf{r}_{1}\right|}.\mathbf{n}_{2}-ik\frac{\mathbf{U}}{c_{2}}.\mathbf{n}_{2}\right]\left[\nabla\overline{T}_{n}.\mathbf{n}_{1}\right](\mathbf{r}_{1})\;{\rm d}^{2}\mathbf{r}_{1}.\label{eq:calcul gradient flow}
\end{equation}
\end{widetext}
A simulation of the temperature field at $t=0$ of the $5^{th}$ resonant
mode of second sound tweezers with aspect ratio $\frac{L}{D}=0.4$,
without lateral shift nor inclination, and with a flow of velocity
$\frac{U}{c_{2}}=0.18$, is displayed in the right panel of Fig. \ref{fig:Temperature-field_sec3.3}.
The effect of the flow can be clearly seen with the upward distortion
of the antinodes of the wave, compared to the reference temperature
profile without flow displayed in the left panel.

\subsection{Quantitative predictions\label{subsec:Quantitative-predictions}}

We present in this section the quantitative results obtained with
the algorithm of sec. \ref{subsec:Numeric-algorithm}. The algorithm
is specifically run in the configuration of second sound tweezers,
but most predictions are relevant for other types of second sound
resonators. We first show that the algorithm can quantitatively
account for the experimental spectra. We then use it to predict the
response in the presence of a flow and a bulk dissipation in the cavity.
The predictions are systematically compared to experimental results
for second sound tweezers. We eventually display some experimental
observations that illustrate the limits of our model.

\subsubsection{Spectral response of second sound resonators}

Given a resonator lateral size $L$, the model of sec. \ref{subsec:Numeric-algorithm}
has three geometrical parameters : the gap $D$, the inclination $\gamma$
and the lateral shift $X_{sh}$ (see notations in Fig. \ref{fig:notations}).
We first sketch qualitatively the importance of those three parameters. 

The gap $D$ is the main parameter: it sets the location of the resonant
frequencies, and the quality factor of the resonances at low mode
numbers. For second sound tweezers, the value of $D$ can be usually
obtained within a precision of a few micrometers ($D$ is of the order
of 1 millimeter). The relative inclination of the plates $\gamma$
is responsible for the saturation of the resonant magnitude and its
decrease at large mode numbers. It is typically smaller than a few
degrees. Contrary to the gap, only the order of magnitude of $\gamma$,
not its precise value, can be determined from the tweezers' spectrum.
The lateral shift $X_{sh}$ has very little impact on the spectrum
if the value $\frac{X_{sh}}{L}$ remains small enough (we can typically
reach $\frac{X_{sh}}{L}<0.1$ in the tweezers' fabrication). However,
the effect of this parameter is of paramount importance to understand
open cavity resonators response in a flow (such as second sound tweezers),
and will be investigated in sec. \ref{subsec:Response-with-a flow}.
We consider the case $X_{sh}=0$ in the present section. The tweezers
size $L$ is known from the probe fabrication process.\\

The method goes as follows: we first find a gap rough estimation $\widetilde{D}$,
for example from the average spacing between the experimental resonant
peaks. Then we can run a simulation for parallel plates ($\gamma=0$),
unit gap $D=1$, and aspect ratio $\frac{L}{\widetilde{D}}$, in the
range $0<k^{*}<n\pi$ (where $n$ is the number of modes to be fitted,
and $k^{*}=kD$ is the non-dimensional wave number). This gives a
function $f_{L/\widetilde{D}}(k^{*})$. The experimental spectrum
can then be fitted with the function $T(f)=Af_{L/\widetilde{D}}(\frac{2\pi fD}{c_{2}})$,
where $A$ and $D$ are the two free parameters to be fitted, provided
the experimental value of $c_{2}$ is known. The high sensitivity
of the location of the resonant frequencies makes this method very
accurate to obtain the gap $D$.

\begin{figure}
\includegraphics[width=0.5\textwidth]{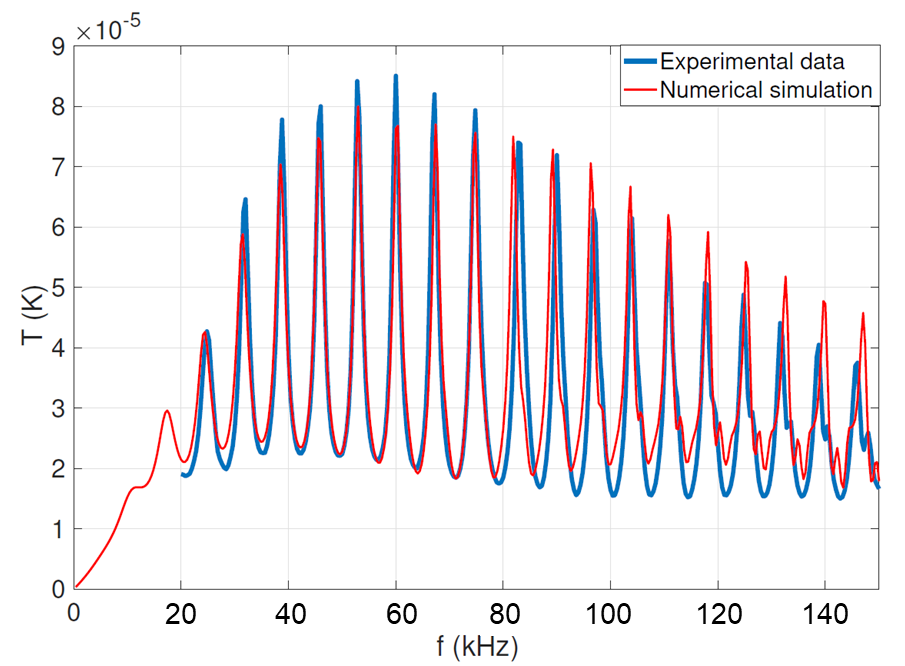}
\caption{The second sound tweezers experimental spectrum of
Fig. \ref{fig:Experimental-spectrum_sec3.1} (blue curve) and the result of the numerical algorithm (red curve).
The remaining fitting parameters of the model are the gap $D$, the inclination $\gamma$ and the total heating power.\label{fig:Prediction-of-the_spectrum}}
\end{figure}

Once $D$ has been found, new simulations have to be run to find the
order of magnitude of $\gamma$. As was previously said, $\gamma$
controls the saturation and the decrease of the resonant magnitudes
for large mode numbers. Its value can thus be approximated from a
fit of the resonant modes with the largest magnitude. A fit of an
experimental tweezers spectrum is displayed in Fig. \ref{fig:Prediction-of-the_spectrum}.
The values of the fitting parameters for this spectrum are $D=1.435\pm0.003$
mm and $\gamma=4.2\pm0.5$ deg. Given the simplicity of the model
assumptions, in particular the assumptions of perfectly insulating
and infinitely thin plates without support arms, the agreement with
experimental results is very good. \\

Interestingly, the resonators can also be used in some conditions
as thermometers. Once the gap $D$ is known with high enough precision,
the spectrum can be fitted using $c_{2}$ as a fitting parameter instead
of $D$. Away from the second sound plateau of the curve $c_{2}(T_0)$
located around $1.65$ K, the value of $c_{2}$ obtained from the
spectrum gives access to the average temperature with a typical accuracy
of one mK, simply by inverting the function $c_{2}(T_0)$.

\subsubsection{Response with a flow\label{subsec:Response-with-a flow}}

\begin{figure}
\includegraphics[width=0.45\textwidth]{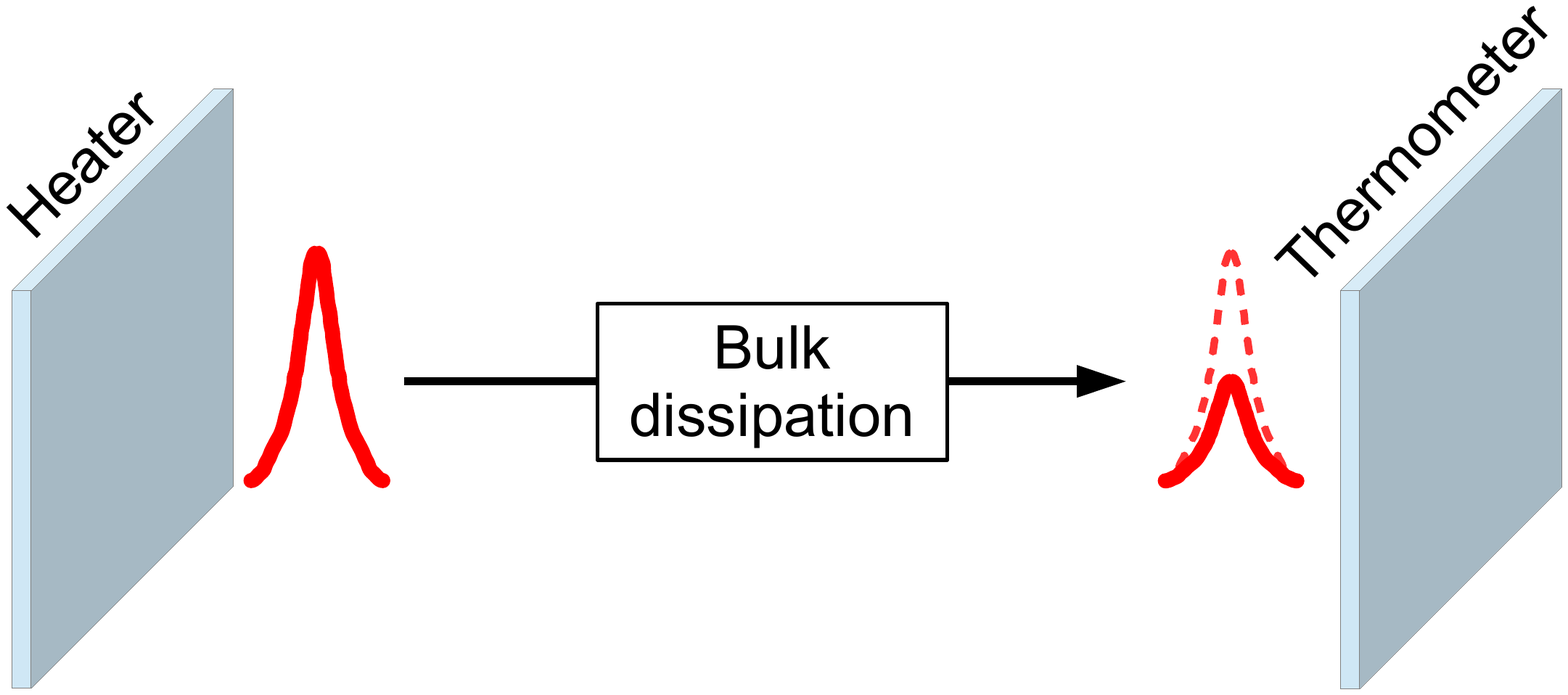}
\includegraphics[width=0.45\textwidth]{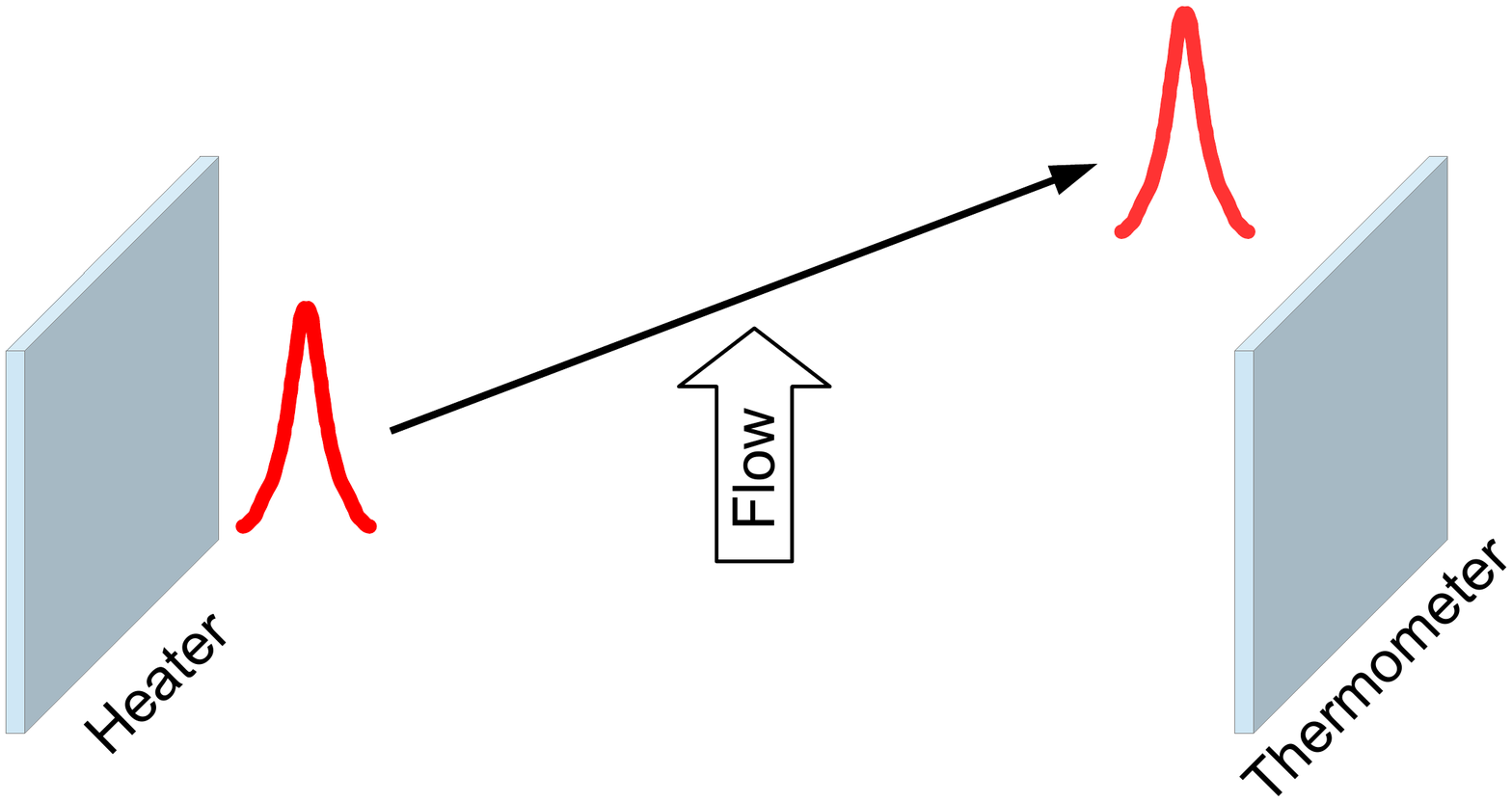}
\caption{Schematic representation of the two attenuation mechanisms. The top
panel illustrates a bulk dissipation of the wave, due for example
to the presence of quantum vortices. The bottom panel illustrates
the ballistic deflection of the wave by a flow directed parallel to
the plates. \label{fig:mechanisms}}
\end{figure}

Once the characteristics of the resonator have been determined in
a background medium at rest, their response in a flow can be studied 
using the modified algorithm presented in sec. \ref{subsec:In-the-presence of flow}.
We experimentally observe that the tweezers response is attenuated
in the presence of a superfluid helium flow. This
attenuation is related to two physical mechanisms, illustrated in
Fig. \ref{fig:mechanisms}: first, the thermal wave crossing the cavity
is damped by the quantum vortices carried by the flow. This type of
damping is usually considered as being proportional to the density
of quantum vortex lines between the plates. Secondly, the flow mean
velocity is responsible for a ballistic advection of the thermal wave
outside the cavity. The thermal wave emitted by the heater partly
``misses'' the thermometer plate, and, even if the wave is not attenuated,
a decrease of the tweezers' response will be observed. Both mechanisms
described above exist in experimental superfluid flows, and cannot
be observed independently: once there is a superfluid flow, quantum vortices
are created. One key objective is to be able to separate the attenuation of the experimental signal  due to bulk attenuation inside the cavity, from the attenuation due to ballistic advection of the wave outside the cavity. We will introduce a mathematical procedure to perform such a separation for a fluctuating signal.

What cannot be experimentally achieved can be simulated with the tweezers
model developed in sec. \ref{subsec:Numeric-algorithm}. The bulk
dissipation can be implemented in the algorithm with a wave number
complex part $\xi$ (see Eq. (\ref{eq:wave number})), and the flow
ballistic deflection can be implemented with a non-zero velocity $\mathbf{U}$
(see Eqs. (\ref{eq:modified Green}-\ref{eq:calcul gradient flow})).
Both effects can be independently studied, by alternatively setting $\xi$
or $\mathbf{U}$ to zero. We first detail below the respective effects
of $\xi$ and $\mathbf{U}$ for perfectly aligned plates ($X_{sh}=0$).\\

\begin{figure*}
\includegraphics[width=0.45\textwidth]{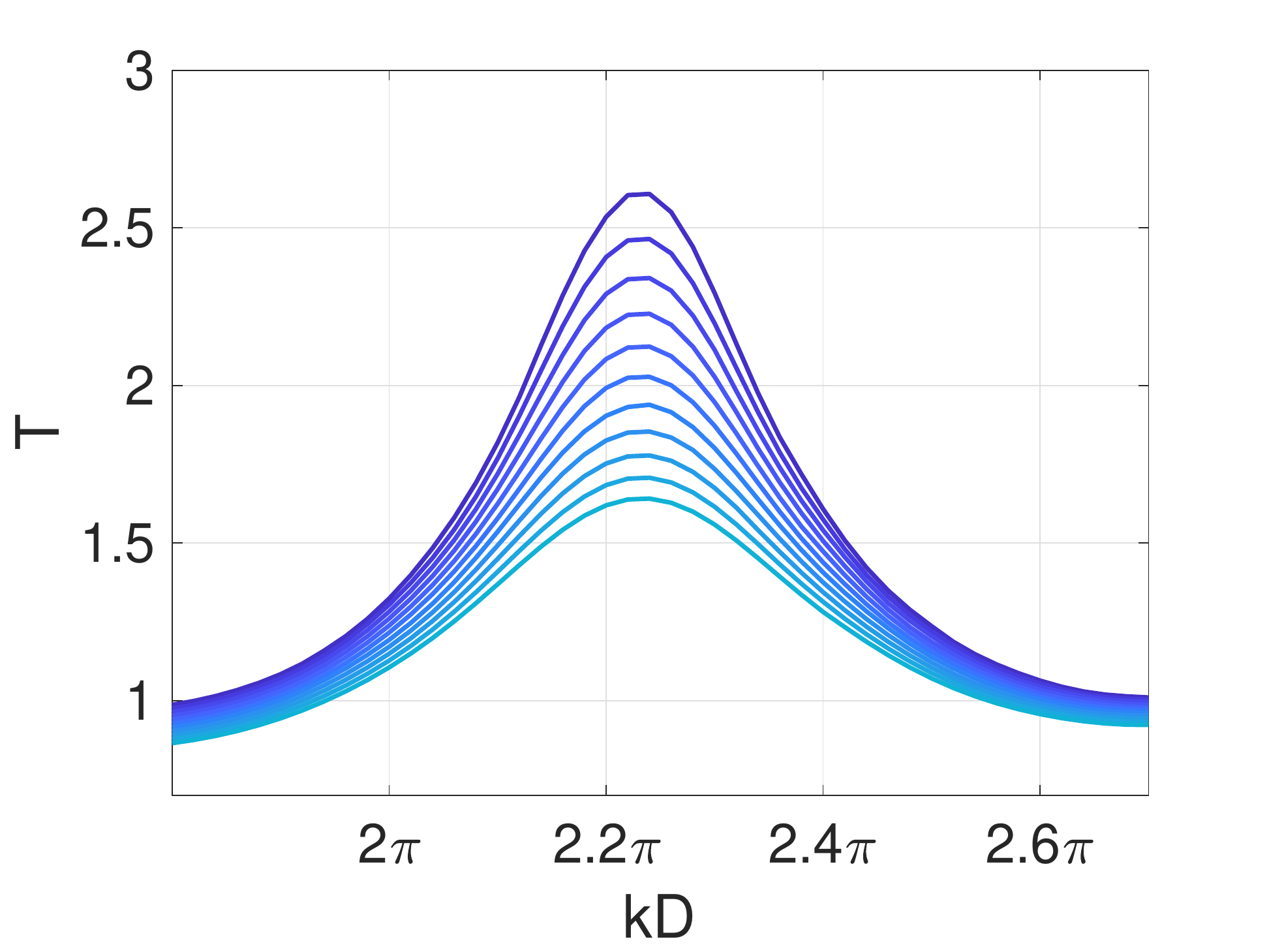}
\hspace{1cm}
\includegraphics[width=0.45\textwidth]{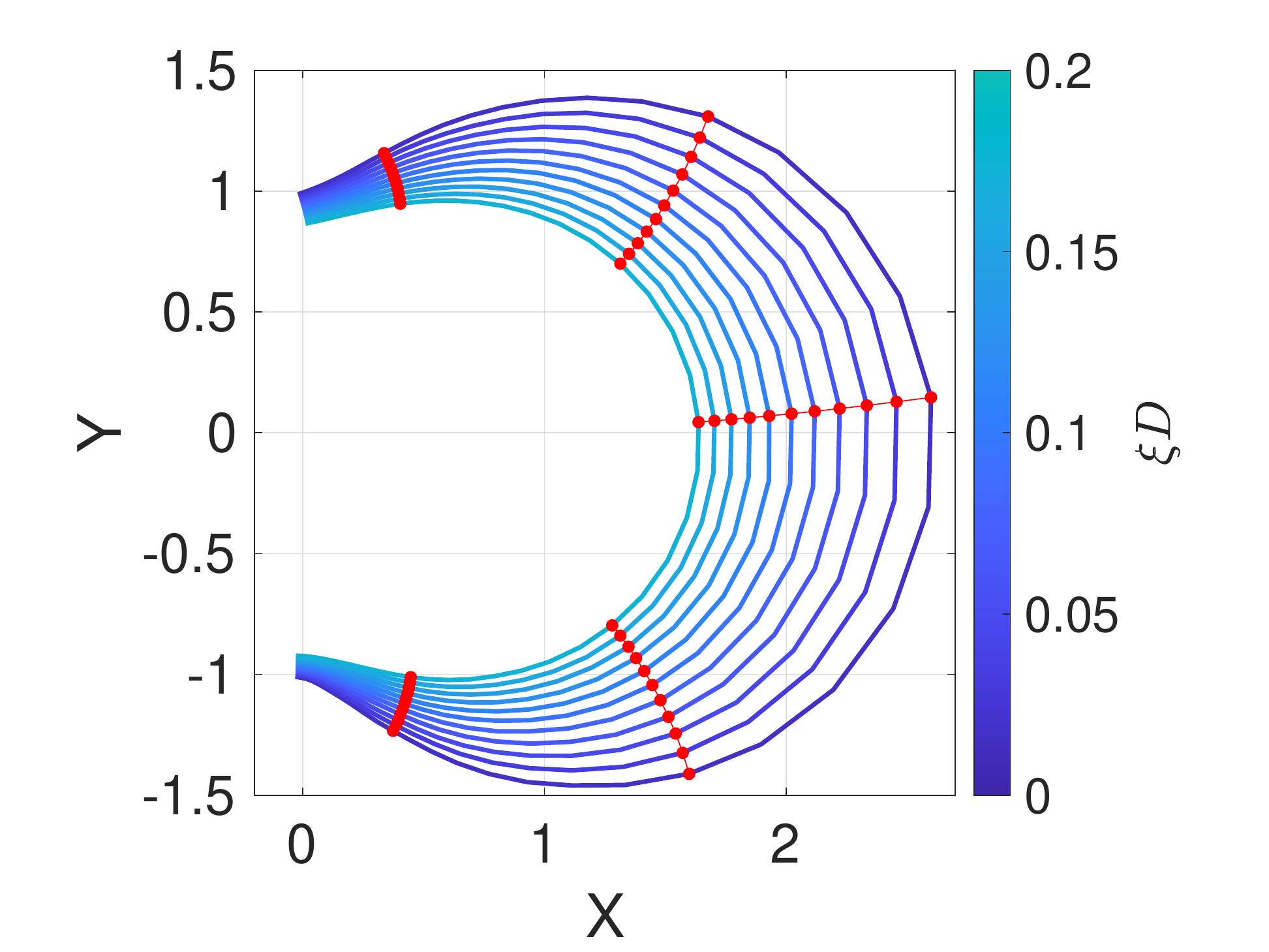}
\caption{\textbf{Numerical simulation:} collapse of a resonance due to increasing
values of the bulk attenuation $\xi$. The left panel display the
magnitude of the thermal wave as a function of the wavevector $k$,
and the right panel display the same resonance in the phase-quadrature
plane. It can be seen that the bulk attenuation results in a homothetic
collapse of the resonance, that means, without global phase shift.
For a given value of $k$, the model predicts that attenuation is
directed toward the center of the resonant Kennelly circle (red curves
of right panel). \label{fig:Collapse-of-a-resonanceXi_sec3.4}}
\end{figure*}

Fig. \ref{fig:Collapse-of-a-resonanceXi_sec3.4} display the result
of a numerical simulation for second sound tweezers of aspect ratio
$L/D=1$, $\gamma=0$ and increasing values of bulk dissipation in
the range $0<\xi D<0.2$. The left panel display the magnitude of
the second resonant mode as a function of the wave number, and the
right panel display the same resonant mode in the phase-quadrature
plane. More precisely, if we call $T(k)$ the thermal wave magnitude
recorded by the thermometer, and $\varphi(k)$ its phase, the right
panel displays the curve $Y(k)=T(k)\sin(\varphi(k)),X(k)=T(k)\cos(\varphi(k))$.

The resonant curve $(Y(k),X(k))$ is called in the following the resonant
\textbf{``Kennelly circle''} (see also sec. \ref{subsec:The-simple-method}), because the curve is very close to a circle
crossing the origin. Furthermore, the resonant curve becomes closer to a perfect circle for increasing resonant quality
factors. The major characteristic to be observed in Fig. \ref{fig:Collapse-of-a-resonanceXi_sec3.4}
is that the collapse of the resonant Kennelly circle due to bulk attenuation
is homothetic. It means that the different curves have no relative
phase shift between each other, when the bulk attenuation increases.
The red curves in the right panel display the displacement in the
phase-quadrature plane for a fixed value of the wavevector. The model
predicts that the displacement is directed toward the Kennelly circle
center, which implies that the path at a fixed wavevector approximately
follows a straight line. By comparison, the left panel of Fig. \ref{fig:shift resonance}
display an experimental resonance in the phase-quadrature plane, for
second sound tweezers of size $L=1$ mm in superfluid Helium at $1.65$
K. The global orientation of the resonant Kennelly circles is simply
due to a uniform phase shift introduced by the measurement devices,
and should be overlooked. It can be seen that the resonance collapse
with increasing values of the flow velocity follows the predictions
of Fig. \ref{fig:Collapse-of-a-resonanceXi_sec3.4}: it is homothetic.
The red paths correspond to the tweezers signal at fixed heating frequency.
Those paths follow approximately a straight line directed to the Kennelly
circle center. The slight deviation in the path orientation from the predictions of Fig. \ref{fig:Collapse-of-a-resonanceXi_sec3.4}
can be explained by a second sound velocity reduction and will be
discussed in sec. \ref{subsec:Limits-of-the}.\\

\begin{figure*}
\includegraphics[width=0.45\textwidth]{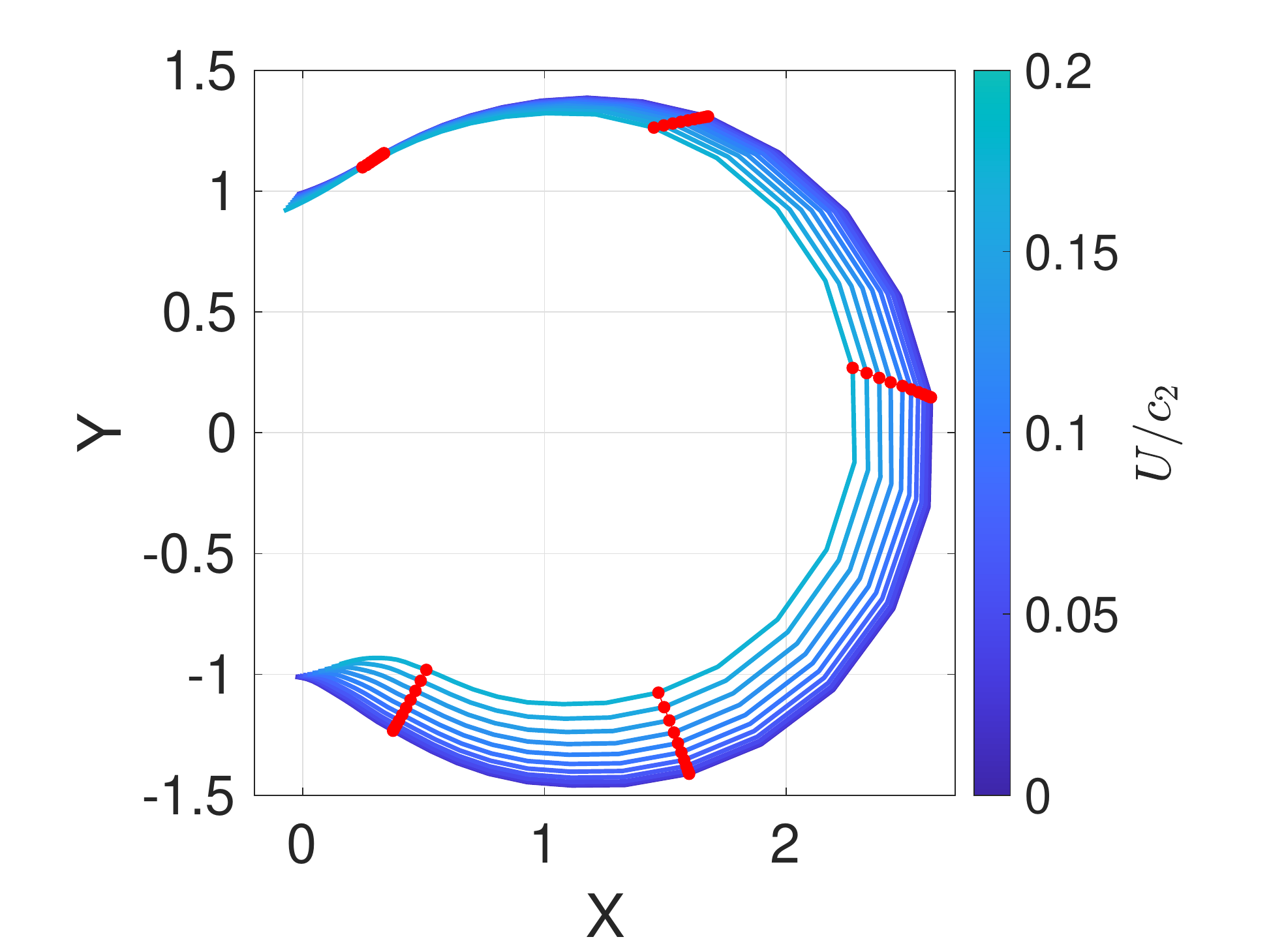}
\hspace{1cm}
\includegraphics[width=0.45\textwidth]{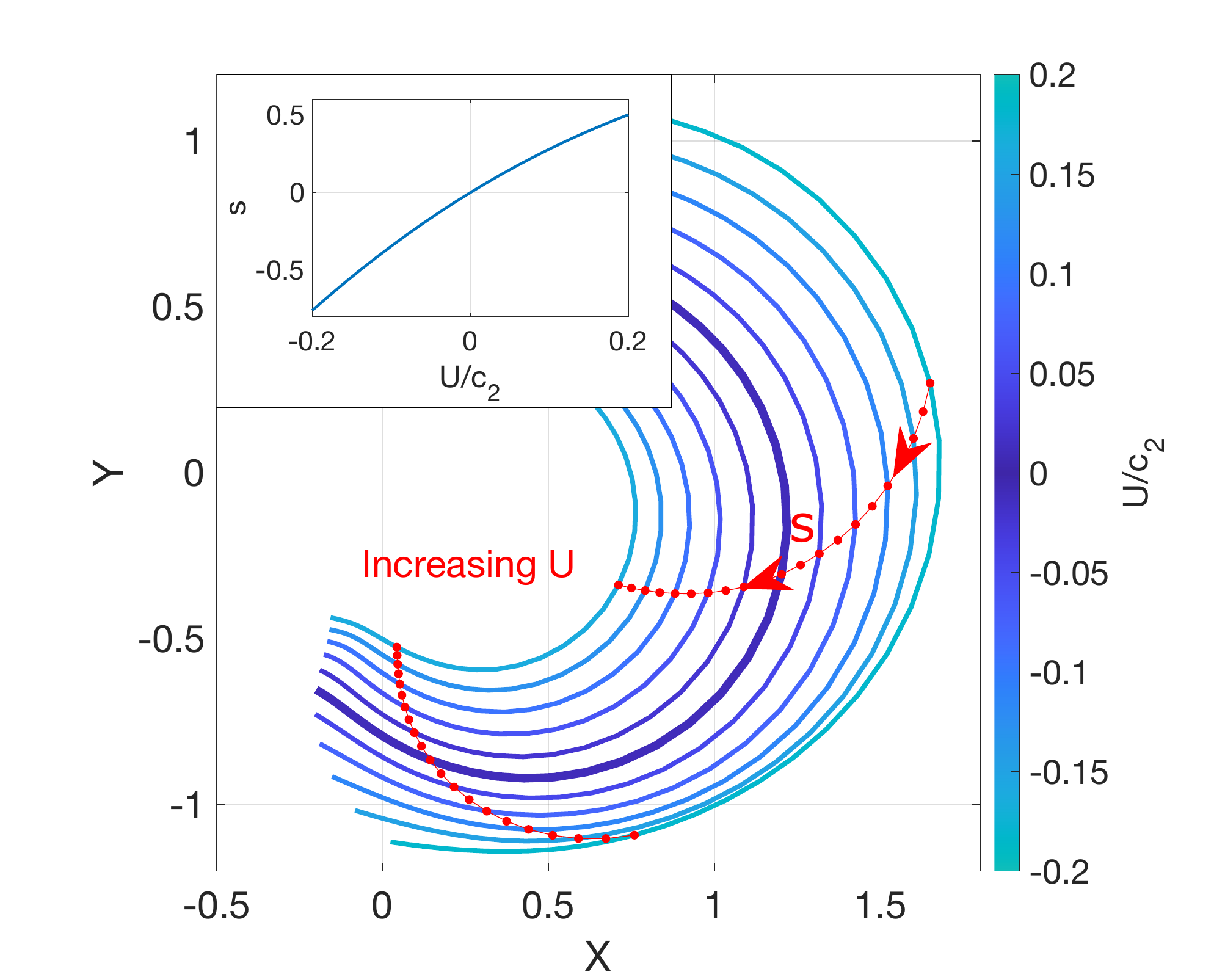}
\caption{\textbf{Numerical simulation:} collapse of a resonance due to ballistic
deflection of the thermal wave in the presence of a flow of velocity
$U$, without bulk attenuation ($\xi$=0). The left panel display
the result for tweezers without lateral shift ($X_{sh}=0$). Contrary
to the results of Fig. \ref{fig:Collapse-of-a-resonanceXi_sec3.4},
it can be seen in the present case that the collapse is associated
with a global anti-clockwise phase shift of the resonant Kennelly
circle. Each red curve represents the attenuation at a given value
of the wavevector $k$. The right panel displays the result for tweezers
with a strong lateral shift $X_{sh}=0.5\times L$ (where L is the
tweezers size). Such tweezers are very sensitive to the velocity
$U$, with both an attenuation of the resonance and a strong clockwise
angular shift of the Kennelly circle. We note $s$ the curvilinear
abscissa of the curve obtained at a given value of $k$ (red curve).
The inset displays the function $s(U)$. This shows that, once calibrated,
second-sound tweezers can be used as anemometers.\label{fig:Numerical-simulation:-collapse ballistic}}
\end{figure*}

Fig. \ref{fig:Numerical-simulation:-collapse ballistic} display the
result of a numerical simulation for second sound tweezers of aspect
ratio $L/D=1$, $\gamma=0$, with $\xi=0$ and a flow mean velocity
$0<\frac{U}{c_{2}}<0.2$. As there is no tweezers lateral shift $X_{sh}=0$,
negative velocities would lead to the same result from symmetry considerations.
The figure illustrates the effect of pure ballistic advection on a
resonance in the phase-quadrature plane. First, it can be seen that
the collapse of the resonant Kennelly circle is accompanied by a relative
anti-clockwise phase shift of the curves when the velocity increases.
Also, the displacement of the tweezers signal at fixed wavenumber
follow the red straight paths directed anti-clockwise. This type of
signal strongly contrasts with the one of Fig. \ref{fig:Collapse-of-a-resonanceXi_sec3.4}
obtained for a pure bulk attenuation. The prediction of the left panel
in Fig. \ref{fig:Numerical-simulation:-collapse ballistic} cannot
be directly compared to experiments because, as stated before, a superfluid
flow always carries quantum vortices that overwhelm the tweezers signal
for tweezers satisfying $X_{sh}\approx0$.

\subsubsection{Effect of lateral shift of the emitter and receiver plates}

We discuss in this section the consequences of a lateral shift, that
is $X_{sh}\neq0$ with the notations of Fig. \ref{fig:notations}.
Contrary to the previous sections, the present discussion is restricted
to second sound tweezers, for which a lateral shift has major quantitative
effects. A lateral shift would not be as important, for example in
the case of wall embedded resonators.\\

The lateral shift has a marginal effect on the tweezers' spectrum when
the background fluid is at rest. An effect only appears in the presence
of a nonzero velocity specifically oriented in the shifting direction
$\mathbf{U}=U\mathbf{e}_{x}$, because of the mechanism of ballistic
advection of the thermal wave by the flow (see the representation
of the mechanism in Fig. \ref{fig:mechanisms}). The importance of
this effect depends on the tweezers' aspect ratio, on the reduced velocity
$\beta=\frac{U}{c_{2}}$, and on the lateral shift $X_{sh}$. The
lateral shift in the plates' positioning magnifies the signal component
related to ballistic advection. This property opens the opportunity
to build second sound tweezers for which ballistic advection of the
wave completely overwhelms bulk attenuation from the quantum vortices,
which means that the tweezers signal is in fact a measure of the velocity
component in the shifting direction. We illustrate this mechanism
in Fig. \ref{fig:Numerical-simulation:-collapse ballistic}.

The right panel displays a numerical simulation of a tweezers resonant
mode in the phase-quadrature plane, for the parameters $\frac{L}{D}=1$,
$\gamma=0$ and $X_{sh}=0.5$, for positive and negative values of
the flow velocity in the range $-0.2<\frac{U}{c_{2}}<0.2$. As can
be seen at the first sight, the deformation of the resonant curve
- that we equivalently call the Kennelly circle - is very different
from a deformation due to a bulk attenuation (see Fig. \ref{fig:Collapse-of-a-resonanceXi_sec3.4}).
First, we observe that the deformation can result in an \emph{increase}
of the magnitude of the thermometer signal, when the velocity is negative.
This can be explained in this configuration, because the thermal wave
emitted by the heating plate is redirected toward the thermometer
plate: less energy is scattered outside the cavity when the wave is
first emitted by the heater, and the signal magnitude increases. On
contrary, the signal magnitude decreases when the velocity is positive
because the flow advects the emitted thermal wave further away from
the thermometer plate and more energy is scattered outside the cavity.
Second, the deformation of the Kennelly circle is associated to a
global clockwise rotation, a phenomenon that is not observed for bulk
attenuation in Fig. \ref{fig:Collapse-of-a-resonanceXi_sec3.4}. Coming
back to Fig. \ref{fig:Numerical-simulation:-collapse ballistic},
the red curve displays the displacement in the phase-quadrature plane
for a fixed wave frequency value. The displacement follows a very
characteristic curved path, always directed clockwise. Let $s(U)$
be the curvilinear abscissa of the red path. Once calibrated, the
value of $s$ can be used as a measure of the flow velocity component
in the $\mathbf{e}_{x}$ direction. 

\begin{figure*}
\includegraphics[width=0.45\textwidth]{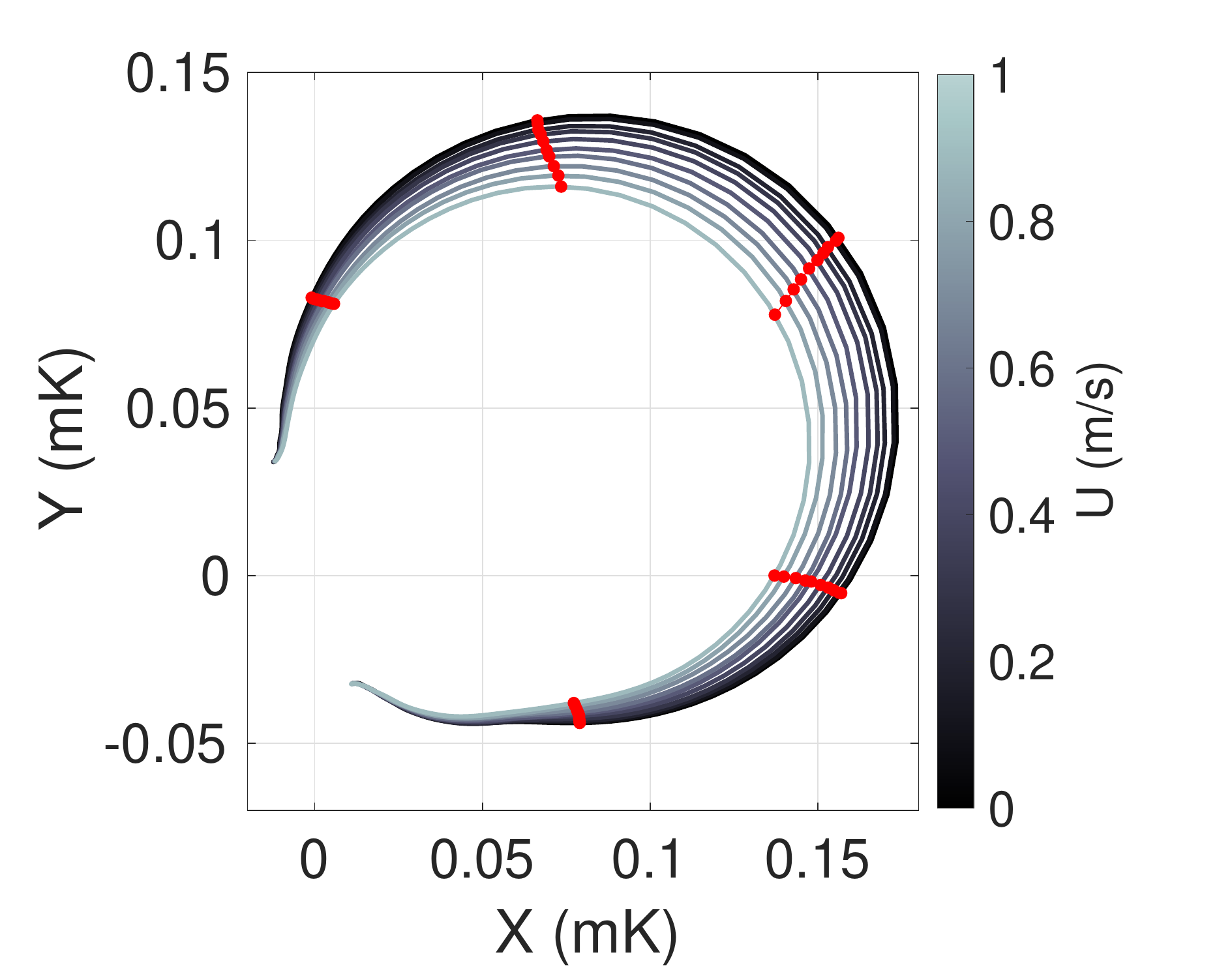}
\hspace{1cm}
\includegraphics[width=0.45\textwidth]{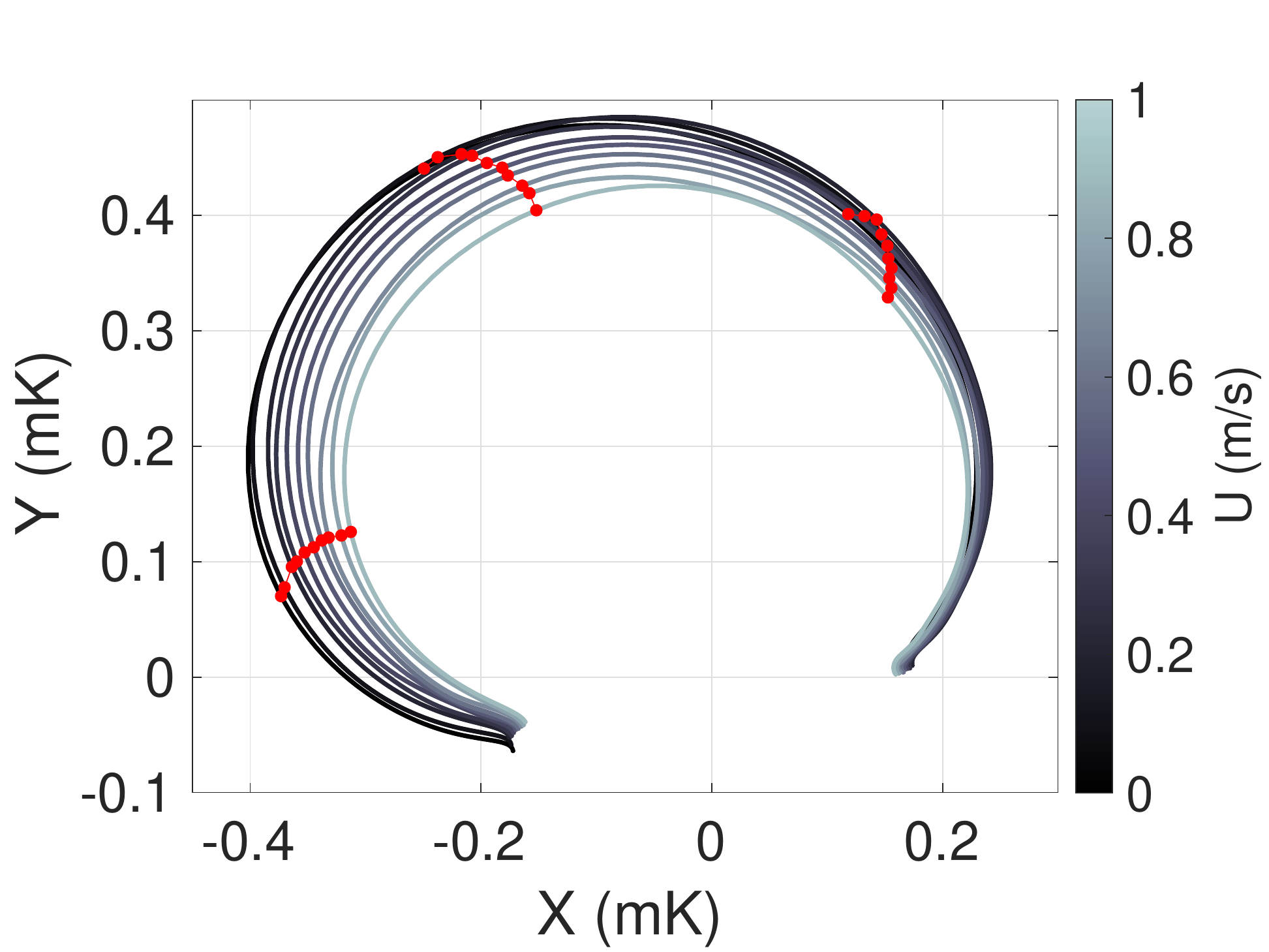}
\caption{\textbf{Experiment: }Collapse of a second sound resonance for increasing
values of the flow mean velocity $U$, in superfluid Helium at $T_{0}\approx1.65$
K. The right panel display the result for tweezers of size $L=1$
mm and minor lateral shift $X_{sh}<0.1\times L$. The figure shows
a homothetic collapse of the resonant Kennelly circle without global
phase shift, as predicted by the model of Fig. \ref{fig:Collapse-of-a-resonanceXi_sec3.4}.
The red curves display the displacement in the phase-quadrature plane
at a fixed value of the second sound frequency $f$. The right panel
displays the experimental data obtained with shifted second sound
tweezers with parameters $L=250$ $\mu{\rm m}$ and $X_{sh}=0.5\times L$.
The figure qualitatively confirms the clockwise angular shift with
increasing values of $U$, predicted by the numerical simulations
of Fig. \ref{fig:Numerical-simulation:-collapse ballistic}.\label{fig:shift resonance}}
\end{figure*}

The right panel of Fig. \ref{fig:shift resonance} displays the experimental
signal observed with second sound tweezers of size $L=250$ $\mu{\rm m}$,
$D=431$ $\mu{\rm m}$ and $X_{sh}\approx125$ $\mu{\rm m}$, for
a positive velocity range $0<U<1$ m/s. The main characteristics of
a ballistic advection signal can be observed: the Kennelly circle
are attenuated with a clear clockwise rotation, and the signal at
fixed frequency follows a curved path in the clockwise direction.
This is a strong indication that those type of tweezers can be used
as anemometers. The signal fluctuations of those type of tweezers
were recently characterized in a turbulent flow of superfluid helium
\cite{woillez2021vortex}. It has been shown in particular that both
the signal spectra and its probability distributions indeed display
all the characteristics of that of turbulent velocity fluctuations. 

\subsubsection{Limits of the model\label{subsec:Limits-of-the}} %XXX mettre à la fin ??? XXXX

Although the model of sec. \ref{subsec:Numeric-algorithm} gives excellent
experimental predictions, we still observe some unexpected phenomena
with real second sound tweezers. We discuss two of them in this section.

\begin{figure}
\includegraphics[width=0.45\textwidth]{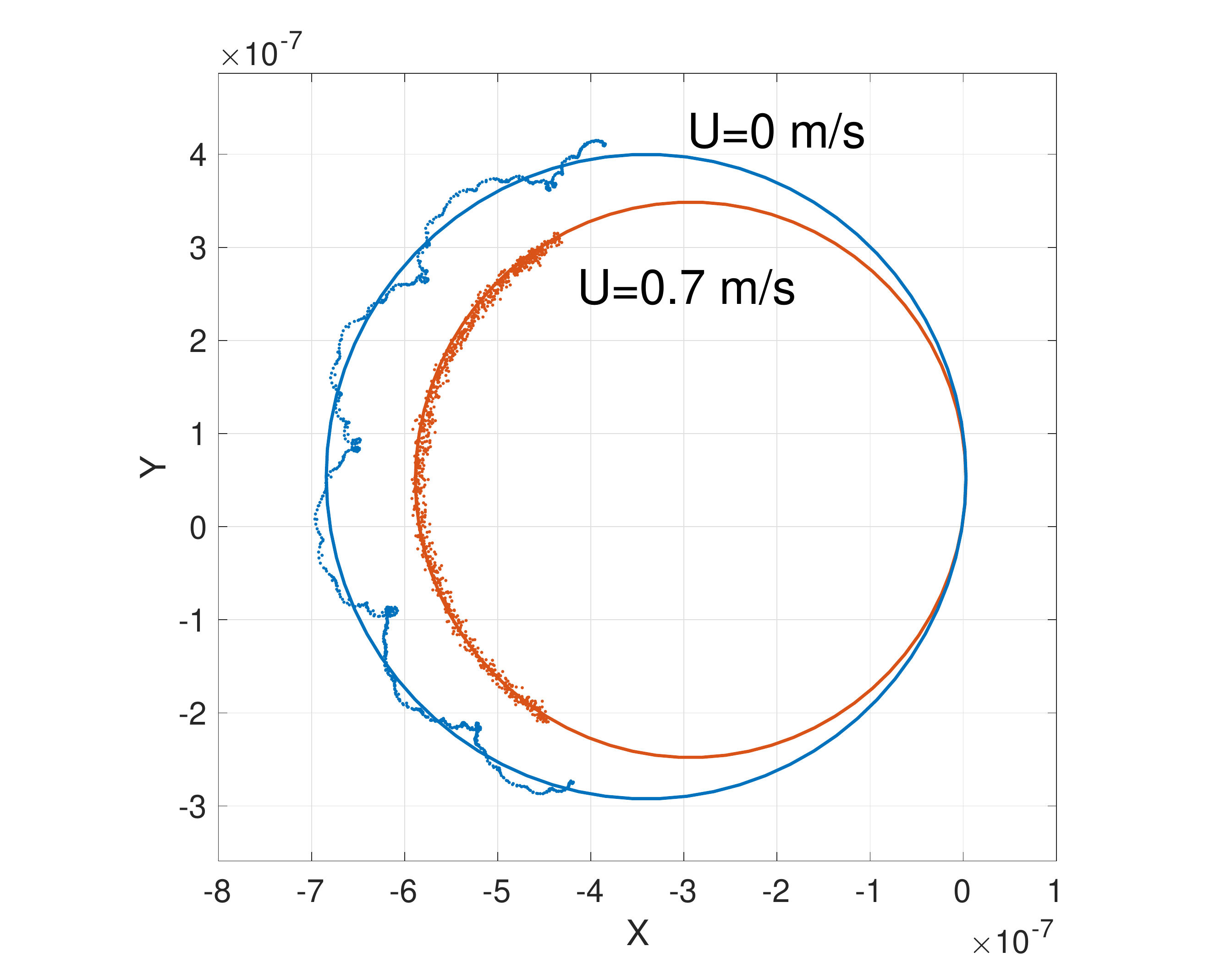}
\caption{Experimental resonance obtained with a second sound tweezers at 1.98
K, for two values of the He flow mean velocity $U$. The blue curve
displays a periodic perturbation of the resonance that we refer to
as the ``daisy effect''. The circle is a fit of the Kennelly osculating
circle for this resonance. This effect is not predicted by our model,
and we interpret it as a secondary resonance in the experimental setup.
The daisy effect perturbs the measurements at low values of $U$,
but it can be seen on the red curve that the effect disappears for
higher values of $U$. \label{fig:paquerette}}

\end{figure}

We have seen in secs. \ref{subsec:Response-with-a flow} that the
thermal wave complex amplitude $\overline{T}(f)$ can be represented
in the phase-quadrature plane by a curve $(X(f),Y(f))$ very close
to a circle crossing the origin. This osculating circle will be called
hereafter the resonant ``Kennelly circle'' (a more precise definition is given in sec. \ref{subsec:The-simple-method}). The wave is damped
in the presence of a superfluid flow, which can be seen in the phase-quadrature
plane as a homothetic shrink of the Kennelly circle toward the origin.
Fig. \ref{fig:paquerette} displays an experimental resonance in the
phase-quadrature plane, for $U=0$ m/s and $U=0.7$ m/s, together
with the fitted Kennelly circles. As can be seen in the figure, the
resonant curve at $U=0$ has periodic oscillations in and out of the
Kennelly circle. We call this phenomenon the ``daisy effect''. The
daisy effect progressively disappears for increasing values of $U$,
and cannot be seen any more on the resonant curve at $U=0.7$ m/s.
We interpret the daisy effect as a secondary resonance in the experimental
setup with a typical acoustic path of a few centimeters. We assume
that the flow kicks out the thermal wave from this secondary resonant
path when $U$ is increased. The daisy effect alters the attenuation
measurements close to $U=0$, and should be considered with care before
assessing the vortex line densities for very low mean velocities.\\

\begin{figure}
\includegraphics[width=0.45\textwidth]{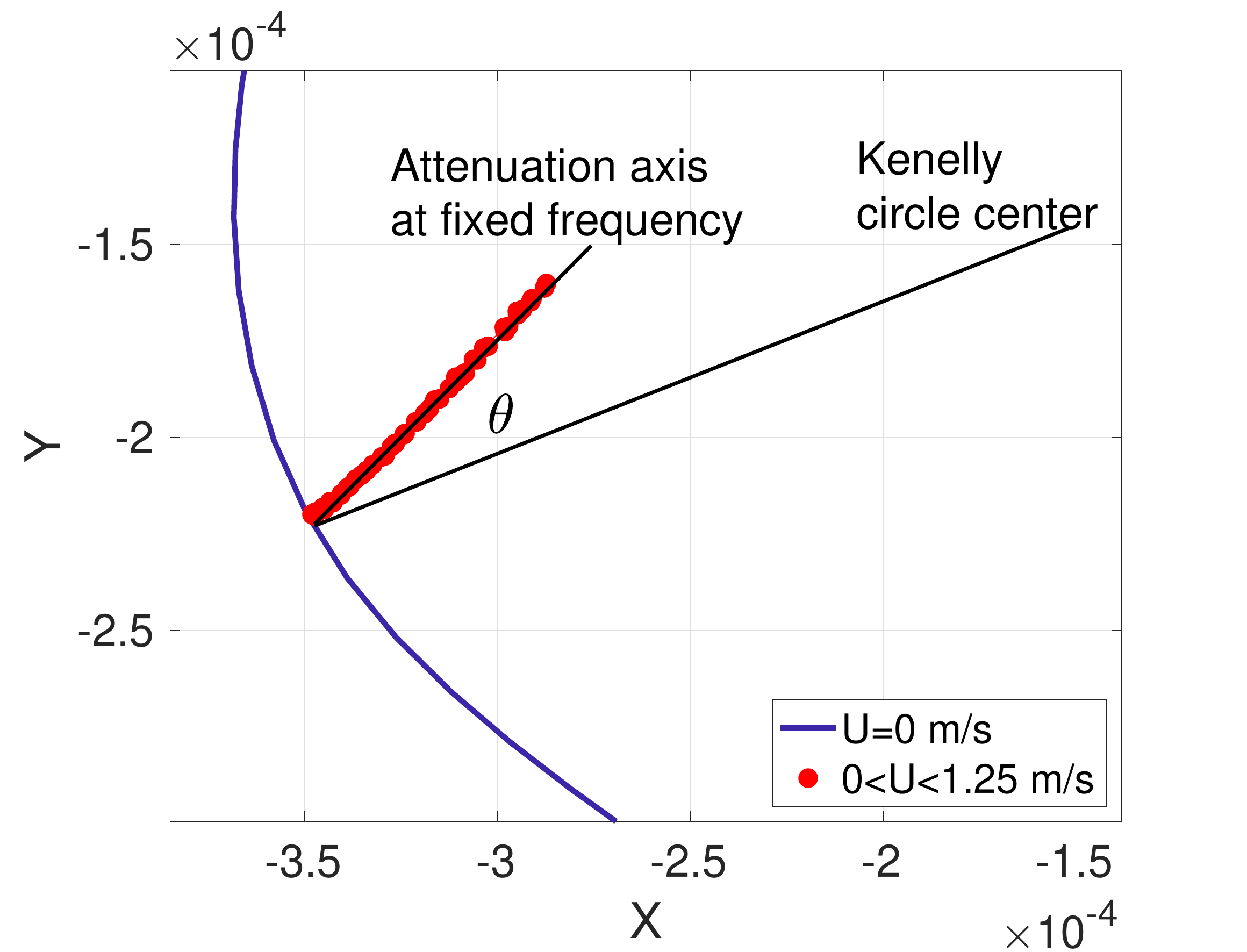}
\caption{Attenuation of a resonance in the presence of a flow mean velocity
$U$, obtained with second sound tweezers in superfluid Helium at
1.65 K. The blue curve is the resonance at $U=0$ in the phase-quadrature
plane. Contrary to the prediction of the model of section \ref{subsec:Quantitative-predictions}
(see also Fig. \ref{fig:Collapse-of-a-resonanceXi_sec3.4}), the attenuation
signal at fixed heating frequency is not directed exactly toward the
centre of the Kennelly osculating circle. We observe a systematic
clockwise angular shift $20{^\circ}<\theta<30{^\circ}$. We still
have no definite explanation for this observation. \label{fig:angle attenuation axis}}
\end{figure}

It has been shown in sec. \ref{subsec:Response-with-a flow} that
the displacement of the tweezers signal in the phase quadrature plane,
for a fixed wave frequency, follows a straight line. We call ``attenuation
axis'' the direction of this straight path. The model predicts that
the attenuation axis should always be directed toward the center of
the resonant Kennelly circle. Fig. \ref{fig:angle attenuation axis}
displays a zoom on a part of the Kennelly circle at $U=0$, together
with the signal displacement at fixed frequency and for increasing
flow velocity. It can be seen that the displacement is indeed a straight
line, but not exactly directed toward the Kennelly circle center.
An angle between $20^{\circ}$ and $30^{\circ}$
is systematically observed between the attenuation axis and the circle
center direction (see Fig. \ref{fig:angle attenuation axis}). Moreover,
the angle is always positive (with the figure convention) and cannot
be interpreted as a ballistic advection, that would give a negative
angle instead. This effect is thus very likely been attributed to
a decrease of the second sound velocity in the presence of the quantum
vortices. Whereas a second sound velocity reduction has previously
been observed in the presence of quantum vortices \cite{lhuillier1974temperature,mehl1974new,miller1978velocity},
the exact value of this reduction turns to be difficult to assess
in particular experimental conditions. We therefore keep the second
sound velocity reduction as a qualitative explanation, and we do not
try to assess quantitative result from the attenuation axis angle.

\subsection{Quantum vortex or velocity measurements ? \label{subsec:Summary-of-the}}

We have shown that second sound resonators are sensitive to two physical
mechanisms. The first one is the thermal wave bulk attenuation inside
the tweezers' cavity, due to the quantum vortices. The second one is
thermal wave ballistic advection perpendicular to the plates\footnote{Advection of second sound by velocity is illustrated e.g. in \cite{Dimotakis:1977p408}.}.
Both mechanisms exist for all the second sound resonators, but depending
on their geometry, they can preferentially be sensitive to the one
or the other mechanism. We call \emph{selectivity} the fraction of
the signal due to quantum vortices or to ballistic advection. Let
$\overline{T}\left(\xi,U\right)$ be the probe signal as a function
of the bulk attenuation coefficient $\xi$ (m$^{-1}$) and flow velocity
$U$(m/s), we define the vortex selectivity as
\begin{equation}
\mathcal{R}_{\xi}=\frac{\left|\overline{T}\left(\xi,0\right)-\overline{T}\left(0,0\right)\right|}{\left|\overline{T}\left(\xi,U\right)-\overline{T}\left(0,0\right)\right|}.\label{eq:selectivity}
\end{equation}
and by symmetry we define the velocity selectivity as
\begin{equation}
\mathcal{R}_{U}=\frac{\left|\overline{T}\left(0,U\right)-\overline{T}\left(0,0\right)\right|}{\left|\overline{T}\left(\xi,U\right)-\overline{T}\left(0,0\right)\right|}.\label{eq:selectivity-U}
\end{equation}

Further investigations in second sound tweezers experiments have shown
that the velocity/vortex selectivity process only weakly depends on
the aspect ratio $\frac{L}{D}$. Indeed, for a given resonator lateral
size $L$, ballistic advection of the wave outside the cavity increases
when the gap $D$ increases, but the number of quantum vortex lines
inside the cavity also increases linearly with $D$. Altogether, both
the ballistic advection and the bulk attenuation due to the quantum
vortices have similar dependence with $D$, that's why changing the
gap has no significant effect on selectivity. For second sound tweezers,
we observe that the selectivity neither depends strongly on the mean
temperature (that controls the superfluid fraction and the second
sound velocity). 

\begin{figure*}
\includegraphics[width=0.4\textwidth]{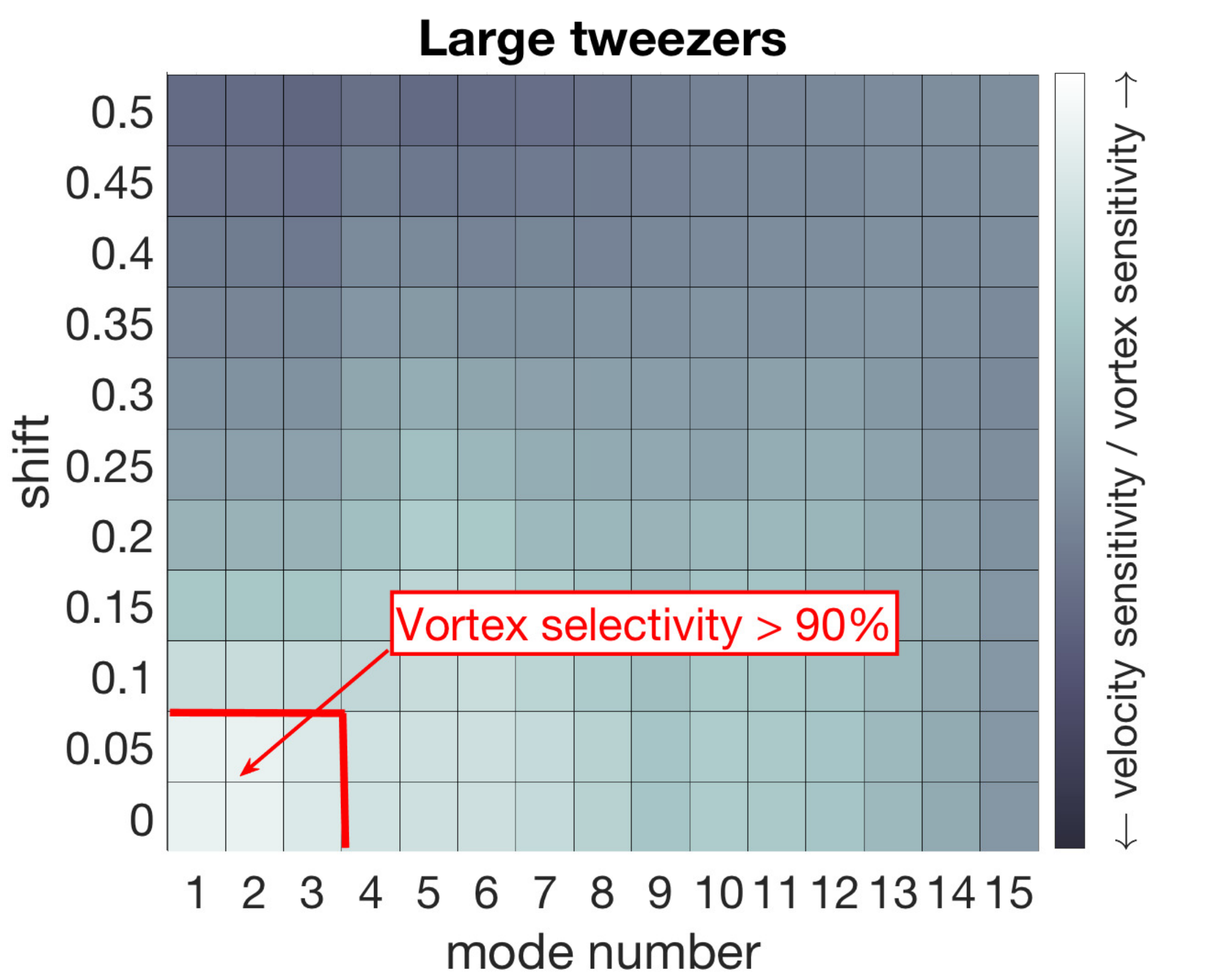}
\hspace{1cm}
\includegraphics[width=0.4\textwidth]{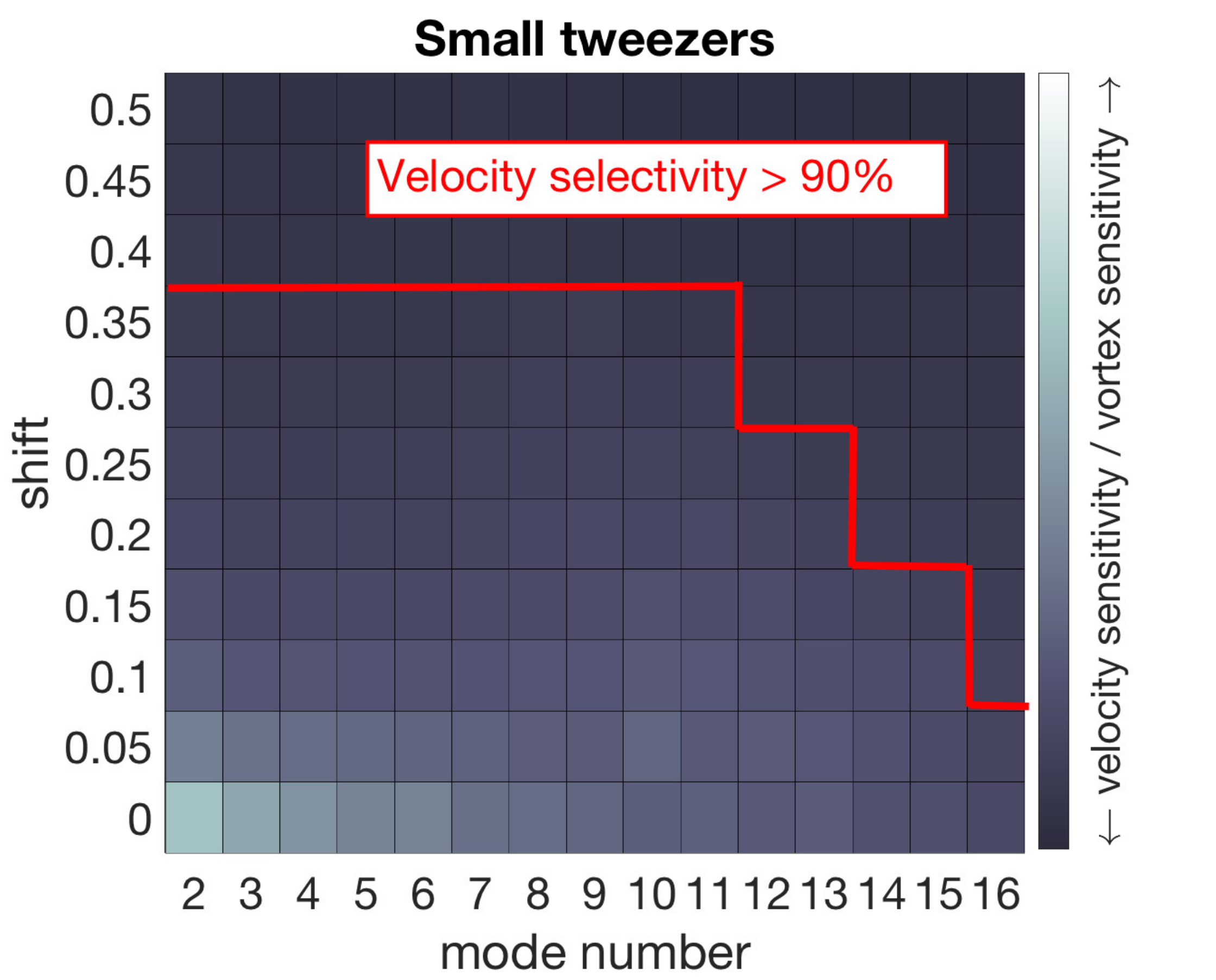}\caption{Selectivity to quantum vortices or velocity advection for two second
sound tweezers, obtained with numerical simulations. The color code
indicates the fraction of the signal due to bulk attenuation by quantum
vortices (see Fig. \ref{fig:mechanisms}). The selectivity of the
tweezers mainly depends on the lateral shift of both plates one from
another, and the resonant mode number excited in the cavity. The left
panel shows that large tweezers ($L=1$mm) are mainly sensitive to
quantum vortices. Almost pure quantum vortex signal can be achieved
with carefully aligned tweezers excited at low mode numbers ($\mathcal{R}_{\xi}>90\%$).
On the reverse, small tweezers ($L=250\mu{\rm m}$) are mainly sensitive
to the velocity. Almost pure velocity signal can be achieved by shifting
the heater and the thermometer plates and by exciting the cavity at
large mode numbers ($\mathcal{R}_{U}>90\%$). The present simulation
was run with a vortex line density $\mathcal{L}=2\times10^{10}$ m$^{-2}$
and $U=1$ m/s, in accordance with the typical values observed in
our experiments.\label{fig:Selectivity}}
\end{figure*}

As explained in sec. \ref{subsec:Response-with-a flow}, the velocity
selectivity is more important for open cavity resonators such as second
sound tweezers. For a given tweezers size, we find that the selectivity
depends mainly on the shift $X_{sh}$ and on the wave mode number.
Perfectly aligned tweezers excited with low mode numbers are preferentially
sensitive to quantum vortices. Increasing $X_{sh}$ or choosing larger
mode numbers leads to a larger velocity sensitivity and changes the
signal balance from vortex selectivity to velocity selectivity. The
tweezers' selectivity also strongly depends on the size $L$: smaller
tweezers can encompass less quantum vortices in the cavity, which means that the total wave attenuation from one plate to the other is smaller for small tweezers. By contrast, the attenuation fraction due to velocity advection does not depend on the tweezers size. The velocity selectivity is thus larger when the
tweezers are smaller. Fig. \ref{fig:Selectivity} displays the selectivity
of two second sound tweezers of size $L=250$ $\mu{\rm m}$ and $L=1$
mm respectively, depending on the shift $X_{sh}$ and the mode number
$n$ (where $k=n\pi$). The simulation was run with a quantum vortex
line density $\mathcal{L}=2\times10^{10}$ m$^{-2}$ and $U=1$ m/s,
in accordance with the typical values observed in the experiments
of \onlinecite{woillez2021vortex}. It can be seen that large tweezers ($L=1$
mm) can reach a vortex selectivity $\mathcal{R}_{\xi}>90\%$ for a
small shift and low mode number, which means that they can be used
for direct quantum vortex measurements\footnote{This result confirms the analysis of the first dataset measured using a second sound tweezer, in 2007\cite{roche2007vortex}.}. On contrary, small tweezers
($L=250$ $\mu{\rm m}$) can reach a velocity selectivity $\mathcal{R}_{U}>90\%$
for large shift or high mode number, and can thus be used as anemometers,
as confirmed by the experiments reported in \onlinecite{woillez2021vortex}.

\section{Measurements with second sound tweezers\label{sec:Measurements-with-second}}

Second sound tweezers are singular sensors in the sense that they
can measure two degrees of freedom at the same time, whereas most
of hydrodynamics sensors only measure one (e.g. Pitot tubes, cantilevers,
hot wires). The tweezers record the magnitude and phase of the thermal
wave averaged over the thermometer plate. Both quantities contain
physical information about the system. To summarize it shortly, magnitude
variations give information about quantum vortices in the cavity,
whereas phase variations give information about the local mean temperature
and pressure. The local mean velocity has an impact on both magnitude
and phase, and will be specifically treated in sec. \ref{subsec:Velocity-measurement}.
The aim of the following sections is to explain how properly separate
quantum vortices signal from other signal components. 

In the following, we call $\mathcal{L}_{\perp}$ the density of projected
quantum vortex lines density (projected VLD)
\begin{equation}
\mathcal{L}_{\perp}=\frac{1}{\mathcal{V}}\int_{\mathcal{V}}\sin^{2}\theta(l){\rm d}l,\label{eq:VLD_def}
\end{equation}
where $\mathcal{V}$ is the tweezers' cavity volume, $l$ is the curvilinear
abscissa along the vortex lines inside the cavity , $\theta(l)$ is
the angle between the quantum vortex line and the direction perpendicular
to the plates (vector $\mathbf{e}_{z}$). Assuming isotropy of the
vortex tangle, the total quantum vortex lines density (VLD) is

\begin{equation}
\mathcal{L}=\frac{3}{2}\mathcal{L}_{\perp}.
\end{equation}

A second sound wave is damped in the presence of a tangle of quantum
vortices. Let $\xi_{VLD}$ (in m$^{-1}$) be the bulk attenuation
coefficient of second sound waves, it has been found\cite{Hall:1956a,Hall:1956b,Tsakadze1962,Snyder1966,Mathieu:1984p381}
that $\xi_{VLD}$ is proportional to $\mathcal{L}_{\perp}$ according
to the relation 
\begin{equation}
\xi_{VLD}=\frac{B\kappa\mathcal{L}_{\perp}}{4c_{2}},\label{eq:attenuation_VLD}
\end{equation}
where $B$ is the first Vinen coefficient and $\kappa\approx9.98\times10^{-8}$
m$^{2}$/s (for $^{4}$He) is the quantum of circulation around one
vortex.

Therefore, Eq. (\ref{eq:attenuation_VLD}) shows that a measure of
the bulk attenuation coefficient gives access to the projected VLD
defined by Eq. (\ref{eq:VLD_def}). We recall in sec. \ref{subsec:The-vortex-line classical}
the standard methods to measure the bulk attenuation coefficient from
a second sound resonance, and we propose in sec. \ref{subsec:The-elliptic-method}
two new analytical methods and we give in sec.
\ref{subsec:Applications-of-the} some examples to apply our elliptic
method to the experimental data.

\subsection{The vortex line density from the attenuation coefficient\label{subsec:The-vortex-line classical}}

We assume that a single second sound resonance can be accurately represented
by the following expression (see Eq. (\ref{eq:local fit}))
\begin{equation}
\overline{T}(f)=\overline{T}_{0}\frac{\sinh\left(\xi_{0}D\right)}{\sinh\left(i\frac{2\pi(f-f_{0})D}{c_{2}}+(\xi_{0}+\xi_{VLD})D\right)},\label{eq:local fit sec 4.1}
\end{equation}
where $f_{0}$ is the second sound frequency of the local amplitude
maximum, $D$ is the resonator gap, $c_{2}$ is the second sound velocity,
$\xi_{0}$ is the attenuation coefficient without flow and $\xi_{VLD}$
is the additional bulk attenuation in the presence of quantum vortices
given by Eq. (\ref{eq:attenuation_VLD}). $\xi_{VLD}=0$ without flow.

A standard method to measure $\xi_{VLD}$ goes as follows: we fix
the second sound frequency at the resonant value $f_{0}$, and we
measure the thermal wave amplitude with, and without flow. Eq. (\ref{eq:local fit sec 4.1})
then shows that $\xi_{VLD}$ is given by
\begin{equation}
\xi_{VLD}=\frac{1}{D}\mathrm{asinh}\left(\frac{\overline{T}_{0}}{\overline{T}(f_{0})}\sinh\left(\xi_{0}D\right)\right)-\xi_{0}.\label{eq:standard xi_VLD}
\end{equation}
Eq. (\ref{eq:standard xi_VLD}) shows that beside the value of $D$, that can be determined from a fit of the tweezers' spectrum (sec.
\ref{subsec:Quantitative-predictions}), the value of $\xi_{0}$ has
to be accurately measured. This is usually done by the measurement
of the resonant half width. With some algebra manipulations, it can
be found from Eq. (\ref{eq:local fit sec 4.1}) that the resonant
magnitude satisfies 
\begin{equation}
\left|\overline{T}\right|^{2}=\left|\overline{T}_{0}\right|^{2}\frac{\sinh^{2}\left(\xi_{0}D\right)}{\sinh^{2}\left(\xi_{0}D\right)+\sin^{2}\left(\frac{2\pi(f-f_{0})D}{c_{2}}\right)}.\label{eq:square amplitude}
\end{equation}
Let $\Delta f$ be the frequency half-width defined by the relation
$\left|\overline{T}(f_{0}\pm\frac{\Delta f}{2})\right|^{2}=\frac{1}{2}\left|\overline{T}_{0}\right|^{2}$,
it can be shown from Eq. (\ref{eq:square amplitude}) that $\xi_{0}$
and $\Delta f$ are related by 
\begin{equation}
\sin\left(\frac{\pi\Delta fD}{c_{2}}\right)=\sinh\left(\xi_{0}D\right).\label{eq:classical xi0}
\end{equation}

We note in particular that the relation (\ref{eq:classical xi0})
can be used to find $\xi_{0}$ as long as the resonance quality factor
is high enough, that means, for $\sinh\left(\xi_{0}D\right)<1$. The
linear approximations of Eqs. (\ref{eq:standard xi_VLD}) and (\ref{eq:classical xi0})
are usually used when $\xi_{0}D \ll 1$, and they give the well-known approximation:
\begin{equation}
\mathcal{L}_{\perp} \simeq \frac{4 \pi \Delta f}{B \kappa} \left( \frac{\overline{T}_{0}}{\overline{T}(f_{0})} -1 \right) ,
\end{equation}

For low
quality factor resonances, another method should be used instead
of the resonant half width. The elliptic method presented in sec.
\ref{subsec:The-elliptic-method} allows determining $\xi_{0}$ for
resonances of any quality factors. \\

\begin{figure*}
\includegraphics[width=0.8\textwidth]{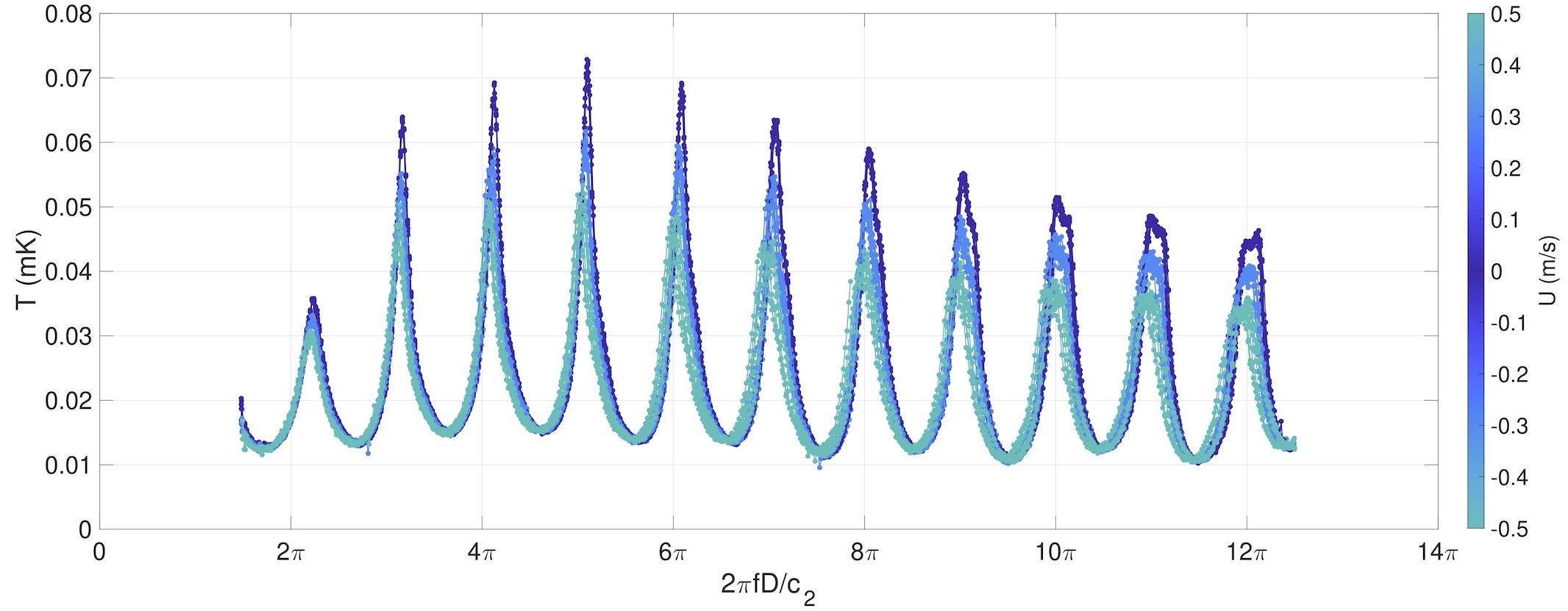}
\caption{Spectral response of tweezers in a turbulent boundary layer of the SHREK facility,
right below the superfluid transition temperature $T_{\lambda}$,
where second sound velocity $c_{2}$ is very sensitive to temperature
(here $D\simeq500\:\mu$m and $c_{2}$ drifts around a mean value of $5.4\:m/s$).
% Données Shrek nuit 13-14 février p144. XXX VERIFIER CHIFFRES XXX $T_{0}-T_{\lambda}\simeq 27\:mK$???,
The frequency axis is adimensionalize
using the second sound velocity $c_{2}$ calculated from the temperature
recorded near the sidewall of the flow. While the temperature of the
bath is regulated, frequency sweeps are repeated half a dozen of times
for different turbulent flow conditions flagged by colors. A systematic
drift of resonance frequencies versus flow conditions is observed
; it is interpreted as an under-estimation of temperature, and therefore
over-estimate of $c_{2}$, due to turbulent dissipation in the core
of the flow. At a given mean flow, some scatter of the resonance frequencies
is apparent ; it is interpreted as noise from the temperature regulation.
The elliptic method introduced in section \ref{subsec:The-elliptic-method}
allows separating such temperature artifacts from the attenuation
due to second sound attenuation by quantum vortices.}
\label{fig:DeriveTemp}
\end{figure*}

\begin{figure*}
\includegraphics[width=0.45\textwidth]{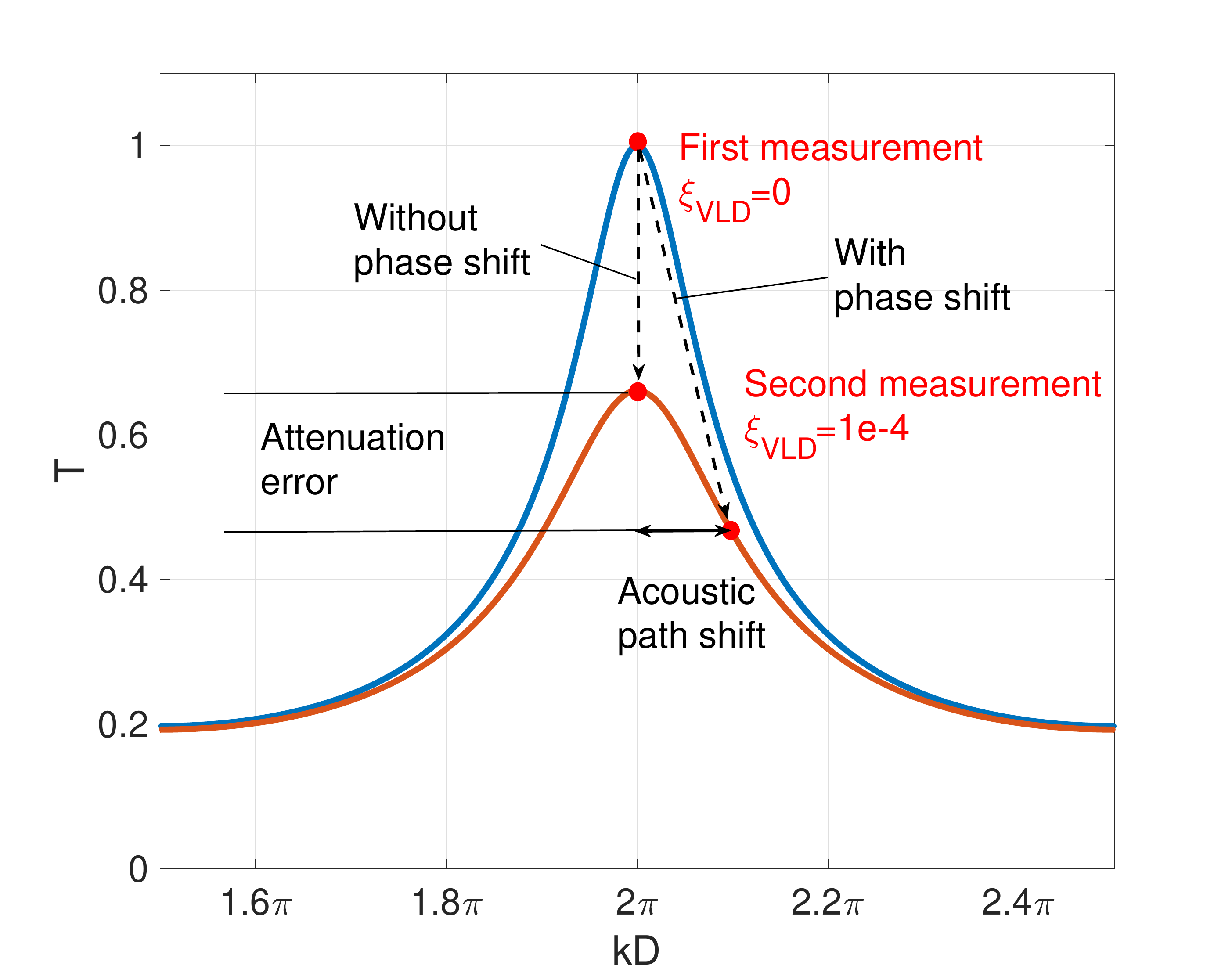}
\hspace{1cm}
\includegraphics[width=0.45\textwidth]{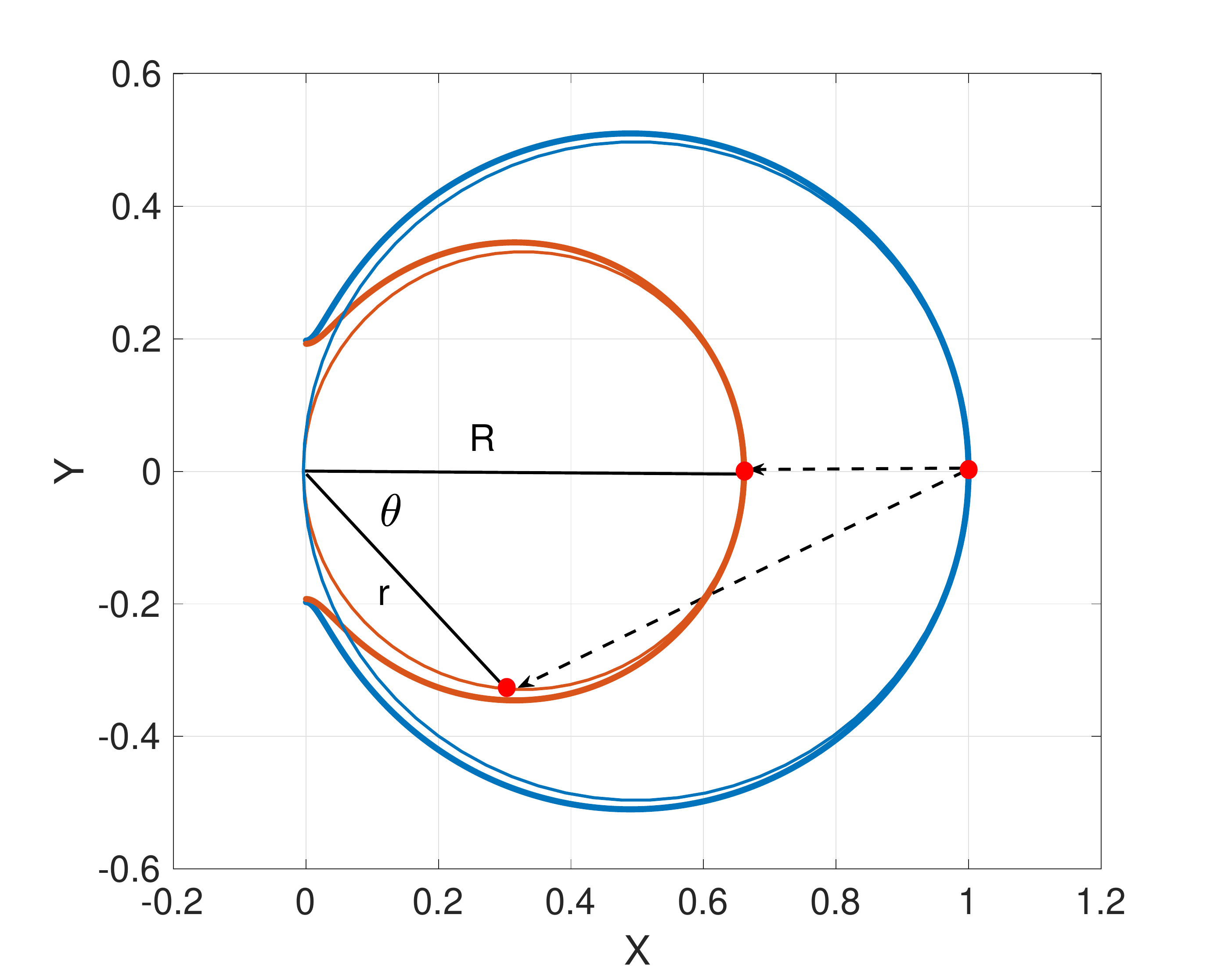}
\caption{An illustration of an attenuation measurement. \textbf{Left: }the
resonant mode without (blue curve), and with quantum vortex attenuation
(red curve). The figure shows that an acoustic path shift can lead
to an important error in the attenuation measurement. The wave vector is given by $k=\frac{2\pi f}{c_2}$.  \textbf{Right:
}the resonant mode represented in the phase-quadrature plane, together
with the fitted Kennelly circle (defined later in sec. \ref{subsec:The-simple-method}). The figure shows that the acoustic
path shift creates a phase shift $\theta$. Using the phase measurement
$\theta$, the maximal magnitude can be recovered using Eq. (\ref{eq:kennelly correction}).\label{fig:An-illustration-of measurement}}
\end{figure*}

The main problem of the method presented above is that it implicitly
assumes that there is no variation of the acoustic path value $\frac{2\pi f_{0}D}{c_{2}}$
during the measurement. In particular, as $c_{2}$ depends on temperature
and pressure, it means that the experiment should have an excellent
temperature and pressure regulation. This can become increasingly
difficult when the second sound derivatives versus temperature become steep, close to
the superfluid transition. Moreover, measurements in the presence
of a flow are necessary done out of equilibrium as the flow dissipates
energy. As an example, measurements in such conditions are illustrated by figure \ref{fig:DeriveTemp}, which shows second sound resonances measured close to the superfluid transition in the turbulent Von Kármán experiment SHREK \cite{Rousset:RSI2014}.
Furthermore, we observe that a measurement with a non-zero value
of $\xi_{VLD}$ can be associated with an acoustic path shift (i.e. a variation
of the factor $\frac{2\pi fD}{c_{2}}$). The situation is illustrated
in the left panel of Fig. \ref{fig:An-illustration-of measurement},
where the acoustic path shift leads to an overestimation of the attenuation
and an important error on $\xi_{VLD}$.

Passive and active approaches have been reported in the literature to handle the most common cause of acoustical path shift during second sound measurement: the temperature drift and its resulting shift of the resonance frequency.

A passive approach consists in performing a sweep of the second sound frequency, across the resonance curve. Afterward, with proper modelling of the resonance, the attenuation and the phase shift can be fitted separately, e.g. as done in \onlinecite{Mathieu1976}. A limit of this approach is its time resolution, that is restricted by the duration of frequency scan. Another passive approach consists in performing systematic calibration of the full frequency responses of the resonator in various conditions, and subsequently interpolating measurements obtained at a fixed working frequency onto this mapping \cite{varga:JLTP2017}.

A standard example of active approaches consist in controlling the helium bath temperature. An alternative or complementary approach consists in controlling the second-sound frequency so that it always matches the resonance peak, despite possible drift of the temperature. The resonator itself can provide the feedback signal of these control loops, for example by monitoring the thermometer or locking the phase of the second sound signal. An even more direct approach has been recently proposed: the resonator is driven by a self-oscillating circuit, which frequency adapts dynamically to the drift of the second sound  velocity \cite{yangRSI2017}.

Below, we introduce two analytical methods to separate  phase shift from attenuation. The first method is relevant for simple cases (sec. \ref{subsec:The-simple-method}), while the second one has a broader range of validity (sec. \ref{subsec:The-elliptic-method}).

\subsection{Analytical method in an idealized case\label{subsec:The-simple-method}}

%Données Shrek nuit 18-19 février 2021 (p144) T=1.6K env
\begin{figure*}
\includegraphics[width=0.45\textwidth]{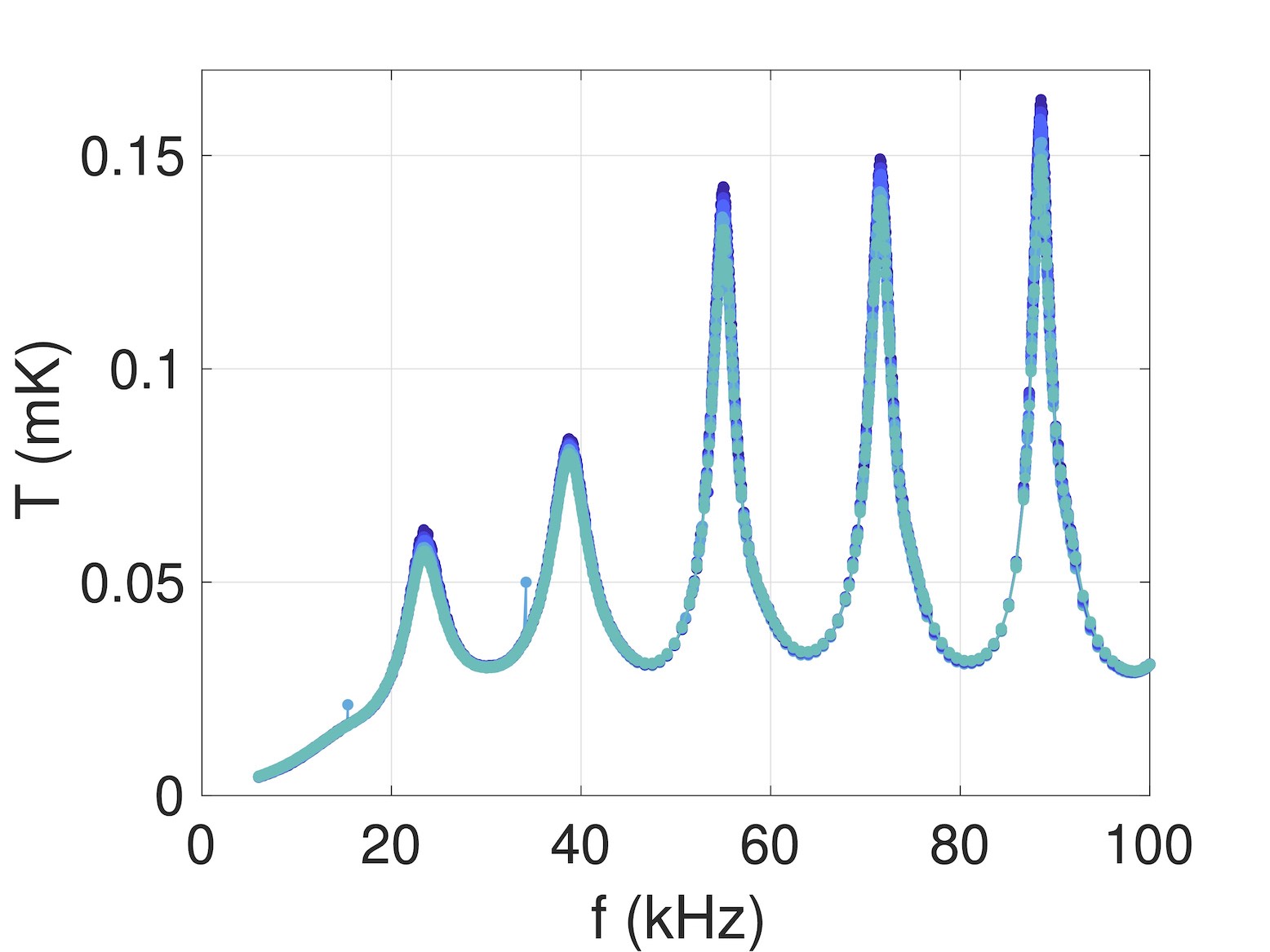}
\hspace{.5cm}
\includegraphics[width=0.45\textwidth]{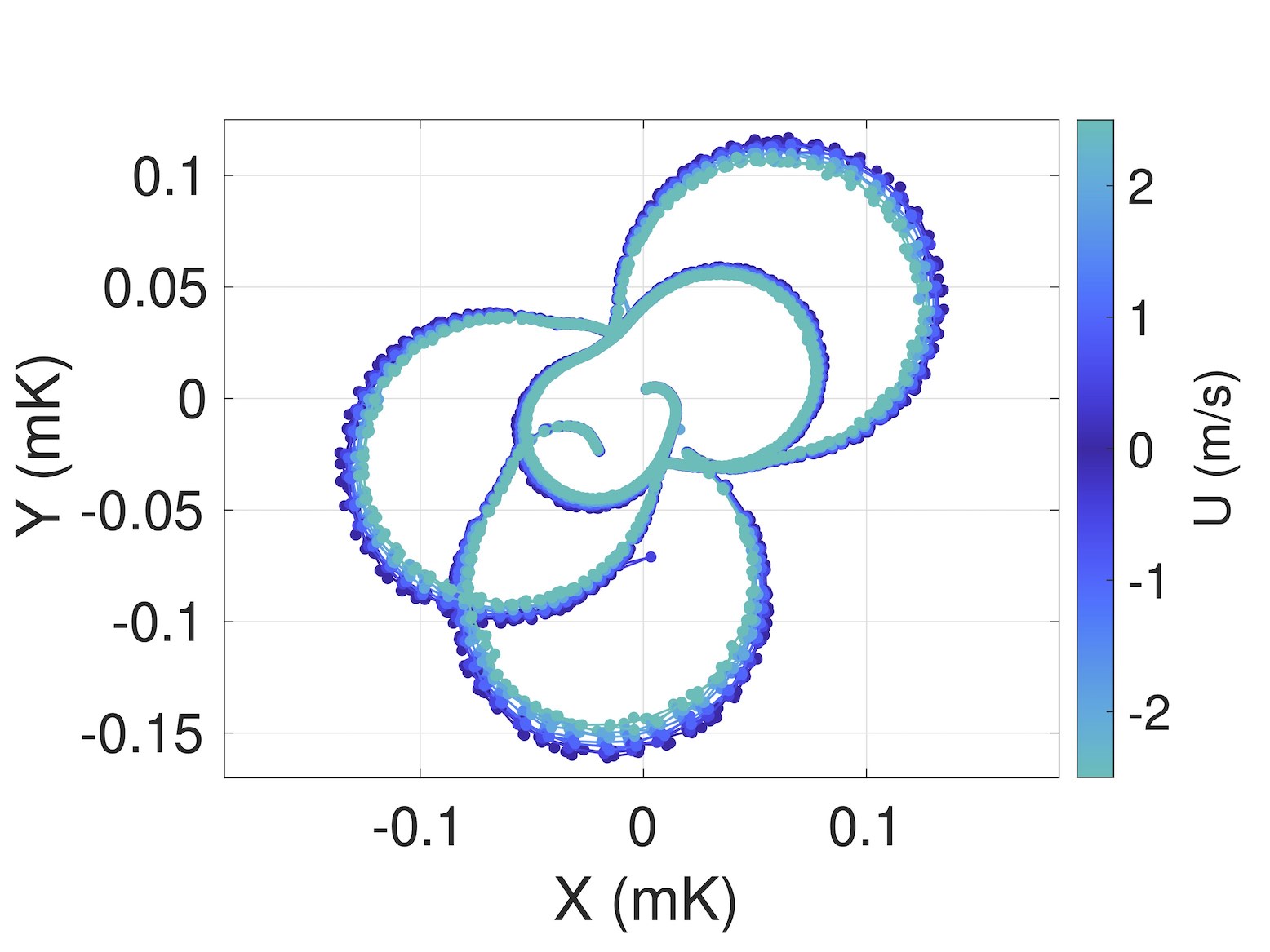}
\caption{Sequence of five resonances of second sound tweezers at 1.6K. The flow is weakly turbulent: the velocity standard deviation is a few percents of the mean velocity displayed on the color bar. The right side plot illustrates that each resonance can be approximated by a circle in the complex plane. The global phase shift of the last resonance (lower circle) compared to the others is attributed to a cut-off of the measurement electronics at high frequency.}\label{fig:jolies}
\end{figure*}

%Before presenting the elliptic method, we present a simple mathematical projection allowing to correct the attenuation from acoustical path shifting. This method is relevant when the amplitude of the second sound resonance can be approximated by circles in the complex plane. This approximation of the resonance curve in the complex plane has been shown to be relevant for some macroscopic second sound resonators (eg. \cite{vidal1971ideal,DHumieres:1980p448}). As illustrated in figure \ref{fig:jolies} and figures from section 3, this approximation is also relevant to most resonances of second-sound tweezers.

The acoustic path shift can be corrected using the resonant representation
in the phase-quadrature plane. Let $X(f)$ and $Y(f)$ be respectively
the real and imaginary parts of $\overline{T}(f)$, the curve $(X(f),Y(f))$
in the phase-quadrature plane is very close to a circle crossing the
origin. It can even be proved (see sec. \ref{subsec:The-elliptic-method})
that the resonant curve converges to a circle when the quality factor
increases, or equivalently when $\xi_{0}$ decreases\footnote{
In this case, the complex amplitude of the n$^{th}$ temperature resonance is often approximated by the Lorentzian formula \cite{vidal1971ideal,DHumieres:1980p448}
\begin{equation}
\overline{T}(f) \simeq \frac{\overline{T_n}}{1+i Q_n\frac{f-f_n}{f_n}}  
\end{equation}
\noindent where $\overline{T_n}$, $Q_n$ and $f_n$ are the amplitude at resonance, the quality factor and resonance frequency of the mode of interest. To highlight that this Lorentzian approximation describes a circle in the complex plane $X$-$Y$, it can be written as:
\begin{equation*}
 \frac{\overline{T}(f)}{\overline{T}_n/2} =  1+ e^{i \phi(f)}
\end{equation*}
\noindent where $\tan{\phi}=2F/\left( {F^2-1} \right)$ and $F=Q_n \left( f-f_n \right)/{f_n}$.
}. 
An illustration of two resonant curves with their osculating 
circles is displayed in the right panel of Fig. \ref{fig:An-illustration-of measurement}.
The acoustic path shift translates in a phase shift $\theta$ in the
phase-quadrature plane, such that the amplitude of the second measurement
(with $\xi_{VLD}>0$) does not correspond to the maximal amplitude
$R$ of the attenuated resonant peak (see Fig. \ref{fig:An-illustration-of measurement}
for the notations). Using the geometric properties of the (Kennelly)
circle, $R$ can be approximately recovered from the measured amplitude
$r$ and phase slip $\theta$ with
\begin{equation*}
R=\frac{r}{\cos\theta}.\label{eq:kennelly correction}
\end{equation*}

Thus, a modified version of Eq. (\ref{eq:standard xi_VLD}) can be written
to find the VLD attenuation coefficient in the presence of a phase
shift
\begin{equation}
\xi_{VLD}=\frac{1}{D}\mathrm{asinh}\left(\frac{\overline{T}_{0}\cos\theta}{\overline{T}(f_{0})e^{-i\theta}}\sinh\left(\xi_{0}D\right)\right)-\xi_{0}.\label{eq:modified_xi_VLD}
\end{equation}

Using this equation, Eq. \ref{eq:attenuation_VLD} and 
and Eq. \ref{eq:classical xi0}, the VLD $\mathcal{L}$ can then be derived.

\subsection{The elliptic method \label{subsec:The-elliptic-method}}

We present in this section an original method to obtain the values
of the acoustic path shift and the attenuation coefficient, from experimental
data. The method, that we call the ``elliptic method'', is much
simpler to implement, and much more reliable, than the fit of the
Kennelly resonant circle and the use of Eq. (\ref{eq:modified_xi_VLD}).
Besides, the method can be used for resonances with very low quality
factors.

\begin{figure}
\includegraphics[width=0.45\textwidth]{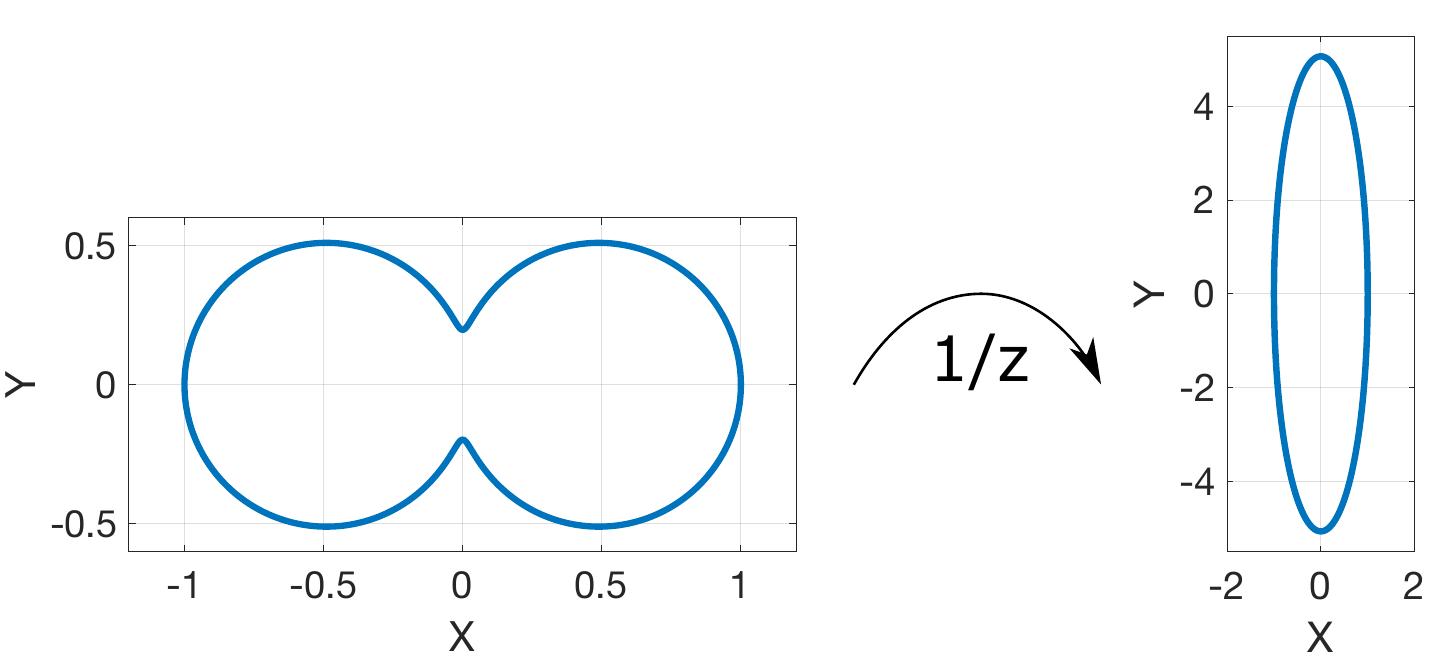}
\caption{Transformation of a pair of consecutive resonances to an ellipse using
the inversion of the complex plane $z\rightarrow1/z$. \label{fig:elliptic transform}}
\end{figure}

The method comes from the observation that a pair of two ideal consecutive
resonances is transformed into an ellipse with the complex inversion
$z\rightarrow\frac{1}{z}$ in the phase-quadrature plane. The inversion
is represented in Fig. \ref{fig:elliptic transform}. To prove this
assertion, consider the inversion of the classical Fabry--Perot expression
Eq. (\ref{eq:classical Fabry-Perot})
\begin{equation}
\frac{1}{\overline{T}}=\frac{\sinh\left(i\frac{2\pi fD}{c_{2}}+\xi D\right)}{A}.\label{eq:inv Fabry}
\end{equation}
Expanding the $\sinh$ in the previous expression gives
\begin{equation}
\frac{1}{\overline{T}}=\cos\left(\frac{2\pi fD}{c_{2}}\right)\frac{\sinh\left(\xi D\right)}{A}+i\sin\left(\frac{2\pi fD}{c_{2}}\right)\frac{\cosh\left(\xi D\right)}{A}.\label{eq:expand Fabry}
\end{equation}
Finally, let $X_{l}=\mathcal{R}e\left(\frac{1}{\overline{T}}\right)$
and $Y_{l}=\mathcal{I}m\left(\frac{1}{\overline{T}}\right)$ be respectively
the real and imaginary parts of Eq. (\ref{eq:expand Fabry}), the
coordinates $(X_{l},Y_{l})$ satisfy the equation
\begin{equation}
\left(\frac{X_{l}}{\sinh\left(\xi D\right)/A}\right)^{2}+\left(\frac{Y_{l}}{\cosh\left(\xi D\right)/A}\right)^{2}=1\label{eq:ellipse Fabry}
\end{equation}
which is exactly the Cartesian equation of an ellipse with semi-major
axis $a=\frac{\cosh\left(\xi D\right)}{A},$ and semi-minor axis $b=\frac{\sinh\left(\xi D\right)}{A}.$
In particular, we note that the attenuation coefficient $\xi$ can
be recovered from the ratio of the semi-major and semi-minor elliptic
axes using the formula
\[
\xi=\frac{1}{D}{\rm atanh}\left(\frac{b}{a}\right).
\]
When the quality factor increases (equivalently when $\xi$ decreases)
the ellipse is flattened. The limit of infinite quality factor ($\xi\rightarrow0$)
corresponds to two parallel straight lines in the complex plane. \\

\begin{figure*}
\includegraphics[width=0.8\textwidth]{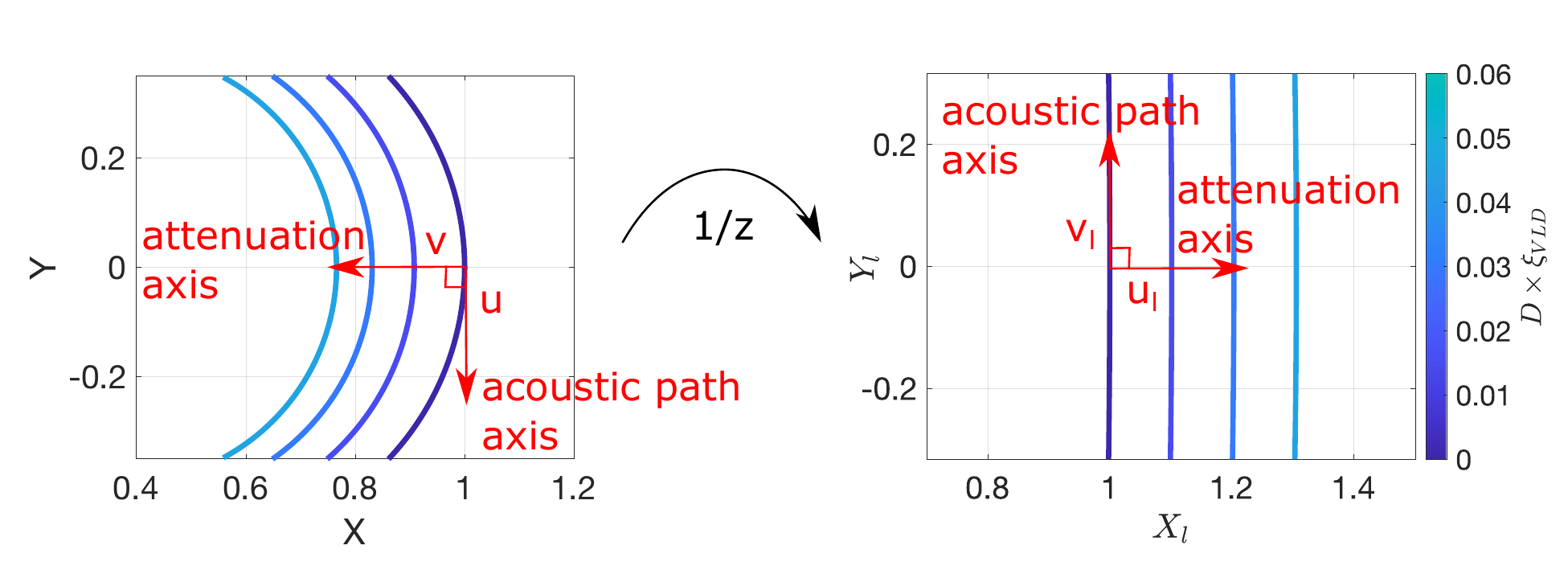}
\caption{A resonant mode in the phase-quadrature plane and its elliptic transform,
for an ideal Fabry--Perot resonance with $\xi_{0}=0.2$ and $1.95\pi<kD<2.05\pi$.
\label{fig:local transform}}
\end{figure*}

Second sound tweezers resonances are not ideal Fabry--Perot resonances.
Yet, we have argued in sec. \ref{subsec:Analytical-approximations}
that a single second sound resonance can be locally fitted by the
following Fabry--Perot equation (see also Eq. (\ref{eq:local fit}))
\begin{equation}
\overline{T}=\frac{A}{\sinh\left(i\frac{2\pi(f-f_{0})D}{c_{2}}+(\xi_{0}+\xi_{VLD})D\right)}.\label{eq:local fit-1}
\end{equation}
 This in particular means that the resonant curve in the vicinity
of its maximal amplitude is transformed into a part of an ellipse
with the complex inversion $z\rightarrow\frac{1}{z}$. The curve
$\left(X_{l}(f),Y_{l}(f)\right)$ is very close to a straight line,
for frequencies $f$ close to the resonant frequency $f_{0}$. The
situation is illustrated in Fig. \ref{fig:local transform}. The figure
shows a part of ideal Fabry--Perot resonances given by Eq. (\ref{eq:local fit-1}),
in the range $1.95\pi<kD<2.05\pi$ (where $k=\frac{2\pi f}{c_{2}}$),
and for increasing values of the VLD attenuation coefficient $\xi_{VLD}$.
The left panel displays the different resonant curves close to their
maximal amplitudes, in the phase-quadrature plane. Using the complex
inversion, those curves become almost parallel straight lines, as
can be seen in the right panel. The transformation of the resonant
Kennelly circles into parallel straight lines has very nice applications
that we explain in the following. \\

In the phase-quadrature plane, a variation of the acoustic path value
$\frac{2\pi fD}{c_{2}}$ corresponds to a displacement along the Kennelly
circle, whereas a variation of the bulk attenuation $\xi$ corresponds
to a displacement orthogonal to the Kennelly circle. The acoustic
path direction and the attenuation direction thus form a local orthogonal
basis. Such a basis is displayed by the red arrows in the left panel
of Fig. \ref{fig:local transform}. When the reference point where
the basis is defined moves along the Kennelly circle, the basis rotates
and the acoustic path and attenuation axes have to be redefined. This
can become a tiresome task while analyzing the experimental data.
Fortunately, the complex inversion is a conformal mapping, which means
that it preserves locally the angles: the local basis composed of
the acoustic path and attenuation axes is transformed into an orthogonal
basis (see the red arrows in the right panel of Fig. \ref{fig:local transform}).
More precisely, let $z_{0}$ be the complex position in the phase-quadrature
plane, and $(\mathbf{u},\mathbf{v})$ be the two complex (unit) vectors
defining the local basis at $z_{0}$, then the local basis $(\mathbf{u}_{l},\mathbf{v}_{l})$
at the point $\frac{1}{z_{0}}$ is given by
\begin{equation}
\begin{cases}
\mathbf{u}_{l} & =-\mathbf{u}\frac{\left|z_{0}\right|^{2}}{z_{0}^{2}}\\
\mathbf{v}_{l} & =-\mathbf{v}\frac{\left|z_{0}\right|^{2}}{z_{0}^{2}}
\end{cases}.\label{eq:global basis}
\end{equation}

The major advantage of defining the local basis $(\mathbf{u}_{l},\mathbf{v}_{l})$
with the elliptic transform, is that it becomes a \emph{global} basis:
when the reference point $\frac{1}{z_{0}}$ moves because of a change
in the acoustic path value or the attenuation value, the basis is
simply translated in the plane, but the vectors $(\mathbf{u}_{l},\mathbf{v}_{l})$
do not change. One can find the global basis $(\mathbf{u}_{l},\mathbf{v}_{l})$
for a given resonance and use it to find the local basis $(\mathbf{u},\mathbf{v})$
at every point in the phase-quadrature plane. We will see in sec.
\ref{subsec:Suppression-of-temperature} how the global basis $(\mathbf{u}_{l},\mathbf{v}_{l})$
can be easily used to suppress temperature and pressure drifts during
second sound attenuation measurements.\\

We finally explain how the elliptic method can be used to measure
the bulk attenuation coefficient $\xi_{0}$. We have seen in sec.
\ref{subsec:The-vortex-line classical} that the standard methods
to find $\xi_{0}$ are based on the measure of the half-width $\Delta f$
and on Eq. (\ref{eq:classical xi0}). As was said previously, the
classical method can only be applied to resonances satisfying $(\xi_{0}D)<1$.
It is a global method, in the sense that one has to sweep the frequency
to measure a large part of the resonant curve. The method is only
accurate provided the resonance does not deviate too much from an
ideal Fabry--Perot resonance, which is often not satisfied for the
first modes of second sound tweezers (see for example the first mode
of Fig. \ref{fig:Prediction-of-the_spectrum}). The alternative method
consists in expanding $(X_{l},Y_{l})$ given by Eq. (\ref{eq:ellipse Fabry})
to leading order in $f-f_{0}$
\begin{equation}
\begin{cases}
X_{l} & \sim\frac{\sinh\left(\xi_{0}D\right)}{A},\\
Y_{l} & \sim\frac{2\pi(f-f_{0})D}{c_{2}}\frac{\cosh\left(\xi_{0}D\right)}{A},
\end{cases}\label{eq:elliptic coordinates}
\end{equation}
where $f_{0}$ is the resonant frequency. Using Eq. (\ref{eq:elliptic coordinates}),
we get
\begin{equation}
\frac{Y_{l}(f)}{X_{l}(f)}\sim\frac{2\pi D}{\tanh\left(\xi_{0}D\right)c_{2}}(f-f_{0}).\label{eq:find xi0}
\end{equation}
$\frac{Y_{l}(f)}{X_{l}(f)}$ is proportional to $f$ in the vicinity
of $f_{0}$, with the proportionality factor $\frac{2\pi D}{\tanh\left(\xi_{0}D\right)c_{2}}$.
The attenuation coefficient $\xi_{0}$ can be found by a linear fit
of the function $\frac{Y_{l}}{X_{l}}$ provided $D$ and $c_{2}$
are known.

\subsection{Applications of the elliptic method\label{subsec:Applications-of-the}}

The motivation to develop and use the elliptic method has come from
experimental constrains: in experiments done in the vicinity of the superfluid transition, where $c_2(T_0)$ sharply varies, or in large superfluid experiments where it can be very difficult to control the values of mean temperature and pressure.
This is even more the case if the superfluid experiment dissipates
energy. One then expects a drift of the thermodynamics conditions
during the measurement. Regarding second sound resonators, the critical
parameter is the second sound velocity $c_{2}$, because variations
lead to uncontrolled acoustic path shifts. The elliptic method has
been designed to easily filter those variations from experimental
data. This includes filtering temperature and pressure drifts (sec.
\ref{subsec:Suppression-of-temperature}) and the vibration of the
tweezers arms (sec. \ref{subsec:Filtering-the-vibration}), and properly
extract the quantum vortex lines fluctuations (sec. \ref{subsec:Measure-of-vortex}).

\subsubsection{Suppression of temperature and pressure drifts\label{subsec:Suppression-of-temperature}}

This section presents an example of the elliptic method implementation
for second sound tweezers, to find the relation between $\left\langle \xi_{VLD}\right\rangle $
and the mean velocity $U$ in the presence of a superfluid flow.

\begin{figure*}
\begin{centering}
$\vcenter{\hbox{\includegraphics[width=6.6cm]{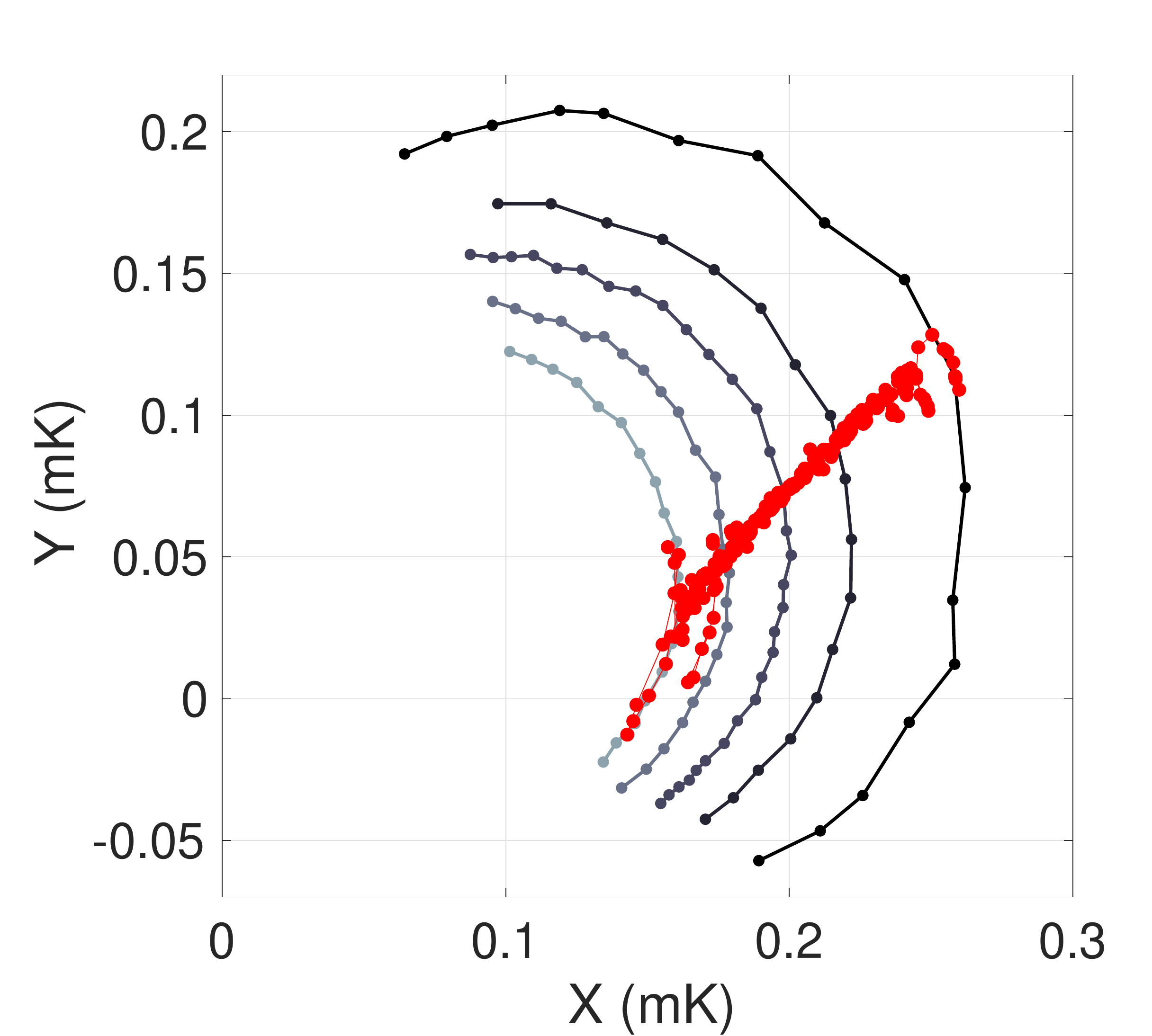}}}$$\vcenter{\hbox{\includegraphics[width=2cm]{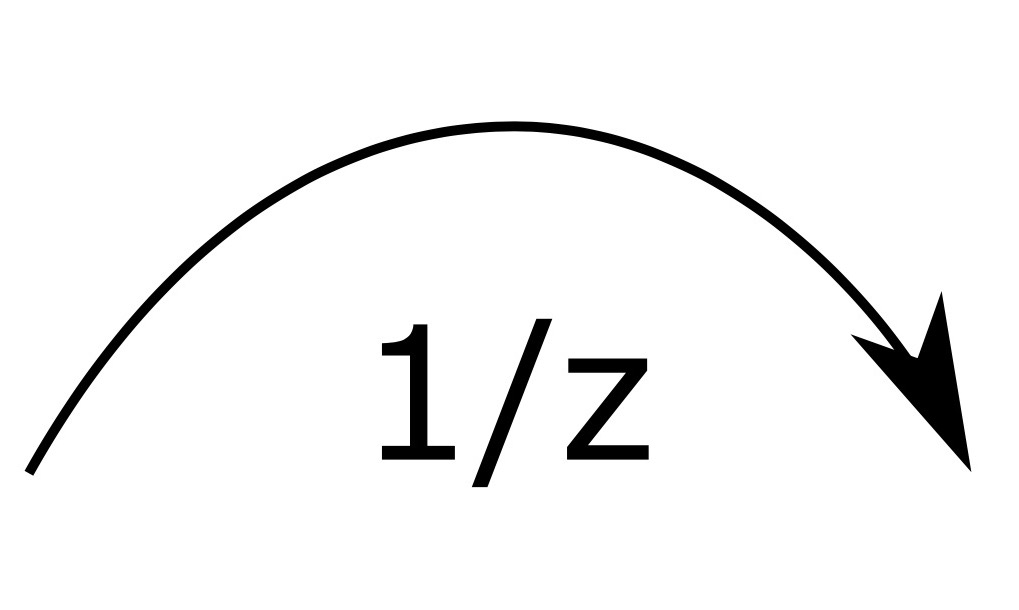}}}$$\vcenter{\hbox{\includegraphics[width=8cm]{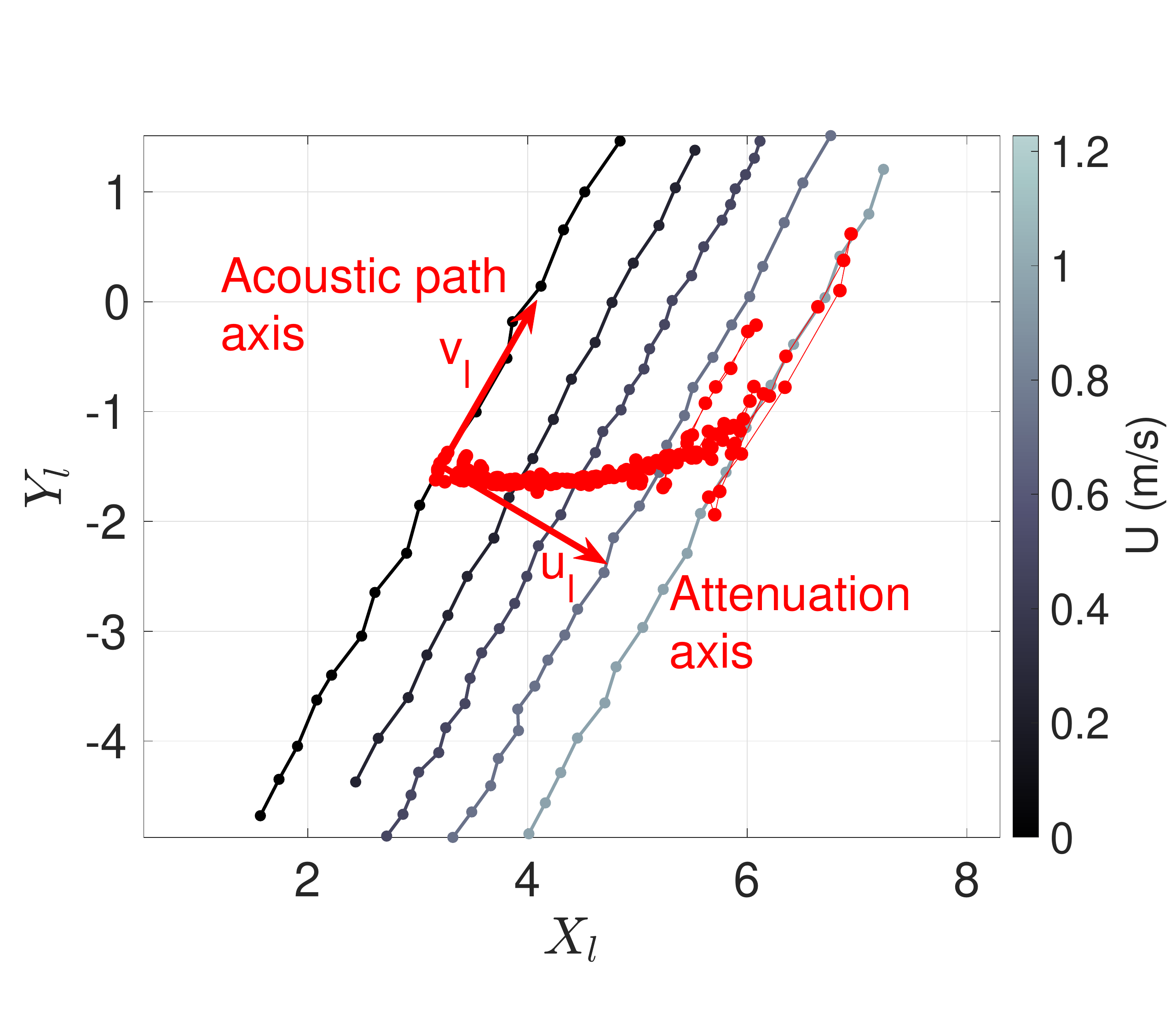}}}$
\par\end{centering}
\caption{\textbf{Experiment}: measurement of a resonance with second sound
tweezers at $T_{0}\approx2.14$ K, for different values of the mean
flow velocity $U$. The left panel presents the data in the phase-quadrature
plane, and the right panel presents the same data after the complex
inversion. The red points are obtained for a fixed value $f=6.35$
kHz, and sweeping the flow mean velocity between $0$ and $1.2$ m/s.
\label{fig:data mean }}

\end{figure*}

As before, we note $(X,Y)$ the temperature signal obtained from the
second sound tweezers in the phase-quadrature plane, and $\left(X_{l},Y_{l}\right)$
the Cartesian coordinates obtained by the complex inversion given
by 
\begin{equation}
\begin{cases}
X_{l} & =\mathcal{R}e\left(\frac{1}{X+iY}\right),\\
Y_{l} & =\mathcal{I}m\left(\frac{1}{X+iY}\right).
\end{cases}\label{eq:elliptic data}
\end{equation}
The coordinates $(X_{l},Y_{l})$ will be called ``elliptic coordinates''
for convenience. Fig. \ref{fig:data mean } displays experimental
data obtained from second sound tweezers of size $L=1$ mm, in a saturated
bath at mean temperature $T_{0}\approx2.14$ K, and for different
mean flow velocities $0<U<1.2$ m/s. At such a temperature close to
the superfluid transition, it was difficult to regulate the mean temperature well enough to prevent measurable variation of the second sound velocity $c_2$. Uncontrolled acoustic path
variations can be observed, for example, in the red points of Fig. \ref{fig:data mean }. 

The first step consists in sweeping the second sound frequency $f$
in the vicinity of the resonant frequency $f_{0}$. The data $(X,Y)$
obtained, displayed by the black curve of the left panel of Fig. \ref{fig:data mean },
form a part of the Kennelly circle. As explained in sec. \ref{subsec:The-elliptic-method},
the elliptic coordinates $(X_{l},Y_{l})$ given by Eq. (\ref{eq:elliptic data})
form a straight line (see the right panel of Fig. \ref{fig:data mean }).
Using a linear fit, it is then straightforward to obtain the unit
vector $\mathbf{v}_{l}$ parallel to the line defining the acoustic
path axis, and the orthogonal vector $\mathbf{u}_{l}$ defining the
attenuation axis. We then call $\mathbf{Z}_{l}=\left(X_{l},Y_{l}\right)$
the vector of the elliptic coordinates. The attenuation coefficient
$\xi_{0}$ can be found by the relation (see Eq. (\ref{eq:find xi0}))
\[
\frac{\mathbf{Z}_{l}(f).\mathbf{v}_{l}}{\mathbf{Z}_{l}(f).\mathbf{u}_{l}}\sim\frac{2\pi D}{\tanh\left(\xi_{0}D\right)c_{2}}(f-f_{0}).
\]
Fig. \ref{fig:data mean } displays experimental resonant curves obtained
for non-zero mean velocities $U>0$ , to illustrate the robustness
of the elliptic method. However, we emphasize that only the resonant
curve with $U=0$ is necessary to find the global basis $(\mathbf{u}_{l},\mathbf{v}_{l})$
in the plane of elliptic coordinates. \\

The second step consists in fixing the second sound frequency to $f_{0}$
and vary the mean velocity $U$ to look at the resonance attenuation.
The experimental data are displayed by the red points in Fig. \ref{fig:data mean }.
In can be seen in the figure that the bulk attenuation is accompanied
by a systematic acoustic path deviation as the mean velocity increases.
For mean velocities $U\approx1$ m/s, energy dissipation in the experiment
leads to a data dispersion along the acoustic path direction. To properly
recover the mean VLD attenuation coefficient $\left\langle \xi_{VLD}\right\rangle $
, we use the elliptic coordinates $\mathbf{Z}_{l}=\left(X_{l},Y_{l}\right)$
, and we project it on the attenuation axis $\mathbf{u}_{l}$. We
get from Eq. (\ref{eq:elliptic coordinates})
\[
\frac{\mathbf{Z}_{l}(U).\mathbf{u}_{l}}{\mathbf{Z}_{l}(0).\mathbf{u}_{l}}=\frac{\sinh\left(\left(\xi_{0}+\left\langle \xi_{VLD}\right\rangle \right)D\right)}{\sinh\left(\xi_{0}D\right)}.
\]
And $\left\langle \xi_{VLD}\right\rangle $ is then given by
\begin{equation}
\left\langle \xi_{VLD}\right\rangle =\frac{1}{D}{\rm asinh}\left(\frac{\mathbf{Z}_{l}(U).\mathbf{u}_{l}}{\mathbf{Z}_{l}(0).\mathbf{u}_{l}}\sinh\left(\xi_{0}D\right)\right)-\xi_{0}.\label{eq:elliptic mean attenuation}
\end{equation}
We note that the previous expression remains accurate even if the
second sound frequency chosen for the measurement is close but not
exactly equal to the resonant frequency $f_{0}$. The average VLD
attenuation can then be converted to the average projected vortex
line density $\left\langle \mathcal{L}_{\perp}\right\rangle $ using
Eq. (\ref{eq:attenuation_VLD}).

\subsubsection{Measure of vortex line density fluctuations\label{subsec:Measure-of-vortex} }

Second sound tweezers are designed to directly probe locally
vortex line density fluctuations, not only its average value. The
method to probe fluctuations slightly differs from the method used
to probe the average value explained in sec. \ref{subsec:Suppression-of-temperature}.
The average VLD value can be directly computed using the complex inversion
of experimental data, but this is no longer possible for its fluctuations.
Indeed, the tweezers signal have different sources of noise, like
e.g. thermal white noise, interfering frequencies, electromagnetic
bursts, etc... Those signals can usually be considered as independent
additive noises in the signal data, and easily filtered out or attenuated
by an appropriate post-processing. On the contrary, the complex inversion
is a non-linear transformation. Using the latter on noisy data can
lead to an overestimation of the signal fluctuations closest to zero,
perturb the additivity of noise sources and make them much more difficult
to filter out. We thus choose to compute the VLD fluctuations only
using linear transformations.\\

\begin{figure}
\includegraphics[width=0.5\textwidth]{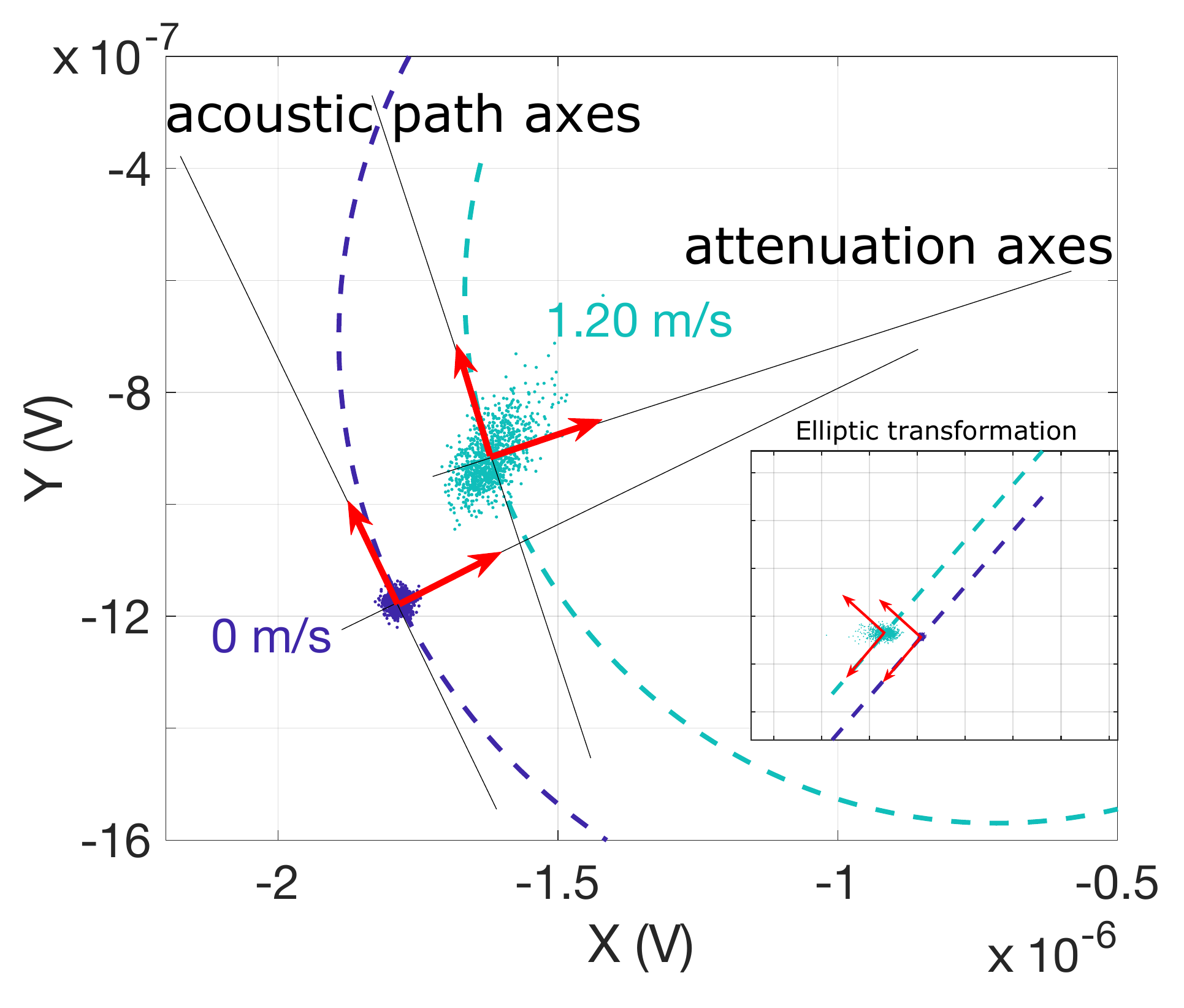}
\caption{Illustration of the method used for fluctuations decomposition. The two patches display the fluctuating signals obtained at the mean flow velocities 0 m/s and 1,2 m/s in the X-Y plane. The inset panel displays the signal after the elliptic transformation of Fig. \ref{fig:local transform}. The method to obtain the two projection axes is detailed in the main text.  \label{fig:fluctuation method}}
\end{figure}

The first step is similar to that of sec. \ref{subsec:Suppression-of-temperature}.
We sweep the second sound frequency $f$ close to the resonant frequency
$f_{0}$, in order to measure a part of the Kennelly circle. We then
transform this Kennelly circle into a straight line using the complex
inversion, and we find the global basis $(\mathbf{u}_{l},\mathbf{v}_{l})$
in the plane of elliptic coordinates. A fit of the Kennelly circle,
and its transformation into a straight line, can be seen in Fig. \ref{fig:fluctuation method}.
This experimental step has to be done just before the fluctuations'
measurement.

We then record the signal fluctuations $\left(X(t),Y(t)\right)$,
for different values of the flow mean velocity $U$. Fig. \ref{fig:fluctuation method}
displays the fluctuating signal for $U=0$ and $U=1.2$ m/s in the
form of clouds of data points. It can be in particular observed that
the $U=1.2$ m/s data are shifted compared to the $U=0$ m/s data
because of both an average attenuation and an acoustic path shift
(see sec. \ref{subsec:Suppression-of-temperature}). Let us define
$\left\langle \mathbf{Z}(t)\right\rangle =\left\langle X(t)\right\rangle +i\left\langle Y(t)\right\rangle $
the average complex position in the phase-quadrature plane, for a
given value of $U$. Following Eq. (\ref{eq:global basis}), the local
basis $(\mathbf{u},\mathbf{v})$ of the attenuation and acoustic path
axes can be computed from the global elliptic basis $(\mathbf{u}_{l},\mathbf{v}_{l})$
by 
\begin{equation}
\begin{cases}
\mathbf{u} & =-\mathbf{u}_{l}\frac{\left\langle \mathbf{Z}\right\rangle ^{2}}{\left|\left\langle \mathbf{Z}\right\rangle \right|^{2}}\\
\mathbf{v} & =-\mathbf{v}_{l}\frac{\left\langle \mathbf{Z}\right\rangle ^{2}}{\left|\left\langle \mathbf{Z}\right\rangle \right|^{2}}
\end{cases}.\label{eq:local basis}
\end{equation}
Fig. \ref{fig:fluctuation method} shows that the local basis $(\mathbf{u},\mathbf{v})$
depends on $\left\langle \mathbf{Z}\right\rangle $ and thus on the
value of $U$: for different mean velocities, the bases are rotated
from one another. If there is a significant drift of the mean signal
value during the measurement (as can e.g. be observed in Fig. \ref{fig:paquerette}),
the local basis will also depend on time, with a typical timescale
that should be much larger than the fluctuation timescale.

The acoustic path fluctuations and the attenuation fluctuations can
then be recovered using a projection on the $(\mathbf{u},\mathbf{v})$
basis. More precisely, let $x$ be the average acoustic path value
and $\delta\xi=\xi-\left\langle \xi\right\rangle $ be a small fluctuation
of the attenuation coefficient, a leading order expansion of expression
Eq. (\ref{eq:local fit-1}) shows that
\begin{widetext}
\begin{equation}
\overline{T}\approx\frac{A}{\sinh\left(ix+\left\langle \xi\right\rangle D\right)}-\delta\xi\frac{AD}{\sinh\left(ix+\left\langle \xi\right\rangle D\right)\tanh\left(ix+\left\langle \xi\right\rangle D\right)}.\label{eq:expansion_tempo}
\end{equation}
\end{widetext}
We then do the approximation $\left\langle \mathbf{Z}\right\rangle \approx A/\sinh\left(ix+\left(\xi_{0}+\left\langle \xi_{VLD}\right\rangle \right)D\right)$
which is equivalent to neglecting non-linear corrections in Eq. (\ref{eq:expansion_tempo}).
We finally get
\begin{equation}
\delta\xi(t)= 
\frac{1}{D}\mathbf{u}.\left(\mathbf{Z}(t)-\left\langle \mathbf{Z}\right\rangle \right)\times\left|\frac{\tanh\left(ix+\left(\xi_{0}+\left\langle \xi_{VLD}\right\rangle \right)D\right)}{\left\langle \mathbf{Z}\right\rangle }\right|,\label{eq:formule fluctuations}
\end{equation}
where the value of $\xi_{0}$ can be found with Eq. (\ref{eq:find xi0})
and $\left\langle \xi_{VLD}\right\rangle $ with Eq. (\ref{eq:elliptic mean attenuation}).

\begin{figure}
\includegraphics[width=0.45\textwidth]{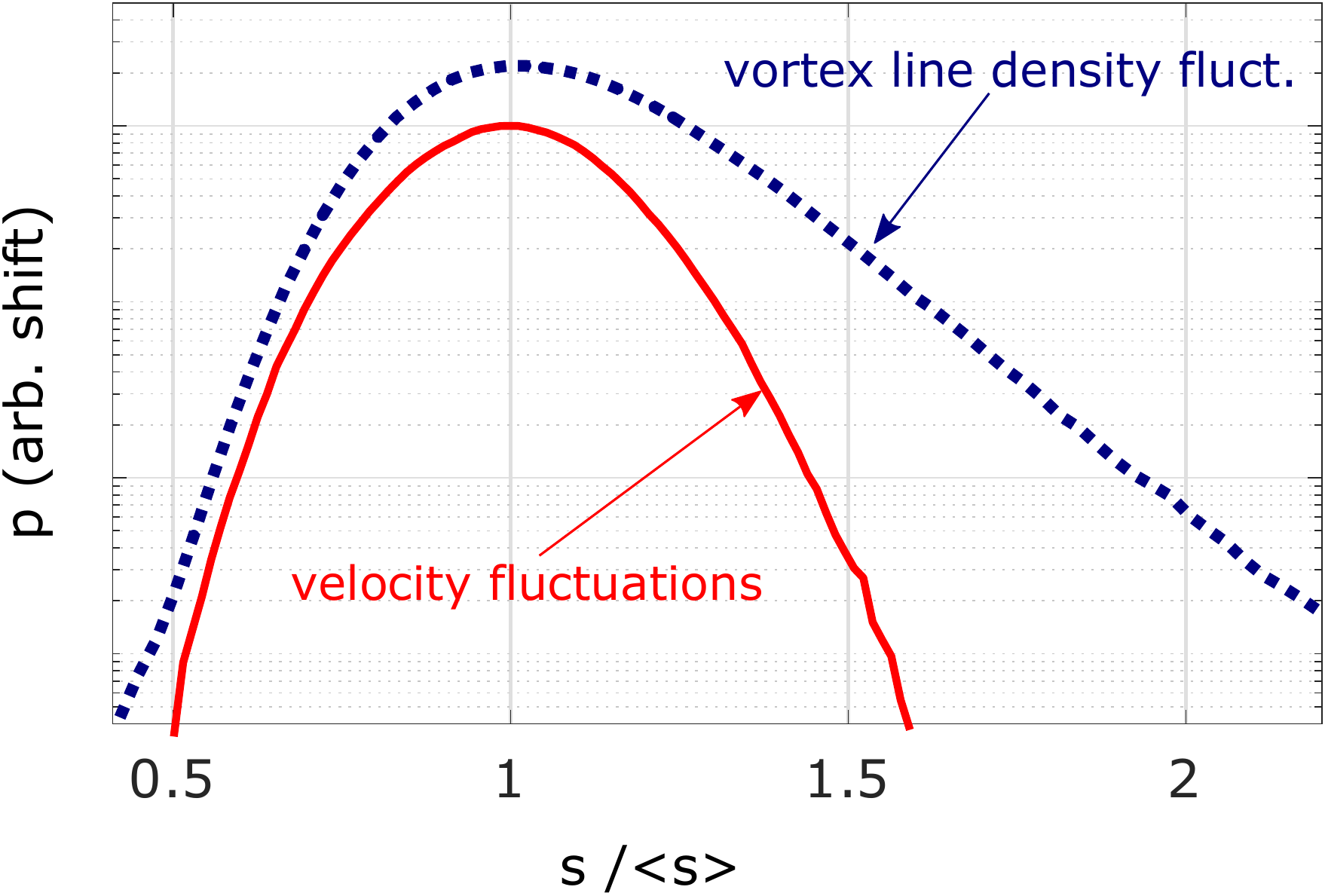}
\caption{Example of the probability density function of vortex line density (dashed line) and velocity (continuous line) measured simultaneously by second sound tweezers in the same superfluid turbulent flow \cite{woillez2021vortex}. In abscissa, each signal $s$ is normalized by its mean value $<s>$. The nearly Gaussian velocity statistics and skewed vorticity statistics are reminiscent of those in classical turbulence. \label{fig:histogramme}}
\end{figure}

The use of the elliptic method is illustrated by Figure \ref{fig:histogramme}, which reports the probability density function of the fluctuations of the quantum vortex density in a nearly isotropic superfluid turbulent flow
The data of this plot were reported together with spectra of vorticity fluctuations. For details about the setup and analysis of these results, see \onlinecite{woillez2021vortex}.

\subsubsection{Filtering the vibration of the plates\label{subsec:Filtering-the-vibration}}

One possible source of noise for second sound resonator measurements is the vibration of the plate arms whenever $U\neq0$. The signature of those vibrations can be very clearly identified in the form of two thin peaks in the fluctuation power spectrum. Those two peaks are located at the two arm resonant frequencies: their exact values can vary for different tweezers, but we always observe them around $f\approx1$ kHz (see Sec. \ref{subsec:Applications-of-the}).

\begin{figure}
\includegraphics[width=0.45\textwidth]{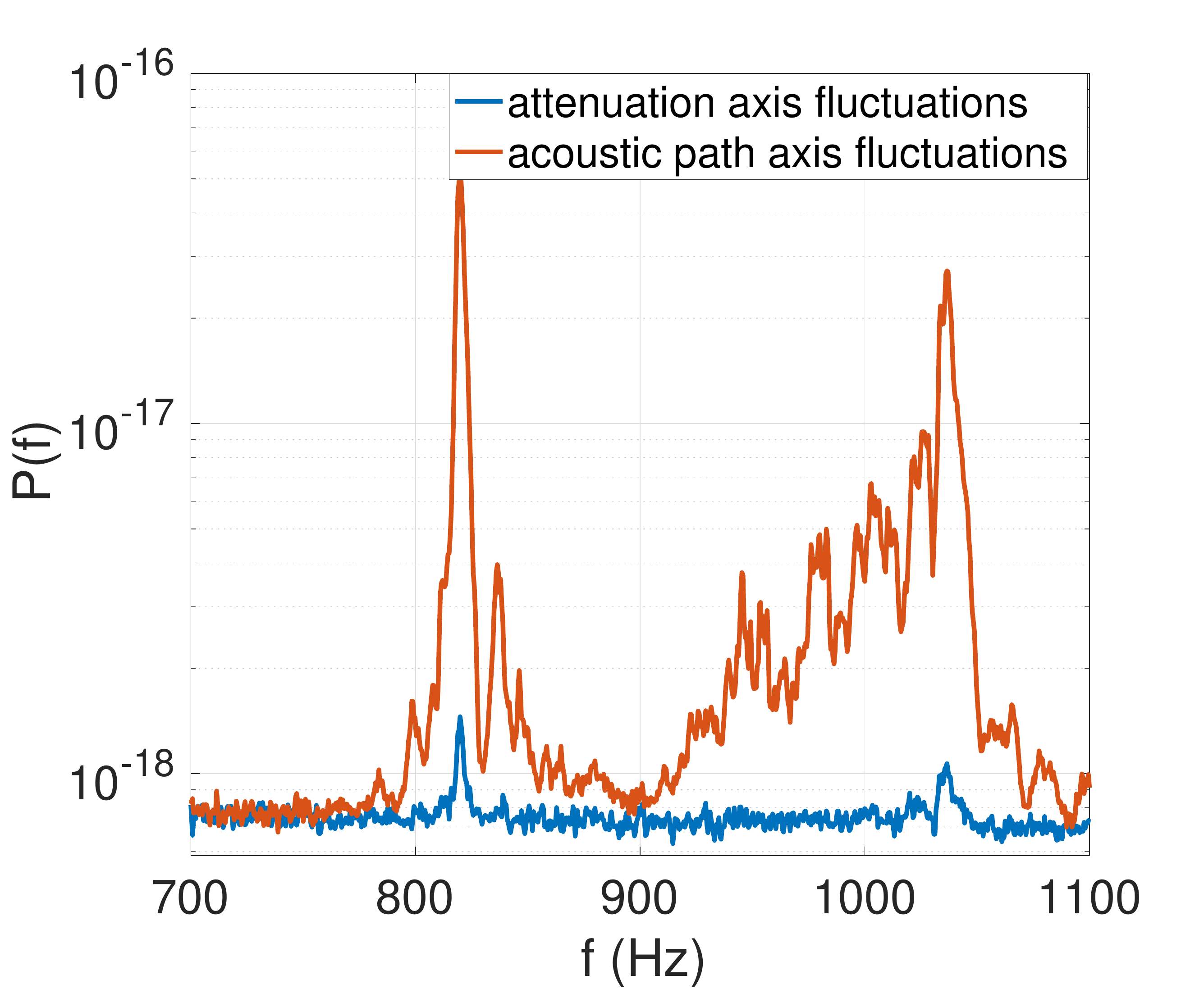}
\caption{\textbf{Experiment: }a part of second sound tweezers fluctuation power
spectrum, with $T_{0}=1.65$\:K, $U=1.2$\:m/s and for a tweezers gap
$D=1.320$\:mm. The tweezers' arms resonances can be clearly identified
in the acoustic path fluctuations. \label{fig:arme vibrations}}
\end{figure}

Fortunately, the tweezers' arm vibrations correspond to a variation of the gap $D$, and thus to acoustic path fluctuations. Fig. \ref{fig:arme vibrations} displays a part of the tweezers' fluctuation power spectrum. The fluctuations are projected along the attenuation axis (blue curve) and along the acoustic path axis (red curve), following the method presented in Sec. \ref{subsec:Measure-of-vortex}. The two peaks located at $f\approx825$ Hz and $f\approx1050$ Hz are identified on the acoustic path axis fluctuation power spectrum, whereas the same peaks are damped by many orders of magnitude on the attenuation axis fluctuations. These spectra illustrate the effectiveness of the elliptic method.

Using the power spectrum of Fig. \ref{fig:arme vibrations}, we can estimate the order of magnitude of the gap standard deviation. We find $\sqrt{\left\langle \left(\delta D\right)^{2}\right\rangle }\approx0.5$ $\mu{\rm m}$, and $\frac{\sqrt{\left\langle \left(\delta D\right)^{2}\right\rangle }}{D}\approx4\times10^{-4}$. This confirms that the arm vibrations have a negligible impact on the measurement.

\subsection{Velocity measurements\label{subsec:Velocity-measurement} }

\begin{figure}
\includegraphics[width=0.45\textwidth]{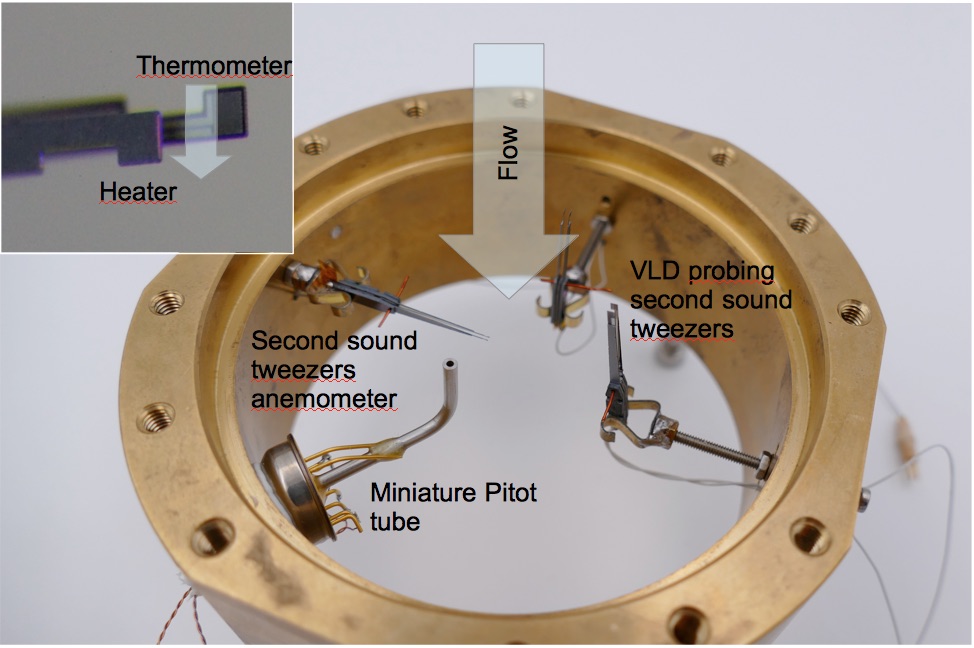}
\caption{Example of an arrangement of three second sound tweezers dedicated to velocity (x1, left side) and vorticity (x2, right side) time series acquisitions, together with a miniature total head pressure tube (loosely labelled "Pitot tube" on the bottom left side of the picture) used for velocity calibration. All probes are mounted on a ring connecting two 76\:mm-inner-diameter coaxial pipes (see Fig.1 of ref. \cite{woillez2021vortex}). The insert is a close-up view of the shifted plates of the anemometer tweezers.\label{fig:anneau}}
\end{figure}

As shown in Sec. \ref{subsec:Summary-of-the}, the geometry of the second sound tweezers can be optimized to sense specifically velocity rather than vortex density. One trick for achieving this is to shift one plate with respect to the other in the flow direction. Figure \ref{fig:anneau} shows three second sound tweezers and one anemometer that is based on the same principle as Pitot tubes. All sensors are positioned across a nearly homogeneous turbulent flow bounded by a cylindrical pipe (not shown here). Ffor details on this set-up, see \onlinecite{rusaouen2017intermittency}. The insert in the figure is a close-up view of the tip of the left-side tweezers, which is dedicated to velocity measurements. This shift of one plate versus the other in the downstream direction is clearly visible.

The projection of the anemometer-tweezers signal in the complex plane is not performed along orthogonal axes. These axes are determined with an in-situ calibration, ramping the mean velocity. The complementary "Pitot tube" signal is used to calibrate the axis in units of m/s.

As an illustration, Figure \ref{fig:histogramme} presents the probability density function of the velocity fluctuations measured by the anemometer tweezers, together with vortex line density fluctuations recorded by the other tweezers in the same set-up (see Fig. \ref{fig:anneau}). As already reported in quantum turbulence \cite{Salort:EPL2012}, the velocity statistics are close to Gaussian when probed at scales significantly larger than the intervortex distance, which is the case here. Velocity spectra derived from the same dataset are reported in \onlinecite{woillez2021vortex}.

\section{Summary and Perspectives}

This study covers three independent topics: (i) the comprehensive analytical modeling of second-sound resonators with a cavity that allows a throughflow of superfluid (Section 3), (ii) new mathematical methods to process the signals provided by such resonators (Section 4), and (iii) the miniaturization of immersed second-sound resonators enabling time- and/or space-resolved flow sensing (Section 2). These miniaturized resonators, named second-sound tweezers, have been used throughout the manuscript to demonstrate the strength and limitations of the modeling and analysis methods.

Two observations remain unexplained: the origin of some noise in the acoustic path length and second-order oscillations in the resonator spectral response in quiescent $^4$He, named the "daisy effect" (section 3.4.4). Fortunately, neither effect affects the measurement of flow vorticity or velocity.

Some unexpected results have arisen from this study, and it is worthwhile to recall and discuss them.

\begin{itemize}
\item the possibility to operate the second-sound tweezers in a non-linear mode by over-driving the standing wave beyond an intrinsic turbulent transition. In this mode, the probe becomes sensitive to velocity, which is interpreted as a signature of the local vortex tangle being swept by the outer flow (section 2.2.3). This operating mode is analogous to hot film and hot wire anemometry in classical fluid dynamics, where sensitivity to velocity is due to the more-or-less pronounced sweeping of the thermal boundary layer around an overheated thermometer.
\item In the linear regime, where the second-sound standing wave has a small amplitude, the  tweezers can be designed and operated to be predominantly sensitive to either the quantized vortices or the throughflow velocity (section 3.5). This theoretical prediction is experimentally verified by comparing the statistics obtained from turbulent flows (section 4.5). The spatial and temporal resolutions of tweezers operated as anemometers, both in nonlinear and linear modes, are comparable to or better than alternative miniaturized anemometers operating in He\:II, such as hot-wires \cite{Duri:RSI2015,diribarne2021cooling}, cantilevers \cite{SalortRSI2012,Rusaouen:intermittencyPoF2017}, Pitot tubes \cite{SalortRSI2012,Rusaouen:intermittencyPoF2017}, and total head-pressure probes \cite{Maurer1998,woillez2021vortex}.
\item In the absence of throughflow, the spectral response of the resonators and its quality factors can be accurately determined by considering only the loss due to diffraction and misalignment of the reflecting plates of the cavity. Other sources of dissipation have negligible contributions under the conditions studied here. However, in the presence of a throughflow carrying quantum vortices, the vortices can significantly contribute to the total dissipation. We have not experimentally explored the production and detection of second-sound by mechanical means, which requires a cavity with rough surfaces such as millipore or nucleopore membranes. In this case, the effect of vortices pinned on the surface may no longer be negligible. For a discussion, see \onlinecite{DHumieres:1980p448}.
\item The elliptic method allows the variations in vortex line density or velocity to be sensed without knowledge of the variations in second sound velocity (or acoustical path). More generally, the method enables a mathematical decoupling of both effects using a projection method in the (inverse) complex plane. This result is of significant practical interest in flows where the second-sound velocity is not accurately controlled due to residual temperature variations or thermal gradients. Such situations can occur, for instance, in flows sequentially driven at various levels of forcing (e.g., to explore Reynolds number dependence), in inhomogeneous dissipative flows, and in flows close to the lambda superfluid transitions where the second-sound velocities strongly depend on temperature.
\end{itemize}

Two applications of second-sound tweezers have been demonstrated. The small size of the probe enables measurements within a turbulent boundary layer (see Fig. \ref{fig:DeriveTemp}), while the high time and spatial resolution allow for time-series measurements in the bulk of quantum turbulent flow (see Fig. \ref{fig:histogramme}). The probe has potential for mapping the velocity or vorticity field of inhomogeneous flows, as demonstrated by recent  mapping of the vorticity in a counterflow jet \cite{Svancara_PoF2023}.

Another potential application of tweezers is to simultaneously measure temperature fluctuations and either velocity or vorticity, enabling exploration of their correlations in turbulent counterflows or co-flows. The tweezers thermometer can provide a direct measurement of temperature within a bandwidth ranging from zero frequency up to a fraction of the frequency of the second sound standing wave, and this signal can be multiplexed with the measurement of the second sound standing wave. Alternating measurements in the linear and non-linear modes can also be useful to locally explore both velocity and vorticity in a given flow. Another possible application is to use a double-tweezers configuration consisting of a heating plate between two thermometer plates, or vice versa. Such a stack can be used to probe joint statistics of vorticity on one side and velocity on the other, or it can be used alternatively to probe transverse gradients of either vorticity or velocity.

\section{Acknowledgements}*

We would like to express our gratitude to our colleague Benoît Chabaud for his support during the cryogenic tests and operation of the TOUPIE wind tunnel. We would also like to acknowledge and thank the staff of the PTA technological platform at the Minatec campus in Grenoble, France, and of the Nanofab technological platform at the Institut Néel. The data in Figure \ref{fig:DeriveTemp} were acquired in the SHREK facility, and we are grateful to all the members of the SHREK collaboration for their contributions, especially Michel Bon Mardion and Bernard Rousset for their specific operation near the superfluid transition. Finally, we acknowledge the contribution of A. Elbakyan for contributing to make our bibliography more complete.

The design, fabrication, cryogenic operation, and theoretical analysis of the probe spanned almost 7 years and were made possible by the following grants: ANR Ecouturb grant (ANR-16-CE30-0016), ANR QUTE-HPC grant (ANR-18-CE46-0013), and the EU Horizon 2020 Research and Innovation Program, "the European Microkelvin Platform" (EMP 824109).

\bibliographystyle{alpha}
\bibliography{mabiblio}

\appendix

\section{Micro-fabrication process\label{sec:appendix A}}

\begin{figure*}
\includegraphics[width=0.85\textwidth]{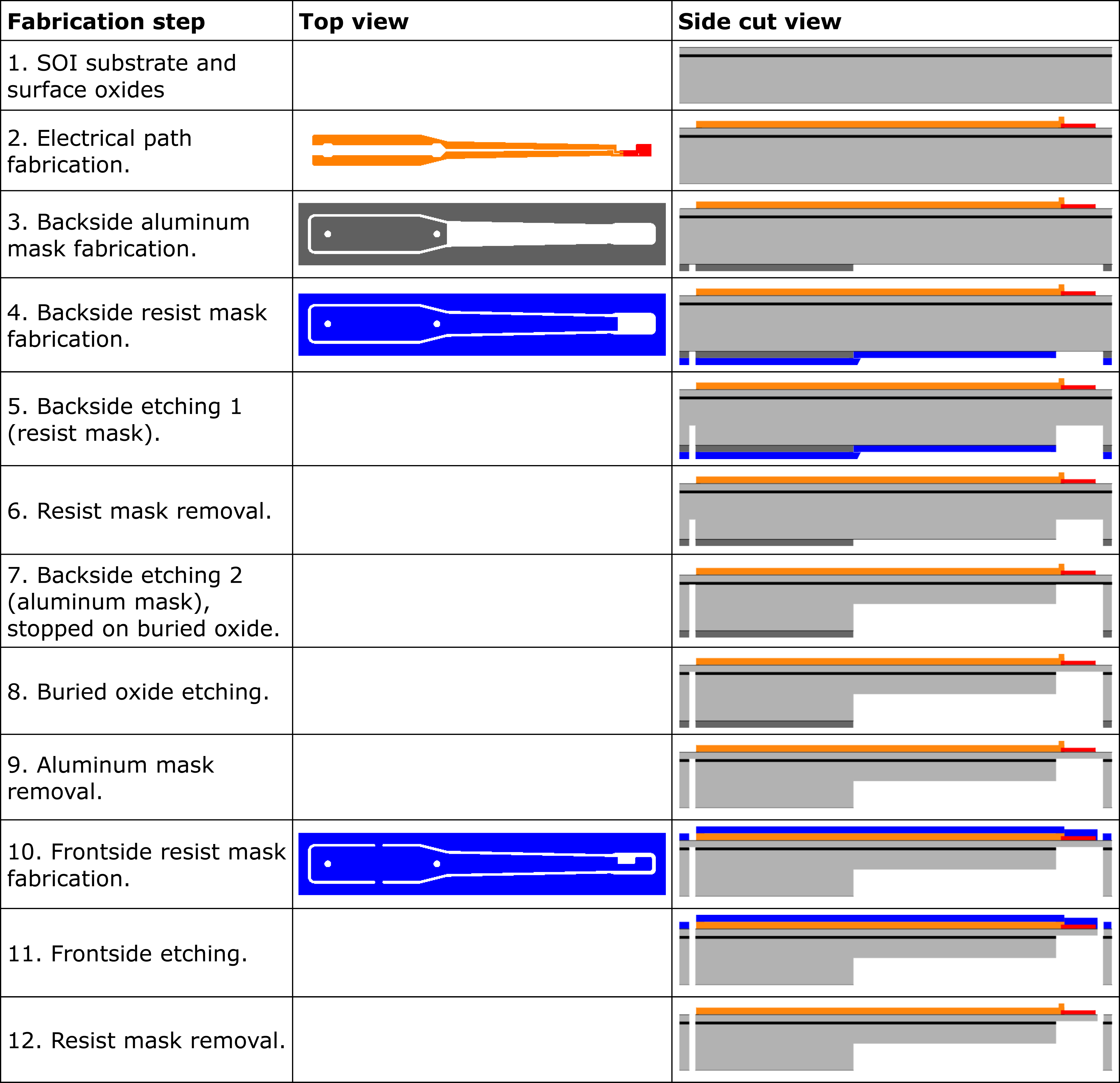}
\caption{Cantilevers' fabrication process.\label{fig:process}}
\end{figure*}

The cantilevers’ fabrication process is presented in Fig. \ref{fig:process} and summarized below.

\textbf{Step 1: Starting substrate}The cantilevers were fabricated by processing SOI (Silicon On Insulator) wafers by microelectronic techniques.
The thicknesses of the device, buried oxide and silicon substrate layers were respectively 20\:µm, 1\:µm and 500\:µm.
The wafers diameters were 100\:mm.
The wafers were double side polished and were oxidized to form a 100\:nm thick SiO$_2$ layer on both sides.

\textbf{Step 2: Electrical path fabrication}The serpentine electrical path (red color) was deposited first on the SOI wafer frontside.
The deposition was done using a standard photolithography, evaporation and lift-off sequence.
The photoresist model was AZ 5214E from Microchemicals GmbH, processed as a negative photoresist.
Depending on the cantilever type, heater or thermometer, two different evaporation sequences were used: Ti 5\:nm + Pt 80\:nm or Au 25\:nm + Sn 100\:nm.
Evaporation was preceded by in situ wafer surface cleaning by an argon ions bombardment during 20\:s.
Lift-off was initiated in an acetone bath during 5\:min and then completed by ultrasounds during a few tens of seconds.

The current leads (orange color) were deposited during a second photolithography, evaporation and lift-off sequence.
The evaporation sequence was Ti 5\:nm + Au 200\:nm + Ti 5\:nm + Pt 50\:nm.
The usage of a platinum layer was found to facilitate lift-off and may also be useful for brazing purpose.
A thin protective resist layer was deposited on frontside in order to protect it during all subsequent operations on the backside.

\textbf{Steps 3-4: Backside masks fabrication}In order to etch silicon from the backside, two superimposed etch masks were fabricated.
First, an aluminum mask was made by a photolithography, evaporation and lift-off sequence on the wafer backside.
A double-side alignment was necessary during photolithography.
The aluminum thickness was 120\:nm.
The protective resist layer on frontside had to be deposited again after lift-off.

Then, a resist mask was deposited over the aluminum one.
This mask was made by photolithography on the positive AZ4562 photoresist spin-coated at 4000\:rpm.

The aluminum mask was identical to the resist mask except within the arm area, as shown in Fig. \ref{fig:process}.
This area was covered by resist, but not by aluminum.

\textbf{Step 5: Backside etching 1}The etch mask during this phase was the resist mask (see step 4 in Fig. \ref{fig:process}).
The area exposed to etching includes the tip area and a 200\:µm wide trench which delimits the piece contour.
First, the thin surface oxide layer was etched by reactive ion etching (RIE) based on SF$_6$ gas.
Then, the silicon was etched in a STS HRM deep reactive ion etching (DRIE) equipment using a standard recipe based on Bosch process.
The etch rate was around 10\:µm/min, the etched depth was around 230\:µm (120 cycles).
The longer this phase, the thicker the cantilever arm at the end of the process.

\textbf{Steps 6-7: Backside etching 2}Following the first silicon etch phase, the resist mask was removed by an oxygen plasma, thus uncovering the arm area and the remaining aluminum etch mask.
The area exposed to etching includes the previous one plus the arm area.
The backside surface oxide was etched into the arm area.
Then, another silicon etch sequence was applied with the aluminum mask.
This sequence was stopped when the buried oxide was reached everywhere in the tip area and in the cantilever contour (NOT in the arm area).
The same recipe was used, 200 cycles were applied.

An additional difficulty appeared at this step due to a parasitic effect during etching. This effect originates from the polymer passivation layer deposited on sidewalls during the first silicon etch phase.
As the arm area was masked during the first silicon etch sequence and unmasked during the second one, the passivation layer located along the arm edges was released and generated locally some irregular micromasking effect.
In order to decrease the micromasking, the etch recipe was interrupted 3 times, every 50 cycles, in order to apply a 1\:min oxygen plasma followed by 20\:s of an isotropic silicon etching recipe.

The final 50 etching cycles ended up reaching the buried oxide layer, at the trench bottom, all around the cantilever.
It is necessary at this step to check that the buried oxide is fully uncovered by silicon everywhere in the trench bottom and in the tip area.
However, etching cycles should not be applied in excess to avoid some mechanical weakening of the wafer.

\begin{table}
\begin{tabular}{|c|c|c|c|c|}
\hline 
Phase & Deposition & \multicolumn{3}{c|}{Etch} \tabularnewline
\hline 
Sub-phase & Main & Delay & Boost & Main\tabularnewline
\hline 
\hline 
Gas & C$_4$F$_8$: 250 sccm & 
\multicolumn{3}{c|}{
\begin{tabular}{@{}lc@{}}SF$_6$: 250 sccm \\ O$_2$: 10 sccm \end{tabular}
}
\tabularnewline
\hline 
Duration & 3 s & 2.0 s & 5.5 s & 22.5 s\tabularnewline
\hline 
Pressure & 14 mTor & 20 mTor & 40 mTor & 40 mTor\tabularnewline
\hline 
Coil power & 300 W & 300 W & 300 W & 1 W\tabularnewline
\hline 
Platen power & 20 W & 100 W & 50 W & 1 W\tabularnewline
\hline 
Electroma-  &  &  &  & \tabularnewline
gnet current & 0 A & 0 A & 2 A & 0 A\tabularnewline
\hline 
Platen \\ frequency & \multicolumn{4}{c|}{RF 13.56 MHz}  \tabularnewline
\hline 
He backside \\pressure & \multicolumn{4}{c|}{10 Tor}\tabularnewline
\hline 
\end{tabular}
\caption{Bosch process recipe used during frontside deep silicon etching.}
\label{jerome7}
\end{table}

\textbf{Steps 8-9: Buried oxide etching}The buried oxide was then removed by plasma etching.
This oxide was fully removed at the trench bottom and in the tip area.
If this layer is not removed, some bending may occur on the tip at the end of the process, due to mechanical stress of oxide.
The aluminum mask was removed in an aluminum etchant solution at 50 °C during a few minutes.

\textbf{Steps 10-12: Frontside etching}The frontside protective resist layer was removed during an acetone cleaning bath.
The 3D cantilever structuration was ended by a third silicon etching made from the frontside.
The etch mask was formed by photolithography on AZ1512HS resist, deposited at 4000\:rpm.
As shown on Fig. \ref{fig:process}, the mask design includes two bridges on the baseplate sides in order to maintain the cantilever after having opened the trench that surrounds it.
The design also includes the tip contour.
The frontside thin surface oxide was etched first, then the silicon of the device layer.
The silicon etching was done with specific conditions.
Due to the wafer mechanical weakness at this step, caused by the multiple deep trenches made on the backside, the processed wafer was attached to a blank silicon wafer by Kapton tape.
This ensemble was loaded into the etching chamber.
As no thermal bridge was present between the two wafers, the recipe was adapted: low RF powers were used and a 22 s idle time was added after each etching cycle (see table \ref{jerome7}).
The objective was to avoid overheating during etching.
The silicon etch duration was 50 cycles.
After this sequence, the trench around the cantilever was fully opened.
The resist was removed by a low power oxygen plasma.

\end{document}